\newif\ifsubmit 
\newif\ifconference

\submitfalse 

\conferencefalse 

\ifconference
    \documentclass{llncs}
\else
\documentclass[11pt]{article} 
\fi

\usepackage[utf8]{inputenc}
\ifconference\else\usepackage[pdftex,left=1in,top=1in,bottom=1in,right=1in]{geometry}\fi
\usepackage{amsmath, amssymb, dsfont, mathrsfs}
\ifconference \else \usepackage{amsthm} \fi
\usepackage{braket}
\usepackage{graphicx}
\usepackage{caption}
\usepackage{bm}
\usepackage{xcolor}
\usepackage[framemethod=tikz]{mdframed} 
\usepackage[colorlinks=true,citecolor=blue,linkcolor=blue]{hyperref}
\usepackage[nameinlink, noabbrev, capitalize]{cleveref}
\usepackage{physics}
\usepackage{verbatim} 
\usepackage{enumitem}
\usepackage{mathtools}
\hypersetup{
    colorlinks,
    linkcolor={red!50!black},
    citecolor={blue!50!black},
    urlcolor={blue!80!black}
}

\ifconference
\else
\usepackage{tocloft}

\fi

\definecolor{darkgreen}{rgb}{0,0.5,0}
\definecolor{darkblue}{rgb}{0,0,0.6}

\allowdisplaybreaks

\usepackage{tabularx}
\newcommand{\addrefs}{\bibliography{0include/bibs/abbrev0,0include/bibs/crypto,0this-project/refs}}

\newcommand{\sch}[1]{\mathsf{#1}}
\newcommand{\gam}[1]{\mathsf{#1}}

\newcommand{\samp}{\leftarrow} 

\newcommand{\C}{\mathds{C}}

\newcommand{\BN}{\mathbb{N}}
\newcommand{\nat}{\BN}

\newcommand{\F}{\mathds{F}}

\newcommand{\R}{\mathbb{R}}
\newcommand{\E}{\mathds{E}}

\newcommand{\zo}{\{0,1\}}
\newcommand{\poly}{\mathsf{poly}}

\newcommand{\hyb}{\mathsf{Hyb}}

\newcommand{\messpa}{\mathcal{M}} 
\newcommand{\outtrue}{1}
\newcommand{\outfalse}{0}
\newcommand{\ora}{\mathcal{O}}

\newcommand{\pke}{\sch{PKE}}
\newcommand{\digsig}{\sch{SIG}}
\newcommand{\io}{i\mathcal{O}}
\newcommand{\prf}{\sch{PRF}}

\newcommand{\setup}{\sch{Setup}}
\newcommand{\ver}{\sch{Ver}}
\newcommand{\ceval}{\sch{Eval}}
\newcommand{\gen}{\sch{Gen}}

\newcommand{\genqkey}{\sch{GenQKey}}
\newcommand{\sign}{\sch{Sign}}

\newcommand{\negl}{\mathsf{negl}}
\newcommand{\adve}{\mathcal{A}}

\newcommand{\lrpkegame}[1]{\gam{LR-PKE-GAME}_{#1}}
\newcommand{\lrsiggame}[1]{\gam{LR-SIGN-GAME}_{#1}}
\newcommand{\lrgame}[1]{\gam{LEAK-GAME}_{#1}}
\newcommand{\learninggame}[1]{\gam{LEARNING-GAME}_{#1}}

\newcommand{\qprot}{\sch{QuantumProtection}}
\newcommand{\prot}{\sch{Protect}}
\newcommand{\ptriv}{p_{triv}}

\newcommand{\oss}{\mathsf{OSS}}
\newcommand{\qfiremini}{\mathsf{QFireMini}}
\newcommand{\qfire}{\mathsf{QFire}}
\newcommand{\qkeyfire}{\mathsf{QKeyFire}}
\newcommand{\spark}{\mathsf{Spark}}
\newcommand{\clone}{\mathsf{Clone}}

\newcommand{\qfireminigame}[1]{\gam{MINI-TELEGRAPH-GAME}_{#1}}
\newcommand{\qfiregame}[1]{\gam{TELEGRAPH-GAME}_{#1}}

\newcommand{\regi}{\mathsf{R}}
\newcommand{\regis}[1]{\regi_{\mathsf{#1}}}

\ifconference
\else
\newtheorem{theorem}{Theorem}
 
 \newtheorem{claim}{Claim}
 \newtheorem{lemma}{Lemma}

 \newtheorem{definition}{Definition}
 
 \newtheorem{remark}{Remark}
 
\fi

\newcommand{\qkeyfiresign}{\mathsf{QKeyFireSign}}
\newcommand{\qfireinteractivemini}[1]{\mathsf{MINI-INTERACTIVE-TELEGRAPH-GAME}_{#1}}

\newcommand{\statdist}[2]{\left|{#1} - {#2}\right|}

\newcommand{\setuporacles}{\mathsf{GenOracles}}

\newcommand{\csrd}{C_\mathsf{SRD}}

\newcommand{\oraimgver}{\ora_{\mathsf{ImageVer}}}

\newcommand{\adhocadv}{\mathcal{P}}

\title{Public-Key Quantum Fire and Key-Fire From Classical Oracles}

\date{}
\ifconference
\author{Anonymous Submission}
\else
\author{Alper \c{C}akan\thanks{Carnegie Mellon University. \texttt{alpercakan98@gmail.com}.} \and Vipul Goyal\thanks{NTT Research \& Carnegie Mellon University.  \texttt{vipul@vipulgoyal.org}} \and Omri Shmueli\thanks{NTT Research. \texttt{omri.shmueli1@gmail.com}.}}
\fi

\ifsubmit
\newcommand{\todo}[1]{}
\newcommand{\alper}[1]{}
\newcommand{\vipul}[1]{}
\newcommand{\omri}[1]{}
\else
\newcommand{\todo}[1]{{\color{blue} \footnotesize(TODO: \uppercase{#1})}}
\newcommand{\alper}[1]{{\color{blue} \footnotesize(ALPER: \uppercase{#1})}}
\newcommand{\vipul}[1]{{\color{blue} \footnotesize(VIPUL: \uppercase{#1})}}
\newcommand{\omri}[1]{{\color{blue} \footnotesize(OMRI: \uppercase{#1})}}
\fi

\begin{document}
\maketitle

\ifconference
\input{112CONFERENCE}
\else
\vspace{-.5em}
\begin{abstract}
  Quantum fire is a distribution of quantum states that can be efficiently cloned, but cannot be efficiently converted into a classical string. First considered by Nehoran and Zhandry (ITCS'24) and later formalized by Bostanci, Nehoran, Zhandry (STOC'25), quantum fire has strong applications and implications in cryptography, along with important connections to physics and complexity. However, constructing and proving the security of quantum fire so far has been elusive. Nehoran and Zhandry showed how to construct quantum fire relative to an \emph{inefficient quantum oracle} (which cannot be instantiated even heuristically). Later, Bostanci, Nehoran, Zhandry gave a candidate construction based on group actions, however, even in the classical oracle model they could only conjecture the security of their scheme, and were not able to give a security proof or argue security.

  In this work, for the first time, we give a construction of public-key quantum fire relative to a classical oracle and prove its security unconditionally. Going further, we introduce two stronger notions that generalize quantum fire, and we give secure constructions for these notions.
  \begin{itemize}
  \item We introduce a notion called \emph{quantum key-fire} where the clonable fire states serve as \emph{keys}: They can be used to evaluate a functionality (such as a signing or decryption key), and for security we require \emph{unbounded leakage-resilience}, which means that given the fire state, the key cannot be efficiently converted into a classical string.
  
  \item We consider the notion of \emph{interactive (i.e. LOCC) security} for quantum (key-)fire. In this setting, instead of trying to convert the state or key into a classical string; the adversary, given the flame state, attempts to transfer it to another adversary interactively over a classical channel.
  \end{itemize}

   We give a construction of quantum \emph{key-fire} relative to a classical oracle and unconditionally prove that it satisfies interactive security for any unlearnable functionality. As a result, we also obtain the \emph{first classical oracle separations} between various notions in physics and cryptography:
   \begin{itemize}
 \item A separation in the computational universe between two fundamental principles of quantum mechanics: No-cloning and no-teleportation, which are equivalent in the information-theoretic setting.
       \item A separation between \emph{copy-protection security} (Aaronson, CCC'09) and \emph{unbounded/LOCC leakage-resilience security} (\c{C}akan, Goyal, Liu-Zhang, Ribeiro, TCC'24).
    \item A separation between \emph{computational no-cloning security} and \emph{computational no-learning security}, two notions introduced recently by Fefferman, Ghosh, Sinha, Yuen (ITCS'26).
   \end{itemize}
  
  In all our constructions, the oracles can be implemented efficiently using one-way functions. Thus, our schemes can be heuristically instantiated in the plain model using one-way functions and indistinguishability obfuscation, which gives us the first constructions in the plain model with concrete evidence for their security.

\end{abstract}

\newpage
\tableofcontents
\newpage

\section{Introduction}
Possibly the two most well-known basic principles of quantum mechanics are the so-called \emph{no-cloning} and the \emph{no-teleportation} principles. The first principle says that no physical process can create two copies of an unknown quantum state when given a single copy, and the second principle (which we will refer to as the \emph{no-telegraphing principle}\footnote{We follow this naming convention that \cite{ITCS:NehZha24} introduced to avoid confusion with \emph{quantum teleportation}}) says that no physical process, given a single copy of a quantum state, can convert it into a classical string from which the state can be reconstructed.

 Apart from being fundamental rules of the quantum universe we live in, these principles have also found a tremendous amount of applications in quantum cryptography. The no-cloning principle has had profound applications, from being utilized to construct \emph{quantum money} in the very first paper of \cite{Wie83} that started the field of quantum cryptography, to fundamental applications like quantum key-distribution and even more modern applications such as \emph{copy-protection (anti-piracy security) for software} (\cite{aaronson2009quantum,C:CLLZ21,cryptoeprint:2025/1197}). The no-telegraphing principle has also found important applications in cryptography, such as the notion of \emph{unbounded/LOCC leakage-resilience} (\cite{TCC:CGLR24,ITCS:NehZha24}).

It is well-known in quantum information theory that the no-cloning and the no-telegraphing principles are equivalent: If one can telegraph a state then they can also clone it, and if one can clone a state they can also telegraph it by performing state tomography after creating a sufficient number of copies of the state. One caveat is that state tomography necessarily requires exponentially many copies in the number of qubits (\cite{flammia2012quantum,o2016efficient}), and the resulting classical string (describing the state) will also be of exponential size. Thus, the equivalency between the two principles relies on an inefficient physical process. However, it is one of the fundamental hypotheses of quantum complexity theory that any physical process we can observe in real-life is \emph{quantum polynomial time}. Thus, one might ask the following  question: Are these two principles of quantum mechanics equivalent through physically-observable processes (or, \emph{reasonable-size experiments} in the terminology of \cite{PhysRevLett.117.120501})? Indeed, Nehoran and Zhandry (\cite{ITCS:NehZha24}, ITCS'24) precisely asked this question and showed that the answer is likely negative, by showing a separation relative to an inefficient quantum oracle. While this is a step towards answering the question, unfortunately, quantum oracle separations might be misleading (\cite{agarwal2025}) and in fact it is even unclear how exactly to model them (\cite{C:Zhandry25}). Later, Bostanci, Nehoran, Zhandry (\cite{STOC:BosNehZha25}, STOC'25) gave a candidate classical oracle separation based on \emph{cryptographic group actions}; however, they were not able to give a proof or argue security. Thus, what we know so far about the relationship between these two principles in a computational universe is unsatisfactory, and thus we ask the following question:

\begin{center}
\textit{
    Can we show a separation between the no-cloning and the no-telegraphing principles in a computational universe?
}
\end{center}
Note that this question has fundamental implications for our understanding of quantum states, as well as to cryptography and complexity (as we explain in detail in the subsequent sections). It is widely believed that quantum states are more \emph{powerful} than classical strings (e.g. the well-known $\mathsf{QMA}$ vs $\mathsf{QCMA}$ question). However, what exactly is the fundamental property of quantum states that makes them more powerful than classical strings? Is it unclonability? If we show a clonable state that satisfies the no-telegraphing principle, then the answer is negative: Even clonable quantum states are more powerful then classical strings.

\paragraph{Cryptographic Applications.}
\cite{STOC:BosNehZha25} defined a notion called \emph{quantum fire}, which can be thought of a cryptographic version of the aforementioned separation\footnote{Indeed, quantum fire is stronger than the separation, as it immediately implies the separation.}. In more detail, quantum fire is defined to be a distribution of quantum states (called flame states) which can be efficiently cloned. For security, we require \emph{untelegraphability}: No efficient adversary who is given a flame state can convert it into a classical string from which the state can be reconstructed. \cite{STOC:BosNehZha25} mention possible applications of quantum fire, such as storing some information in a quantum encoding (i.e. as a flame state) on a server to protect it from \emph{exfiltration} by adversaries. However, the state of affairs for existence of quantum fire is unsatisfactory: Just like above, the only known constructions are the inefficient quantum oracle based construction of  \cite{ITCS:NehZha24} and the candidate construction  of \cite{STOC:BosNehZha25} without a proof of security. Thus, we ask the following question.

\begin{center}
\textit{
    Can we construct quantum fire with provable security?
}
\end{center}

Given the myriad of applications of the no-cloning and the no-telegraphing notions in cryptography, we also ask whether a separation between them also has applications in cryptography, beyond quantum fire.
\begin{center}
\textit{
    Does a separation between no-cloning and no-telegraphing have applications in cryptography? 
}
\end{center}

\subsection{Our Results}
In this work, we answer the above questions in the classical oracle model.

\paragraph{Quantum Fire} We give a construction of quantum fire relative to a quantumly-accessible\footnote{Throughout the paper, we will write classical oracle for brevity, and we always mean quantumly-accessible classical oracle.} classical oracle and prove its correctness and security unconditionally, for the first time.
\begin{theorem}[Informal]
     Relative to a classical oracle, there exists a public-key quantum fire scheme with untelegraphability security.
 \end{theorem}

This also gives the first classical oracle separation between no-cloning and no-telegraphing and resolves the open questions posed by \cite{ITCS:NehZha24} and \cite{STOC:BosNehZha25}. We note that our construction is completely different from previous works.

Our scheme can be implemented efficiently using one-way functions, and thus can be instantiated heuristically using indistinguishability obfuscation to obtain the first quantum fire scheme in the plain model with concrete evidence for its security.
 
This significantly improves upon existing results. As discussed above, previously \cite{ITCS:NehZha24} gave a construction of quantum fire relative to an inefficient unitary quantum oracle (which bakes cloning into the oracle itself and thus it is inherently inefficient), and \cite{STOC:BosNehZha25} gave a \emph{candidate} construction based on cryptographic group actions. However, even in the classical oracle model, \cite{STOC:BosNehZha25} only conjectured security and were not able to prove security. In fact, they even claimed that any quantum fire scheme with proof of security in the classical oracle model would likely lead to a classical oracle separation between $\mathsf{QMA}$ and $\mathsf{QCMA}$\footnote{We note that our construction does not yield a separation between $\mathsf{QMA}$ and $\mathsf{QCMA}$. However, it is an interesting question if our techniques could be useful.}, which is one of the most biggest questions in quantum complexity that has been open for almost 20 years\footnote{Subsequent to our work, a new work by Bostanci, Haferkamp, Nirkhe, Zhandry (STOC'26) solved this problem via a different approach.}. 

\paragraph{Stronger Notions} Going further, we introduce two stronger notions that generalize quantum fire. 
\begin{itemize}
    \item
    First, we introduce \textbf{quantum key-fire}, where the fire states encode a key that can be used to evaluate a functionality (e.g. a signing key, or a decryption key, or a software). For quantum key-fire, as the security notion, instead of untelegraphability, we require the stronger guarantee of \textbf{unbounded leakage-resilience} (\cite{TCC:CGLR24}). Unbounded leakage-resilience is a strengthening of no-telegraphing where not only that the flame state cannot be encoded into a classical string, we further require that no efficiently generated classical string (called \emph{leakage} in this context) can help evaluate the functionality or win a security game related to the functionality. In other words, quantum key-fire is to quantum fire what quantum copy-protection (\cite{aaronson2009quantum,C:ALLZZ21,C:CLLZ21}) is to public-key quantum money, and quantum-key fire implies quantum fire in a straightforward way.
    
    \item
    As the second generalization, we consider the notion of \textbf{interactive security}. Previous work only considered the single message untelegraphability security game where the adversary $\adve_1$ who has the flame state (the sender adversary) produces a classical string, using which the second adversary $\adve_2$ (the receiver adversary) tries to reconstruct the flame state. We consider the case where the sender and the receiver interact over a classical channel for any (polynomial) number of rounds. We call this notion \emph{interactive untelegraphability security} in the case of quantum fire, and \emph{local operations and classical communication (LOCC) leakage-resilience} (\cite{TCC:CGLR24}) in the case of quantum key-fire.
\end{itemize}

To recap, we have four security notions for quantum fire and key-fire: \textbf{Untelegraphability, interactive untelegraphability, unbounded leakage-resilience and LOCC leakage-resilience} - a summary is given in \cref{tablecompare}. Note that LOCC leakage-resilience implies interactive untelegraphability, and unbounded leakage-resilience implies untelegraphability.

\begin{table}[h]
\centering
\renewcommand{\arraystretch}{1.2}
\begin{tabularx}{\linewidth}{|X|X|X|}
\hline
\textbf{Security Notion} & 
\textbf{Interaction Between $\adve_1$ and $\adve_2$} & 
\textbf{Goal of $\adve_2$} \\
\hline

Untelegraphability 
& No
& Reconstruct the exact same state \\

\hline

Interactive Untelegraphability 
& Arbitrary classical interaction 
& Reconstruct the exact same state \\

\hline

Unbounded Leakage Resilience 
& No
& Solve a challenge (e.g. decrypt a challenge ciphertext) \\

\hline

LOCC Leakage Resilience (the strongest notion)
& Arbitrary classical interaction 
& Solve a challenge (e.g. decrypt a challenge ciphertext) \\

\hline
\end{tabularx}
\caption{Comparison of security notions for quantum fire and key-fire.}\label{tablecompare}
\end{table}

\paragraph{Quantum Key-Fire}We give a construction of quantum key-fire relative to a classical oracle and prove that it satisfies LOCC leakage-resilience for all unlearnable functionalities. 
\begin{theorem}[Informal]
     Relative to a classical oracle, there exists a quantum key-fire scheme that satisfies LOCC leakage-resilience for any unlearnable functionality.
 \end{theorem}
 We note that this is the best possible result since even in the oracle model, unbounded/LOCC leakage-resilience for learnable functionalities is trivially impossible. 
 
 As above, our scheme can be instantiated efficiently using one-way functions and indistinguishability obfuscation in the plain model with heuristic security.
 
Previously, even in the quantum oracle model, only provably secure construction of quantum fire \cite{ITCS:NehZha24} had proof of security in the significantly weaker setting of single-message untelegraphability and there was no proof of even 2-message interactive security.

\paragraph{Size of the Leakage} Finally, we note that, unlike previous work, we impose no restriction on the length of the messages between the adversaries (e.g. they can be doubly exponential in the security parameter, or more): The only requirement for security is that the adversaries make polynomially many queries to the oracles.

\paragraph{Applications} As we will discuss later in more detail (see \cref{sec:implications}), quantum key-fire is a strong primitive in the setting of \emph{leakage-resilient cryptography}. It retains the security of the functionality (e.g. a decryption key) even in the face of \emph{unbounded} leakages, which is impossible with classical keys, but at the same time (assuming one has a quantum computer) it is almost as practical as a classical key due to its clonability: It can be later on used in different locations/servers, can be kept alive for long periods by frequently refreshing/cloning it (almost like a classical dynamic RAM) and so on. 

\paragraph{Implications} As we discuss in \cref{sec:implications}, our results have interesting implications in cryptography, physics, and complexity, and we believe it also leads the way for a myriad of interesting new research directions (such as \emph{clonable quantum cryptography}) and questions.

As an example of cryptographic implications, our results imply a separation between the recently introduced cryptographic assumptions of \emph{computational no-cloning assumption} and \emph{computational no-learning assumption}, thus answering an open question of \cite{fefferman2025}. Our results also imply a separation between the security notions of \emph{copy-protection security} and \emph{unbounded/LOCC leakage-resilience security}, answering the open questions of \cite{ITCS:NehZha24} and \cite{TCC:CGLR24}. As an example of physical implications, our result shows that the fundamental property that makes quantum states more \emph{powerful} than classical strings is not unclonability (since even clonable states cannot be converted into classical strings).

\paragraph{Technical Contributions} Our work introduces new techniques in the oracle model for analyzing two quantum adversaries that are communicating interactively for arbitrary number of rounds over a classical channel (i.e. LOCC adversaries), which are in general hard to analyze due to internal unclonable states of the adversaries. Through our new techniques, we also draw interesting connections between LOCC adversaries and a natural property called \emph{incompressibility} of oracles that says that not \emph{too many outputs} of the oracle can be packed inside a classical string (see \cref{sec:incompross,sec:incomprrom} for the formal definitions). We also prove that our incompressibility notion is satisfied for two particular cases: for random oracles and for a modified version of the one-shot signature scheme of \cite{C:ShmZha25}.

We believe our new techniques and our new incompressibility results might be helpful for other problems in similar settings.

\begin{theorem}[Informal]
     Random oracles satisfy classical-message incompressibility.
 \end{theorem}
\begin{theorem}[Informal]
     Relative to a classical oracle, there exists a one-shot signature scheme with incompressibility.
 \end{theorem}

\subsection{Applications, Implications and Future Directions}\label{sec:implications}
In this section, we will discuss the cryptographic applications (e.g. constructions) of our results, and then discuss the implications of our results (e.g. separations we obtain) and finally discuss some new research directions motivated by our results.

We first recall the notion of leakage-resilience.

\paragraph{Primer on Leakage-Resilience} Storing secret keys (e.g. of an encryption scheme or a signature scheme) is one of the oldest and most fundamental problems in cryptography. Leakage-resilient cryptography is a fruitful research area in cryptography that has received significant interest over decades that aims to deal with this problem: It aims to construct schemes which are secure in face of attacks where adversary obtains \emph{leakage} on the honest parties secret keys, through \emph{side-channels} \cite{Koc96,QS01,AARR02}. Side-channels are generally various measurements of the system storing the keys, such as the electromagnetic radiation emitted, or the power consumed during computation.

In classical leakage-resilient cryptography, the leakage output is modeled as as function $f(sk)$ of the secret key $sk$ where $f$ has an output size significantly smaller than the size of $sk$. This is a somewhat artificial restriction that is nevertheless ubiquitous, since otherwise there is no possibility of obtaining security guarantees: the adversary can simply obtain the whole secret key.

The recent work of \cite{TCC:CGLR24} has introduced\footnote{Concurrent  to \cite{TCC:CGLR24}, \cite{ITCS:NehZha24} introduced \emph{exfiltration security}, which is essentially equivalent to non-interactive unbounded leakage-resilience. They did not give a plain model construction from standard cryptographic assumptions.} the notion of \emph{unbounded leakage-resilience}, where the secret key is stored as a quantum key, and the leakage measurement output is allowed to be of any unbounded size\footnote{The only requirements is that leakage is a measurement, i.e., it has a classical output. However, measurements accurately model side-channel attacks and in fact capture all known attacks.}, in particular, it can be much larger than the size of the secret key! They also introduced the notion of \emph{LOCC leakage-resilience} where the adversary adaptively obtains unbounded leakage over any number of rounds. They showed, through a new proof, that the unclonable\footnote{That is, the decryption/signing keys of the schemes satisfy \emph{copy-protection security}} public-key encryption and signing schemes of \cite{C:CLLZ21} also satisfy LOCC leakage-resilience, thus giving a plain model construction assuming indistinguishability obfuscation (iO) and one-way functions.

\paragraph{Application: Unbounded/LOCC Leakage-Resilience with Clonable Keys} The notion of \emph{unbounded/LOCC leakage-resilience} is a strong (and uniquely quantum) security guarantee with important practical ramifications. However, a major issue with the existing scheme of \cite{TCC:CGLR24} is that while the secret quantum key satisfies unbounded/LOCC leakage resilience, the key is also unclonable. This severely limits the practicality of the scheme, since if we ever need to use the decryption key in multiple systems, this will not be possible\footnote{Naturally, we cannot store a classical description of the key either, since that would defeat the whole purpose and we would lose the unbounded leakage-resilience guarantee.}. Further, while there have been significant advances in quantum computation in recent years, quantum memory implementations can still keep states coherent for very short amounts of time.

However, a quantum key-fire scheme gives us the best of both worlds: While the keys satisfy LOCC leakage-resilience (which is impossible classically), they are also efficiently clonable, which makes it almost as useful as a classical key! For example, the key can be used in different computers too (e.g. to parallelize big tasks, or if these are the decryption keys for a company and a new employee joins). Further, they can be frequently cloned in the quantum memory to \emph{refresh} them to keep them alive for much longer and prevent decoherence, much like a classical \emph{dynamic RAM} that are widely used in practice. Thus, we believe quantum key-fire  will have important practical ramifications in the future.

\paragraph{Separating Leakage-Resilience and Copy-Protection} Related to above discussion, our quantum key-fire result also gives a separation between the security notions of unbounded/LOCC leakage-resilience (\cite{TCC:CGLR24}) and copy-protection (\cite{aaronson2009quantum}), answering an open question of \cite{TCC:CGLR24,ITCS:NehZha24}. Note that so far, only known unbounded/LOCC leakage-resilient schemes also satisfy copy-protection security.  Copy-protected schemes are known to imply public-key quantum money (PKQM), and constructing PKQM without obfuscation is one of the biggest problems in quantum cryptography. However, our work for the first time shows that leakage-resilience does not imply copy-protection/unclonability, which means  there might be a hope of achieving unbounded/LOCC leakage-resilience without obfuscation. On the (slightly) negative side, our separation shows that we cannot simply prove unbounded/LOCC leakage-resilience of a scheme and be content that it also satisfies copy-protection security: when the goal is to obtain copy-protection security, we need to prove it separately.

\paragraph{Clonable Cryptography and Separations in MicroCrypt} MicroCrypt is the study of existence of quantum cryptographic primitives (such as quantum commitments, one-way quantum-state generators etc) even in a world where $\mathsf{P} = \mathsf{NP}$, and it has gained significant attention in the recent years. Recent work of \cite{fefferman2025} introduced two new quantum cryptographic assumptions for MicroCrypt: the computational no-learning assumption and
the computational no-cloning assumption, and they show how to build various MicroCrypt primitives
from these assumptions. They also conjecture that the computational no-learning assumption is potentially
weaker (i.e., it is a more conservative assumption). Our work shows that relative to a classical oracle, there exists a family of circuits for which the computational no-learning holds, but the
computational no-cloning assumption does not. This resolves their open question. We believe our techniques might also help separate various MicroCrypt primitives, such as one-way state generators and unclonable state generators.

We believe our work will also lead the way for a new research direction: \emph{clonable quantum cryptography}. Our work shows that even clonable states are more powerful than classical strings. Thus, \textbf{in any context where the purpose of using quantum states is not unclonability itself, it might be possible to obtain a clonable version of that quantum primitive, which would make it more practical} (for example, the state can be kept alive for longer by refreshing it through cloning it, much like a classical dynamic RAM). For example, there are various works \cite{countcrypt,C:MorYam22,C:KMNY24} that show that various primitives (such as public-key encryption, signatures) can exist with quantum keys or quantum ciphertexts/signatures even if their classical-string counterparts do not exist. However, all such candidates so far have unclonable keys/signatures/ciphertexts, which significantly limits their practicality. Thus, it is an interesting question if one can show existence of clonable versions of these primitives in a world their classical counterpart do not exist, and we believe our techniques might help.

\paragraph{Physics in a Computational Universe} In quantum mechanics, various fundamental principles called \emph{no-go theorems} exist, such as the \emph{no-cloning theorem}, \emph{no-hiding theorem}, \emph{no-deleting theorem}, \emph{no-telegraphing theorem} and so on. Most of these no-go theorems have been shown to be equivalent to each other in the information-theoretic setting. However, our result gives the first (classical oracle) separation between the \emph{no-go principles} of quantum mechanics. Our result means that, relative to a classical oracle, no-telegraphing does not imply no-cloning. Our result means that the equivalence of the fundamental no-go principles of quantum mechanics that we take for granted might not hold in a computational universe. An interesting question is if other principles of quantum mechanics can be separated in a computational universe.

Another interpretation of our result is as follows. In an information-theoretic setting, the main difference between a (local) quantum state/memory and a classical string is the unclonability of the quantum state, and clonable states can simply be turned into classical strings: We can clone it and apply a state tomography to obtain a classical description of it, and then work with this classical string. However, our work shows that in a computational world, even clonable states are different from classical memory, thus the fundamental property that distinguishes quantum states from classical memory is not unclonability.

\paragraph{Complexity} Separating the complexity classes $\mathsf{QMA}$ and $\mathsf{QCMA}$ relative to a classical oracle is one of the most important questions in quantum cryptography and has been open since 2007.
Due to the connection established by \cite{ITCS:NehZha24} between quantum fire and separating $\mathsf{clonableQMA}$ from $\mathsf{QCMA}$ (which also separates $\mathsf{QMA}$ and $\mathsf{QCMA}$), \cite{STOC:BosNehZha25} claimed that a quantum fire scheme with proof of security in the classical oracle model would likely lead to a classical oracle separation between $\mathsf{QMA}$ and $\mathsf{QCMA}$, solving this major open question. While our construction does not lead to such a separation, it is an interesting question if our techniques could help\footnote{Subsequent to initial posting of our work, Bostanci, Haferkamp, Nirkhe, Zhandry (STOC'26) solved this problem via a different approach, though the core philosophical principle under their proof is similar to ours: Classical leakage/witnesses are by default clonable for free and reusable, so we can run an adversary many times with it, unlike quantum states.}.

\subsection{Organization}\label{sec:organization}
In \cref{sec:overview}, we give a technical overview of the paper.

In \cref{sec:prot}, we recall some prior definitions, such as quantum protection schemes and LOCC leakage-resilience security, that are relevant to defining and constructing quantum fire. In \cref{sec:defnwork}, we give the quantum fire and quantum key-fire definitions.

In \cref{sec:incomprrom}, we prove an incompressibility result for random oracles. In \cref{sec:incompross}, we define and construct incompressible one-shot signatures.

In \cref{sec:cons}, we give our main construction: quantum key-fire for signing. Then, we prove its verification and cloning correctness.

In \cref{sec:proofsec}, we prove the security of our main construction, except for one major hybrid, which we prove in \cref{sec:bighybrid} for ease of reading.
\section*{Acknowledgments}
A\c{C} thanks Barak Nehoran for answering questions about their work (\cite{STOC:BosNehZha25}), Christian Majenz for the correct citation for the gentle measurement lemma (\cite{awinter}) and Dakshita Khurana for helpful discussions. We thank Vinod Vaikuntanathan and Andrew Huang for helpful comments.

A\c{C} was supported by the following grants of Vipul Goyal: NSF award 1916939, DARPA SIEVE program, a gift from Ripple, a DoE NETL award, a JP Morgan Faculty Fellowship, a PNC center for financial services innovation award, and a Cylab seed funding award.  This work was done in part while OS was a research fellow at the Simons Institute for the Theory of Computing at UC Berkeley.

\newpage
\section{Technical Overview}\label{sec:overview}
In this section, we give an overview of our work, and the end of the of the section we discuss some related work. Formal definitions, constructions and proofs can be found in relevant sections (see \cref{sec:organization} for the organization). Note that for most of the overview, for pedagogical reasons, we will consider the notion of quantum fire mini-schemes with untelegraphability security. However, in the main body of the paper, we instead construct quantum key-fire and prove LOCC leakage-resilience security, which is stronger.

\paragraph{Notation} We follow the common math/cryptography notation - more details can be found in \cref{sec:notation}, but importantly,  we write $\regi$ to denote quantum registers, and variables written in lower-case letters will be classical objects, such as $sk, pk$. Further, \emph{efficient} adversary means quantum polynomial time (QPT) in the plain model and in the oracle model it means an adversary who makes polynomially many quantum queries to the oracles (it can take unbounded time outside oracle calls). We write $\ora$ to denote a classical oracle and write $\adve^\ora$ or $\mathcal{B}^\ora$ to denote an adversary or algorithm with quantum query access to $\ora$. Throughout the paper, oracle will always mean quantumly-accessible classical oracle (even when do not explicitly write so).
 
\subsection{Definitions}
We first start by providing an overview of the primitives and security definitions we consider in this overview - the formal definitions can be found in \cref{sec:defnwork}. The definition of one-shot signatures (which we use as a building block) can be found in \cref{predef:oss}. 

\paragraph{Quantum Fire Mini-Scheme} In the first parts of this overview, we will work with a notion called \emph{quantum fire mini-scheme} (\cref{predef:qfiremini}), which consists of three efficient quantum algorithms: $\spark$, $\ver$, $\clone$. $\qfiremini.\spark(1^\lambda)$ outputs a serial number $sn^*$ and a flame state $\ket{\psi_{sn^*}}$.  $\qfire.\ver(sn, \regi)$ verifies if the state in the register $\regi$ is a valid flame state for the serial number $sn$. Finally, $\qfire.\clone(sn, \regis{flame})$ outputs a register $\regis{clone}$, which contains a copy of the the flame state in the register $\regis{flame}$, along with outputting the register $\regis{flame}$ back.  We require verification correctness and cloning correctness. Verification correctness requires that honestly generated states pass verification of $\ver$ with overwhelming probability. For cloning correctness, we require that the $\clone(sn, \ket{\psi_{sn}})$ outputs $\ket{\psi_{sn}}\otimes \ket{\psi_{sn}}$.

\paragraph{Untelegraphability Security} In this overview, we will consider \emph{untelegraphability security} (\cref{def:qfireminiunteleg,predef:qfireunteleg}), which is defined by the following security game between a challenger and a pair of efficient adversaries $\adve = (\adve_1, \adve_2)$. The sender adversary $\adve_1$ is given $sn^*, \ket{\psi_{sn^*}} \samp \qfiremini.\spark(1^\lambda)$ and produces a classical output $L$. Afterwards, the receiver adversary $\adve_2(L)$ simply outputs a register $\regi'$, which is tested for $sn^*$ by running $\qfiremini.\ver(sn^*, \regi')$. A scheme is said to satisfy untelegraphability security if for any efficient adversary $\adve$, the probability of winning the game is negligibly small. 

\paragraph{Full Schemes} In a full fledged quantum fire scheme $\qfire$ (\cref{predef:qfire}), there is a secret key $sk$ with which $\qfire.\spark$ is run. In the security game (\cref{predef:qfireunteleg}), the adversary $\adve_1$ receives multiple independent flame states, and $\adve_2$ tries to produce a flame state with \emph{any} serial number (not necessarily one received by $\adve_1$). Working with mini-schemes simplifies our discussion, and any mini-scheme can be generically upgraded to a secure full-fledged quantum fire scheme by using a digital signature on the serial number $sn$: $\qfire.\spark(sk)$ outputs $\sign(sk, sn) || sn, \regis{flame}$ where $sn, \regis{flame} \samp \qfiremini.\spark(1^\lambda)$.

\subsection{Construction}
We now give an overview of our quantum fire mini-scheme construction, slowly building it to highlight the main ideas and design choices. The full version of our main construction (a \emph{key-fire} scheme for signing) be found in \cref{sec:cons}.

Intuitively, the notions of clonability and untelegraphability are at odds. The parties need to be able create new copies of the flame state $\ket{\psi_{sn}}$, however, since we have classical oracles, we cannot simply output new copies of the state directly, or output any quantum states for that matter. Since we need to satisfy untelegraphability, we cannot output a description of the state directly either. Indeed, it is overall unclear how a classical oracle can help synthesize a state without simply outputting the description of the state in some form.

\paragraph{Starting Point} As a starting point, we turn our attention to \emph{one-shot signature} (OSS) schemes (\cref{predef:oss,predef:ossora}), which has been recently constructed in the classical oracle model by \cite{C:ShmZha25}. In a one-shot signature scheme, using a classical oracle $\ora$, parties are able to \emph{sample} a single copy of a quantum state (called signing keys) $\ket{\phi_{vk}}$ for some random verification key $vk$, by running the algorithm $\oss^\ora.\genqkey(1^\lambda)$. Then, they can sign any message $m \in \{0,1\}$ using $\ket{\phi_{vk}}$, but the security requirement is that they cannot sign both $0$ and $1$ with the same $vk$ value. In particular,  this requires that $\ket{\phi_{vk}}$ cannot be cloned. Using OSS for quantum fire might seem counterintuitive: After all, the states $\ket{\phi_{vk}}$ are unclonable, which is the opposite of clonability(!), and a main application of OSS is classically-transferable quantum primitives (such as quantum money that can be transferred over classical channels), whereas untelegraphability is the polar opposite of this!

Indeed, while the focus of OSS schemes is unclonability of the states  $\ket{\phi_{vk}}$, we take a new viewpoint: We observe that the OSS security also guarantees that parties can sample $\ket{\phi_{vk}}$ without learning its classical description! Thus, we consider using a one-shot signature signing key $\ket{\phi_{vk}}$ as our flame states, and we hope to use $\genqkey$ as our cloning algorithm. However, there are two major issues with this approach. First, the state generation algorithm $\genqkey$ outputs a different state  $\ket{\phi_{\textcolor{red}{vk'}}}$ each time  (necessarily, since otherwise OSS security would be broken) rather than outputting a copy of our state! While we solve this problem by using the \emph{purified} output $\ket{\eta}$ of $\oss^\ora.\genqkey(1^\lambda)$ (which is essentially a superposition $\sum_{vk} \ket{vk}\ket{\phi_{vk}}$ of all possible keys), the second issue is much more significant: $\oss.\genqkey$ is a public algorithm, and thus anyone can create $\ket{\eta}$! Thus, the sender adversary $\adve_1^\ora$ for quantum fire will not even need to prepare a classical encoding, since $\adve_2^\ora$ can simply recreate the public state $\ket{\eta}$ on its own, even though neither actually knows a classical encoding of it. To solve this issue, one might consider defining a secret key version of one-shot signatures (e.g. $\oss.\genqkey$ requires $sk$), however, this would be of no help: (i) the secret key itself would serve as a classical encoding from which $\adve_2$ can recreate $\ket{\eta}$, and more importantly (ii) the flame state must be publicly clonable, thus, we would need a new way to generate states without revealing the secret key, taking us back to square one.

\paragraph{Self-Encrypted Oracles} To overcome these issues, we consider \emph{encrypting} the oracle $\ora_{\oss}$ of $\oss$ into a new oracle $\ora'$ in a way so that the oracle can be evaluated only if one \emph{proves} that they already have the flame state $\ket{\eta}$ (we will call this a $\ket{\eta}$-verification mechanism). With this approach, we would at least eliminate the trivial attack where the receiver adversary $\adve_2$ can (without any message from $\adve_1$) create $\ket{\eta}$ on their own. Further, (publicly) cloning would still be possible, since (hopefully) one would be able to \emph{simulate} a run of $\oss.\genqkey^{\ora_{\oss}}(1^\lambda)$ by decrypting outputs of $\ora'$ using their initial state $\ket{\eta}$ and passing them along to $\oss.\genqkey$.

However, there are still major challenges with this approach. First, it is not immediately clear how one can \emph{prove} to a \emph{classical} oracle that they posses a quantum state. Second, even if one can achieve a \emph{sound} oracle encryption, the mechanism used to achieve this and implement the encrypted oracle $\ora'$ must be so that we can actually decrypt and evaluate the original oracle $\ora_{\oss}$ \emph{correctly} even for \emph{quantum queries}, since $\oss.\genqkey$ will have quantum queries to the oracle $\ora_{\oss}$. Further, the $\ket{\eta}$-verification mechanism of our encrypted oracle will need to use (in some form) the verification mechanism of $\oss$ itself (such as $\oss.\ver$), however, these algorithms also need access to $\ora_{\oss}$, thus we will need to leave some parts of the oracle unencrypted without revealing too much information. Third, once we try to prove security, we run into a major circularity issue: While we want to say that the receiver adversary $\adve_2$ will not be able to create the state $\ket{\eta}$ since it cannot decrypt the oracle $\ora'$; to show that $\adve_2$ cannot decrypt $\ora'$, (since this oracle is encrypted under $\ket{\eta}$) we need to show that it cannot have $\ket{\eta}$ to begin with!

\paragraph{Our Oracles} We now give an overview of our oracle encryption mechanism that solves all of these issues. We first note that our oracle encryption mechanism will not fully hide the oracle $\ora_{\oss}$ from $\adve_2$: It will be able to evaluate it, but we will show that it can only do so at few points, which will be sufficient to prove that it cannot produce $\ket{\eta}$,

Now we move onto our oracles. First, we require that the OSS scheme's oracle $\ora_{\oss}$ consists of three parts: $\ora_{\oss.\ver}, \ora_{\oss.\sign}$ and $\ora_{\oss.\genqkey}$ where each of the three algorithms ($\oss.\ver$, $\oss.\sign$, $\oss.\genqkey$) uses only the corresponding oracle. Of course, setting all of these oracles to be the full oracle of the scheme would also satisfy this, but as discussed, we require a non-trivial splitting of the oracle so that (i) $\ora_{\oss.\ver}, \ora_{\oss.\sign}$ do not reveal too much information (more details when we discuss the proof) and (ii) $\ora_{\oss.\ver}, \ora_{\oss.\sign}$ are sufficient for our $\ket{\eta}$-verification mechanism. Then, we define the following three oracles $\ora_0, \ora_1, \ora_2$ as our encrypted oracle $\ora'$. On input $vk, isig, z$, $\ora_b$ (for $b \in \zo$) verifies (using $\ora_{\oss.\ver}$) that $isig$ is as valid $\oss$ signature for the message $b$ with verification key $vk$, and then it outputs $H_b(vk || z)$ where $H_0, H_1$ are random functions. We will call the values $H_b(vk || z)$ \emph{attestation strings}. Finally, on input $vk, y_0, y_1, z$, $\ora_2$ verifies that $y_0 = H_0(vk || z)$, $y_1 = H_1(vk || z)$, and then outputs $\ora_{\oss.\genqkey}(z)$. 

Essentially, our encrypted oracle verifies that the party has a signature for both $0$ and $1$ (by verifying the corresponding attestation strings) with the same verification key, and then releases the output of the original oracle $\ora_{\oss.\genqkey}(z)$. Later, we will argue that this is indeed a sound $\ket{\eta}$-verification mechanism.

\paragraph{Design Choices: What Does Not Work} Here, we motivate our oracle design given above by mentioning some other possible choices that do not work. First, note that using two separate oracles $\ora_0, \ora_1$ is required: We cannot have a single oracle that asks for a signature of $0$ and $1$ with the same verification key, since by the security of the one-shot signature scheme, we would not be able to unlock such an oracle encryption. Also note that it would not be sound to accept a signature of $0$ and $1$ with different verification keys because it leads to a simple attack: After cloning $\ket{\eta}$ to get $\ket{\eta}\otimes\ket{\eta}$, the adversary $\adve_1$ can measure the verification key registers to obtain $\ket{\phi_{vk_1}}, \ket{\phi_{vk_2}}$, and sign $0$ and $1$ with these keys respectively, and send the classical signatures to the receiver adversary $\adve_2$. If we were to accept different verification keys in $\ora_2$, the adversary $\adve_2$ would be able to run the oracle $\ora_2$ on any input $z$!

\paragraph{Cloning} Now, we discuss cloning. To clone our flame state $\ket{\eta} = \sum_{vk}\ket{vk}\otimes \ket{\phi_{vk}}$, we simulate $\oss.\genqkey$ while simulating its oracle queries using $\ora_0, \ora_1, \ora_2$. Suppose $\oss.\genqkey$ makes the quantum query $\ket{\tau} = \sum\alpha_z\ket{z}$, which would normally be executed with $\ora_{\genqkey}$, but we simulate it as follows. We first run (a purified version) of $\oss.\sign$ on $\ket{\eta}$ to convert it to a superposition $\ket{\eta_0} = \sum_{vk}\ket{vk}\otimes \left(\sum_{sig \in \mathsf{SIG}(vk, 0)} \ket{sig}\right)$ of signatures for the message $0$. Then, we query $\ora_0$ on this state, which converts it to $\ket{\eta_0'} = \sum_{z,vk}\alpha_z\ket{z}\ket{vk}\otimes \left(\sum_{sig \in \mathsf{SIG}(vk, 0)} \ket{sig}_{\mathsf{sig}}\otimes \ket{H_0(vk || z)}_{\mathsf{out_0}} \right)$ by definition of $\ora_0$. Note that this can be also written as $\ket{\eta_0'} = \sum_{z,vk}\alpha_z\ket{z}\ket{vk}\otimes \left(\sum_{sig \in \mathsf{SIG}(vk, 0)} \ket{sig} \right)\otimes \ket{H_0(vk || z)}_{\mathsf{out_0}}$ since the attestation string $H_0(vk || z)$ in the register $\mathsf{out_0}$ does not depend on the particular signature $sig$ on the signature register $\mathsf{sig}$. Now, since we have isolated output the attestation string, we can actually undo (by applying the inverse unitary) the signing operation to get our original key back $\ket{\eta}$! Then, we can similarly \emph{sign} $1$ and query $\ora_1$ to get the second attestation and end up with $\ket{\eta''} = \sum_{z,vk}\alpha_z\ket{z}\ket{vk}\otimes \left(\sum_{sig \in \mathsf{SIG}(vk, 1)} \ket{sig} \right)\otimes \ket{H_0(vk || z)}_{\mathsf{out_0}}\otimes  \ket{H_1(vk || z)}_{\mathsf{out_1}}$. Finally, we can query $\ora_2$ with both our attestations in place to get output of the actual oracle $\ora_{\genqkey}$. Afterwards, we show that we can again undo the signature to get our original key back, and be ready for decryption of the next query of $\ora_\genqkey$.

\paragraph{More Design Choices} We note that above, for everything to work \emph{smoothly}, it is crucial that (i) our attestation strings do not depend on the particular signature (i.e. it is $H_0(vk||z)$ and not $H_0(vk||isig||z)$) and (ii) we simply sign $0$ and $1$ rather than signing $z$ (which would create an entanglement between the oracle input/output and our flame state because of the quantum queries of $\genqkey$, so we would not be able move onto the second query with a \emph{clean} flame state). Here, we only (informally) discussed how to handle the first query - the full proof of cloning correctness requires carefully formalizing the above arguments and is given in \cref{sec:clonecorrect}. We also note that, due to the design choice above necessitated by cloning correctness, as mentioned before our oracle encryption $\ora_0, \ora_1, \ora_2$ does not actually fully hide $\ora_{\oss.\genqkey}$ from $\adve_2$. For example, $\adve_1$ can actually obtain (for some $vk$) $H_0(vk || z), H_1(vk || z)$ for a list of $z$ values, and send these to $\adve_2$ with which $\adve_2$ can obtain $\ora_{\oss.\genqkey}(z)$. However, we will show that this can happen only for a few points, and we will still be able to prove security.

\paragraph{Construction Summary} To summarize, the oracles of our quantum fire construction $\qfiremini$ are $\ora_0$, $\ora_1, \ora_2$, $\ora_{\oss.\ver}$, $\ora_{\oss.\sign}$, and the flame state is the purified output $\ket{\eta}$ of $\oss.\genqkey(1^\lambda)$. To verify an alleged flame state, $\qfiremini.\ver$ tries signing (using $\oss.\sign$) first the message $0$, and then rewinds and tries signing the message $1$, and accepts if both succeed. We note that $\qfiremini.\ver$ does not only accept $\ket{\eta}$ - it actually accepts any OSS signing key $\ket{\psi_{vk}}$, but we prove that this is not a security issue. Finally, $\qfiremini.\clone$ proceeds by simulating $\oss.\genqkey$, where oracle queries are simulated by forwarding them to $(\ora_0, \ora_1, \ora_2)$ and decrypting as described above. Note that we do not actually give a direct construction for $\qfiremini$, and the scheme summarized above was for exposition. Our full scheme is a key-fire scheme for signing and is given in \cref{sec:cons}.

\subsection{Proving Security}
We now give a high-level overview of our proof strategy: Our full security proof (\cref{sec:proofsec}) is highly delicate and technical. Note that our formal proof is for the stronger notion of quantum key-fire, whereas in this section for ease of exposition we will discuss the case of quantum fire mini-schemes.

As argued before, to prove security, we hope to show that [Step 1] $\ora_2$ has a sound verification mechanism so that the receiver adversary $\adve_2$ will not be able to use the oracle $\ora_{\oss.\genqkey}$ (except at few points) and [Step 2] the unencrypted parts of the original oracle ($\ora_{\oss.\sign}, \ora_{\oss.\ver}$) are not leaking \emph{too much} information, that is, $\adve_2$ is not able to produce $\ket{\eta}$ when it cannot unlock oracle $\ora_{\oss.\genqkey}$ (but can access $\ora_{\oss.\sign}, \ora_{\oss.\ver}$).

We first start with the second point. To isolate [Step 2], assume that the receiver adversary $\adve_2$ has access to the oracles $\ora_{0},\ora_1,\ora_{\oss.\sign},\ora_{\oss.\ver}$, but does not have access to $\ora_2$ (so, no access to $\ora_{\oss.\genqkey}$). Note that $\adve_1$ still has access to all oracles ($\ora_0,\ora_1,\ora_2,\ora_{\oss.\sign},\ora_{\oss.\ver}$) of quantum fire. In this case, it is still not immediately clear if $\adve_2$ is not able to produce $\ket{\eta}$. After all, $\ora_{\oss.\sign}, \ora_{\oss.\ver}$ might be leaking too much information - indeed, the trivial splitting of the oracles of $\oss$ would set them to be the same as the full oracle $\ora_{\oss.\genqkey}$. However, we prove that there is actually an OSS construction such that any $\adve_2$ that has access to only $\ora_{\oss.\sign}, \ora_{\oss.\ver}$ and receives a classical message from $\adve_1$ provably cannot produce a signing key $\ket{\phi_{vk}}$!

\paragraph{Incompressibility} We note that the above security requirement is somewhat unnatural for an OSS scheme, since for one-shot signature schemes, there is no concept of secrecy (indeed, $\ket{\eta}$ is a publicly generatable state!). Thus, instead of proving the above directly, we introduce a stronger yet more natural property for OSS schemes called \emph{incompressibility}.

\begin{definition}[Incompressible One-Shot Signatures (\cref{def:ossincomprnew,def:oprincompr}, informal)]
    A one-shot signature scheme relative to an oracle $(\ora_{\oss.\sign}, \ora_{\oss.\ver}, \ora_{\oss.\genqkey})$ is said to be \emph{incompressible} if it satisfies the following for any efficient adversary $\adve_1, \adve_2$.

     $\adve_1$ gets access to $\ora_{\oss.\sign}, \ora_{\oss.\ver}, \ora_{\oss.\genqkey}$, and produces a classical output $L$.  Then, $\adve_2$ receives $L$ and oracle access to only  $\ora_{\oss.\sign}, \ora_{\oss.\ver}$. We require that there is a small (i.e. polynomial size) list of verification key values such that with overwhelming probability, any valid signature-verification key pair produced by any $\adve_2$ on input $L$ \emph{must} come from $\mathfrak{L}_L$.
\end{definition}
In a sense, incompressibility says that not too many signatures can be packed in to a classical leakage string. Note that a quantum message analogue of this incompressibility notion is not achievable: $\adve_1$ can simply send an equal superposition of exponentially many signatures.

We prove that the OSS construction of \cite{C:ShmZha25}, with a careful splitting of the oracles, satisfies incompressibility. For technical reasons that will be clear later on, we also prove an incompressibility property for random oracles (\cref{sec:incomprrom}). We believe the notion of incompressibility and our results might be of independent interest. Our formal definitions, constructions and proofs are given in \cref{sec:incomprrom} and \cref{sec:incompross}. Before moving on, we note that while incompressibility is a natural property, it is not automatically satisfied by any OSS scheme and thus requires a completely new proof separate from one-shot security. For example, one can think of an OSS scheme where the signatures and verification keys are rerandomizable, or an OSS scheme where $\ora_{\oss.\sign}$ additionally outputs a signature for a random verification key. None of these properties would cause a violation of one-shot signature security, but they would trivially break incompressibility.

\paragraph{Solving [Step 2]} Observe that with the incompressibility property, we complete [Step 2] as follows. Suppose for a contradiction that $\adve_2$ (with no access to $\ora_2$) receives $L$ from $\adve_1$ and produces a state $\rho$ that passes $\qfiremini.\ver$ with non-negligible probability. However, we can then run $\adve_2(L)$ twice to obtain to two copies of $\rho$, and with one of them we can sign $0$ to get $vk_1, isig_1$ and with the other one we can sign $1$ to obtain $vk_2, isig_2$. Since we said $\qfiremini.\ver$ accepts $\rho$ with non-negligible probability, we can then conclude that $(vk_1, isig_1)$ and $(vk_2, isig_2)$ will both be valid with non-negligible probability by Jensen's inequality. However, by incompressibility property, we also know that $vk_1$ and $vk_2$ come from a small set, thus we can show that with non-negligible probability, $vk_1$ and $vk_2$ will be the same! This gives a contradiction to one-shot security, thus completing proof of [Step 2].

\paragraph{Oracle Encryption Security: Proving [Step 1]} We essentially argued (in [Step 2]) that untelegraphability security is satisfied if $\adve_2$ did not have access to $\ora_{\oss.\genqkey}$. Now, we consider [Step 1]: actually showing that our oracle encryption encryption ensures that this is the case. As discussed before, to do this, we want to rely on the fact that $\ora_2$ in some sense verifies $\ket{\eta}$ before producing its output. However, we are trying to show that $\adve_2$ cannot produce $\ket{\eta}$ to begin with! This circularity problem also persists when we try to utilize the incompressibility property: It only applies if $\adve_2$ does not have access to $\ora_{\oss.\genqkey}$, but that's what we are trying to show to begin with!

To solve this issue and eliminate the circularity problem, we take a fine-grained approach: We consider the execution of $\adve_2$ query-by-query. In particular, we consider\footnote{This is without loss of generality} the queries of $\adve_2$ to $\ora_0,\ora_1,\ora_2$ to proceed in triplets (with queries to $\ora_{\oss.\sign}, \ora_{\oss.\ver}$ in between the triplets). We will slowly remove these query triplets and instead simulate them with messages from $\adve_1$, and eventually we will end up with a $\adve_2'$ which makes no queries $\ora_2$, in which case our discussion above shows untelegraphability security.

To this end, first, observe that the part of $\adve_2$ until its first query to $\ora_{\oss.\genqkey}$ is executed can actually be considered as an adversary that does not have any queries to $\ora_{\oss.\genqkey}$ (after all, by definition it does not make any queries to this oracle before its \emph{first} query to this oracle). Thus, we can actually invoke the incompressibility property for this part (call it $\adve_2^{(1)}$). Due to the incompressibility property of OSS, we know that there is a small list $\mathfrak{L}$ such that when $\adve_2'$ passes the verification of $\ora_0$ and $\ora_1$, it must be for some $vk \in \mathfrak{L}$. Thus, essentially, we can remove the access of $\adve_2$ to the regions of the random oracle $H_0(vk || \cdot)$ and $H_1(vk || \cdot)$ for $vk \not\in \mathfrak{L}$. Now, we know that during its query to $\ora_2$, $\adve_2$ will not have made any queries to $H_0(vk || \cdot), H_1(vk || \cdot)$ for $vk \not\in \mathfrak{L}$. Then, we also observe that when $\adve_2$ is querying $\ora_2$, its valid queries cannot have $vk \in \mathfrak{L}$ since otherwise we actually show that we can extract a signature for both $0$ and $1$ with same $vk$ from $\adve_2$! Thus, we can invoke the incompressibility of random oracles $H_0, H_1$ outside the region $vk \not\in \mathfrak{L}$ to say that valid queries of $\adve_2$ to $\ora_2$ must have $z$ from a small set $\mathfrak{L}'$. Thus, before moving onto the next query, we can actually consider the outputs of $\ora_2$ on these $z$ values to be part of a new leakage from $\adve_1'$ (who evaluates the oracle on this set and sends teh outputs), and thus remove the first query of $\adve_2$ to $\ora_2$ altogether and simulate it! One by one, we remove the queries of $\adve_2$. Finally we show how to simulate the protocol between $\adve_1$ and $\adve_2$ using a single message and polynomially many queries by $\adve_1^{sim}$ and no queries by $\adve_2^{sim}$. The full proof is technical and deals with many other challenges (e.g., showing that we can extract a signature for $0$ and $1$ if valid query of $\adve_2$ has $vk \in \mathfrak{L}$).

Another major point we swept under the rug in the discussion above is that, while we argued that we can remove oracle calls of $\adve_2$ and replace them with simulations using classical messages from $\adve_1'$, this actually requires knowing the small sets $\mathfrak{L}, \mathfrak{L}'$ on which $\adve_2$ can actually \emph{decrypt} the oracles. However, these existential sets (whose \emph{existence} are guaranteed by the incompressibility properties of OSS and random oracles) are actually not available to use directly, so we cannot simply say that $\adve_1'$ evaluates the oracles on these sets and sends the outputs to $\adve_2'$. In general, these lists can depend on the full descriptions of the oracles and are not fully determined by the initial message $L$ of $\adve_1$, since $\adve_1$ can use some outputs of the oracle as a one-time pad to encode other outputs, and $\adve_1$ and $\adve_2$ can have a special encoding in their communication protocol. As another example, these adversaries can even use global properties of the oracles (such finding a preimage for a value using Yamakawa-Zhandry \cite{FOCS:YamZha22}) in their encoding process. Thus, in general we will not have access to these sets, and giving these sets directly to the adversaries might break one-shot security (which we crucially rely on multiple points), or even break incompressibility property itself(!) since it is a global information about the oracles. To solve this issue, we show an estimation procedure where $\adve_1$ estimates these sets by making polynomially many queries, and then uses the estimations, which we prove makes only a small difference. We refer the reader to \cref{sec:proofsec} for details.

\subsection{Interactive Security}
In this work, we also initiate the study of \emph{interactive} untelegraphability security for quantum fire. We first note that the previous definitions of untelegraphability security is very limited. In fact, while we do not show this explicitly, using our techniques it is easy to construct a scheme that satisfies (1-message) untelegraphability security, but its security breaks down fully when $\adve_2$ first sends a message to $\adve_1$ and $\adve_1$ replies afterwards. Further, interactive adversaries better capture real-life leakage attacks: For example, when the flame states encode the secret key of a PKE scheme (see \emph{key-fire} below), the adversary $\adve_2$ will choose the leakage function depending on $pk$, which is not captured by previous definitions of untelegraphability.

\paragraph{Security Game} We now give an overview of the interactive security game for mini-schemes, and the case of full schemes is defined similarly (see \cref{defn:interactive} for the formal definition). The challenger samples a flame state as $sn^*, \regis{flame} \samp \qfiremini.\spark(1^\lambda)$, and sends $sn^*, \regis{flame}$ to $\adve_1$. Then, $\adve_1$ and $\adve_2$ interact over a classical channel for any number of rounds. Finally, outputs a register $\regi'$, which is tested for $sn^*$ by running $\qfiremini.\ver(sn^*, \regi')$.

\paragraph{Proving Security} For interactive security, our construction stays the same as before, however, our proof is significantly different. Here we give a high-level overview, and but the full proof is highly delicate and technical - we refer the reader to \cref{sec:proofsec} for details.

First, recall that our proof of untelegraphability security as discussed above crucially relied on running $\adve_2$ twice on the message $L$ from $\adve_1$ to obtain two signatures, then we invoked incompressibility to obtain a contradiction to the one-shot security of OSS. However, once we move to the interactive setting, our argument completely breaks down: We can no longer run $\adve_2$ twice since it has a (possibly unclonable) internal state, and incompressibility guarantees only apply to a non-interactive adversary $(\adve_1, \adve_2)$. To solve this, we show how to \emph{clone} the internal state of the adversary $\adve_2$ and also round-collapse the leakage protocol between $\adve_1$ and $\adve_2$ into a single message. To this end, we first start with our previous idea of making a fine-grained analysis of the queries of $\adve_2$. We remove the queries of $\adve_2$ to the oracles $\ora_0,\ora_1,\ora_2$ one by one and replace them with estimated ones using the incompressibility. Finally, after removing these calls, we go further and also show how to estimate and remove the calls of $\adve_2$ to all the other oracles! Once we remove these calls, we are able to show that we decrease the number of rounds of interaction by making $\adve_1$ simulate the first round, and then \emph{telling} $\adve_2$ the outcome of the first round, after which $\adve_2$ locally prepares its internal state conditioned on that outcome/transcript so far! While normally it requires exponential many trials to correctly condition, since we have removed the queries of $\adve_2$ (which are instead being simulated by simply reading from the database created by $\adve_1$), we can actually afford to do this procedure. Finally once we collapse down to a single message from $\adve_1$ to $\adve_2$ with no actual oracle calls from $\adve_2$, we can also clone the internal state of $\adve_2$. Thus, we are able to apply our previous argument based on one-shot security. The full proof deals with the challenges in making these arguments actually work, and is given in \cref{sec:proofsec}.

\subsection{Quantum Key-Fire}
In this work, we also introduce a new notion called \emph{quantum key-fire}, where the flame states are \emph{functional} and serve as a \emph{key}. We give an overview, and the formal definitions can be found in \cref{sec:qkeyfire}. A quantum key-fire scheme consists of three algorithms: $\qkeyfire.\prot$, $\qkeyfire.\ceval$, $\qkeyfire.\clone$. For a function $f$ (modeled as a circuit), $\prot(1^\lambda, f)$ outputs a flame state $\ket{\psi_{pp,f}}$ along with public parameters $pp$.  $\qkeyfire.\ceval(pp, \ket{\psi_{pp,f}}, x)$ evaluates the encoded functionality at $x$. Finally, $\qkeyfire.\clone(pp, \ket{\psi_{pp,f}})$ clones the flame state. For correctness, we require that $\ceval(pp, \ket{\psi_{pp,f}}, x)$ outputs $f(x)$, and that $\clone(pp, \ket{\psi_{pp,f}})$ outputs $\ket{\psi_{pp,f}}\otimes \ket{\psi_{pp,f}}$. For security, we require that after an adversary obtains unbounded classical leakage on $\ket{\psi_{pp,f}}$ adaptively over any number of rounds, it still cannot correctly evaluate $f$ at a challenge $x^*$ (\cref{predef:ublrfunc}).

Through known techniques (\cite{C:CLLZ21,C:ALLZZ21,cryptoeprint:2025/1197}), it is easy to see that quantum key-fire for \emph{signing} (\cref{def:qkeyfiresign}) implies quantum key-fire for any unlearnable functionality (in the classical oracle model). Thus, in the rest of the overview, we mainly consider quantum key-fire for signing. In the security game for this primitive (\cref{def:qkeyfiresignsec}), the adversary $\adve$ receives LOCC leakage on the signing key state $\regis{flame}$: For any number of rounds, it (adaptively) specifies a quantum circuit with classical output as a leakage function, which is then applied to $\regis{flame}$ and the outcome is sent to $\adve$. After the leakage, it is asked to sign a random message $m^*$. We require that any efficient adversary can win this game with negligible probability. Note that equivalently, this game can also be modeled as two adversaries $\adve_1,\adve_2$ interacting over a classical channel where $\adve_1$ receives $\regis{flame}$ and after the interaction, $\adve_2$ receives the challenge $m^*$ and produces a signature (see \cref{def:qfireorasec} for this formalization).

\paragraph{Construction and Proving Security} We briefly discuss our construction (we refer the reader to \cref{sec:cons} for our full construction). The starting point of our construction is our key-fire construction, but with the addition of a new oracle that \emph{verifies} the flame states and outputs shares of a random oracle evaluation $H_{\mathsf{sig}}(m)$ which serves a signature on $m$. In the proof, using the techniques discussed before, we show that we can round-collapse the leakage protocol down to a single message, and make the internal state of $\adve_2$ clonable. At this point, we argue (thanks to incompressibility) that running two copies of $\adve_2$ will query the evaluation oracle with the same OSS verification key, and we will be able to extract two OSS signatures to obtain a violation of the one-shot security. The full proof is highly delicate and technical - we refer the reader to \cref{sec:proofsec}.

\subsection{Related Work}
Nehoran and Zhandry (\cite{ITCS:NehZha24}) did not formally define quantum fire, however, it introduced the complexity class $\mathsf{clonableQMA}$ and showed a quantum oracle separation between $\mathsf{clonableQMA}$ and $\mathsf{QCMA}$. Note that $\mathsf{clonableQMA}$ sits between $\mathsf{QMA}$ and $\mathsf{QCMA}$, thus  a separation between $\mathsf{clonableQMA}$ and $\mathsf{QCMA}$ also leads to a separation between $\mathsf{QMA}$ and $\mathsf{QCMA}$ (note that such a separation was already known in the quantum oracle model). They also mentioned possible cryptographic applications, thus hinting at quantum fire. Indeed, their separation is also implicitly a quantum fire construction relative to the same quantum oracle, as observed by \cite{STOC:BosNehZha25}. Unfortunately their techniques bake cloning into the oracle itself, and thus it uses an inherently inefficient quantum oracle and there is no hope even instantiating it even heuristically. 

Bostanci, Nehoran, Zhandry (\cite{STOC:BosNehZha25}) formalized the notion of quantum fire. Then, they gave a candidate construction based on group actions, and proved its cloning correctness. However, even in the classical oracle model, they only conjectured its security and could not give a proof of security or argue security. In fact, as discussed before, they even claimed that any provably secure quantum fire scheme in the classical oracle model would likely lead to a classical oracle separation between $\mathsf{QMA}$ and $\mathsf{QCMA}$, which is one of the biggest questions in quantum cryptography that has been open since 2007. Indeed, since \cite{ITCS:NehZha24} separates both $\mathsf{clonableQMA}$ and $\mathsf{QCMA}$, and at the same time their separation is implicitly a quantum fire construction, it is an interesting question if there is any connection and if our techniques could be useful.

We also discuss a follow-up work, \cite{vinodhuang}. The initial version of our paper on arXiV contained a small gap in only in the last hybrid of our quantum key-fire proof (not quantum-fire proof), which was communicated to us by the authors of \cite{vinodhuang} and at the same time they put their paper online. Within 24 hours of receiving their communication, we fixed the small gap in our proof and let them know. Current version of our paper does not have this gap, however, we include a discussion in \cref{appn:vh} for historical reference.

For leakage-resilience, \cite{TCC:CGLR24} introduced the notion of unbounded leakage-resilience and LOCC leakage-resilience for various cryptographic primitives such as public-key encryption and digital signatures. They also showed that the previous unclonable/copy-protected constructions of \cite{C:CLLZ21,TCC:LLQZ22} also satisfy LOCC leakage-resilience. Therefore, they also asked the question of whether leakage-resilience security implies copy-protection security.
\newpage
\section{Preliminaries}
In this section, we highlight some of our notation and conventions, and recall some basics about classical cryptography and quantum information.

\subsection{Notation and Conventions}\label{sec:notation}
 We refer the reader to \cite{Goldreich_2001,Goldreich_2004} for a preliminary on cryptography and to \cite{NC10} for a preliminary on quantum information and computation. We follow the notations and conventions commonly used in (quantum) cryptography, theoretical computer science and mathematics.  In this section, we non-exhaustively highlight some of them. All of our conventions are implicitly followed unless otherwise specified, e.g. our cryptographic assumptions are always post-quantum even though most of the time we will not write this explicitly, and if there is a rare occasion where we assume classical security only, we will specify explicitly.

\paragraph{Functions} For a function $f: \nat^+ \to \nat^+$, we will say that it is an efficiently computable function if it is polynomially bounded, that is, if there is some $n_0 \in \nat^+$  and $c \in \R^+$ such that $f(n) \leq n^c$ for all $n > n_0$. Sometimes we will say polynomial to also mean polynomially bounded instead (e.g. when we say polynomial, it means $O(\poly(\lambda))$, which includes, say, $\sqrt{n}$) - the precise meaning will be clear from context. $f(\lambda)$ is said to be negligible if for any $c \in \nat^+$, there is $n_0 \in \nat^+$ such that $f(\lambda) \leq \frac{1}{n^c}$ for all $n \geq n_0$. $f(\lambda)$ is said to superlogarithmic if $f = \omega(\log(\lambda))$ or equivalently, if $2^{-f(\lambda)}$ is negligible.

When we describe a function, such as a polynomial, if no input is mentioned, it will implicitly be the security parameter $\lambda$.

\paragraph{Classical and Quantum Information and Oracles} Variables written in lower-case,  such as $pk,sk$, are classical objects. We write $\regi$ to mean a quantum register, which keeps a quantum state that will evolve when the register is acted on, and it can be entangled with other registers. 

We write $\regi \samp \rho$ to mean that the register $\regi$ is initialized with a sample from the quantum distribution (i.e. mixed state) $\rho$. For simplicity, we will usually write non-normalized versions of the quantum states, however, they are always implicitly normalized.

We write $\ora$ to denote a classical oracle, and write $\adve^{\ora}$ to denote an algorithm that has quantum query-access to $\ora$. Similarly, for unitary oracles we write $U$ and $\adve^U$. All adversaries always have quantum access to all of the oracles.

\paragraph{Distributions} When $S$ is a set, we write $x \samp S$ to mean that $x$ is sampled uniformly at random from $S$. When $\mathcal{D}$ is a distribution, we write $x \samp \mathcal{D}$ to mean that $x$ is sampled from $\mathcal{D}$. Finally, we write $x \samp \mathcal{B}(\regi)$ or $x \samp \mathcal{B}(a)$ to mean that $x$ is a sample as output by the (quantum or classical randomized) algorithm $\mathcal{B}$ run on the quantum input register $\regi$ or classical input $a$. We use similar notation for quantum registers, e.g., $\regi \samp \mathcal{B}(a)$. 

\paragraph{Adversaries and Constructions/Algorithms} We write \emph{QPT} to mean \emph{quantum polynomial time} and \emph{PPT} to mean \emph{probabilistic (classical) polynomial time}. We say query-bounded to mean that the algorithm (e.g. construction, adversary) makes polynomially many queries to the oracle, but it can possibly take unbounded time outside the queries. All of these are polynomial in the security parameter $\lambda$. Note that for QPT and PPT, this also means that the inputs to these algorithms are also of polynomial size in $\lambda$. 

We say \emph{efficient} to mean QPT or PPT (will be clear from context) in the plain model, or query-bounded in the oracle model. For constructions, efficient means QPT or PPT, even in the oracle model (though the oracles themselves might not be efficiently sampleable or implementable)

In the plain model, adversaries will be implicitly QPT. In the oracle security model, adversaries will be implicitly \emph{query-bounded}. An admissible adversary means an adversary that conforms to the (implicit) requirements of the security game (e.g. polynomial time or query bounded, has the correct interaction model and has correct input-output format and so on). All of our cryptographical assumptions are post-quantum, e.g., \emph{one-way functions} means \emph{post-quantum secure one-way functions}. Adversaries are stateful uniform QPT.  

When we say that a primitive is subexponentially secure, we mean that any  polynomial time adversary has subexponentially small advantage. Sometimes we require security against subexponential time adversaries, in which case we will specify this explicitly.

All algorithms of a cryptographic scheme are uniform (stateless) PPT or QPT (will be clear from context).

\paragraph{Security Games} A security game is a binary random variable which is the outcome of an experiment between a \emph{challenger} and an adversary. For a security game, we say that the adversary has won if the experiment output is $1$, and otherwise we say that the adversary has lost. When we say that an adversary has negligible advantage, we mean that the probability of the adversary winning (or distinguishing, depending on context) is upper bounded by a negligible function of $\lambda$.

\paragraph{Domains} We write $\messpa$ to denote the message space for various cryptographic primitives, and it will be equal to $\zo^{p(\lambda)}$ for some polynomial $p(\cdot)$ (will be clear from context).

\newcommand{\prelimlev}{\subsection}

\prelimlev{Pseudorandom Functions}
In this section, we recall pseudorandom functions (PRF) and puncturable PRFs.
\begin{definition}[Pseudorandom Functions]\label{predef:prf}
    A pseudorandom function (PRF) scheme $\prf$ is a family of functions $\{F: \zo^{c(
\lambda)} \times \zo^{m(\lambda)} \to \zo^{n(\lambda)}\}_{\lambda \in \nat^+}$ along with the following efficient algorithms.
    \begin{itemize}
        \item $\prf.\gen(1^\lambda):$ Takes in the unary representation of the security parameter and outputs a key $K$ in $\zo^{c(\lambda)}$.
        \item $\prf.\ceval(K, x):$ Takes in a key $K \in \zo^{c(\lambda)}$ and an input $x \in \zo^{m(\lambda)}$, outputs an evaluation of the PRF in $\zo^{n(\lambda)}$.
    \end{itemize} 
    We require the following.
    \paragraph{Correctness.}
    \begin{equation*}
        \Pr[\forall x~\prf.\ceval(K, x) = F(K, x): \begin{array}{c}
              K \samp \prf.\setup(1^\lambda)
        \end{array}] = 1.
    \end{equation*}
    \paragraph{Security} We require that for any QPT adversary $\adve$,
    \begin{equation*}
        \left|\Pr_{H \samp \mathrm{H}}[\adve^{H}(1^\lambda) = 1] - \Pr_{K \samp \prf.\gen(1^\lambda)}[\adve^{F(K,\cdot)}(1^\lambda) = 1] \right| \leq \negl(\lambda).
    \end{equation*}
    where $\mathrm{H}$ is the uniform distribution over the functions $\zo^{m(\lambda)} \to \zo^{n(\lambda)}$.
\end{definition}
 We will usually write $F_K(x)$ or $F(K, x)$ in algorithms and this will implicitly mean $\prf.\ceval(K, x)$.

\begin{definition}[Puncturable Pseudorandom Functions \cite{STOC:SahWat14}]\label{predef:puncprf}
    A puncturable pseudorandom function (PRF) scheme $\prf$ is a pseudorandom function scheme along with the following additional efficient algorithm.
    \begin{itemize}
        \item $\prf.\mathsf{Puncture}(K, S):$ Takes as input a key $K \in \zo^{c(\lambda)}$ and a predicate circuit $C: \zo^{m(\lambda)} \to \zo$, outputs a punctured key.
    \end{itemize}
    
    We require the following.
    \paragraph{Punctured Correctness.} For all efficient distributions $\mathcal{D}(1^\lambda)$ over predicate circuits, we require
    \begin{equation*}
        \Pr[\forall x \in \zo^{m(\lambda)} C(x) = 1 \implies \prf.\ceval(K_C, x) = F(K, x): \begin{array}{c}
              C \samp \mathcal{D}(1^\lambda) \\
              K \samp \prf.\setup(1^\lambda) \\
              K_C \samp \prf.\mathsf{Puncture}(K, C)
        \end{array}] = 1.
    \end{equation*}
    \paragraph{Puncturing Security} We require that any QPT adversary $\adve$ wins the following game with probability at most $1/2 + \negl(\lambda)$.
    \begin{enumerate}
        \item The adversary $\adve$ outputs a predicate circuit $C$.
        \item The challenger samples $K \samp \prf.\setup(1^\lambda)$ and $K_C \samp \prf.\mathsf{Puncture}(K, C)$, and submits $K_C$ to the adversary $\adve$.
        \item The adversary $\adve$ outputs an input $x$ such that $C(x) = 0$.
        \item The challenger samples $b \samp \zo$. If $b = 0$, the challenger submits $F(K, x)$ to the adversary. Otherwise, it submits $y$ to  the adversary where $y \samp \zo^{n(\lambda)}$.
        \item The adversary outputs a guess $b'$ and we say that the adversary has won if $b' = b$.
    \end{enumerate}
\end{definition}

It is easy to see that any puncturable PRF scheme that satisfies puncturing security also satisfies usual PRF security.

\begin{theorem}[\cite{STOC:SahWat14,FOCS:Zhandry12}]\label{prethm:puncprfexists}
Let $n(\cdot), m(\cdot)$ be polynomially bounded.
\begin{itemize}
    \item If (post-quantum) one-way functions exist, then there exists a (post-quantum) puncturable PRF with input space $\zo^{m(\lambda)}$ and output space $\zo^{n(\lambda)}$.

\item If subexponentially-secure (post-quantum) one-way functions exist, then for any $c > 0$, there exists a (post-quantum) $2^{-\lambda^c}$-secure\footnote{While the original results are for negligible security against polynomial time adversaries, it is easy to see that they carry over to subexponential security. Further, by scaling the security parameter by a polynomial and simple input/output conversions, subexponentially secure (for any exponent $c'$) one-way functions is sufficient to construct for any $c$ a puncturable PRF that is $2^{-\lambda^c}$-secure.} puncturable PRF with input space $\zo^{m(\lambda)}$ and output space $\zo^{n(\lambda)}$.
\end{itemize}
\end{theorem}

\prelimlev{Digital Signature Schemes}
In this section we recall the definition of signatures schemes.

\begin{definition}\label{predef:digsig}
A digital signature scheme with message space $\messpa$ consists of the following efficient algorithms that satisfy the correctness and security guarantees below.
\begin{itemize}
    \item $\setup(1^\lambda):$ Takes the unary representation of the security parameter and outputs a signing key $sk$ and a verification key $vk$.
    \item $\sign(sk, m):$ Takes a signing key $sk$ and a message $m \in \messpa$, returns a signature for $m$.
    \item $\ver(vk, m, sig):$ Takes the public verification key $vk$, a message $m \in \messpa$ and an alleged signature $sig$ for $m$, outputs $1$ if $s$ is a valid signature for $m$.
\end{itemize}
\paragraph{Correctness}
We require the following for all messages $m \in \messpa$.
\begin{equation*}
    \Pr[\ver(vk, m, s) = 1 : \begin{array}{c}
         sk, vk \samp \mathsf{Setup}(1^\lambda) \\
         s \samp \mathsf{Sign}(sk, m)
    \end{array}] = 1.
\end{equation*}
\paragraph{Adaptive existential-unforgeability security under chosen message attack (EUF-CMA)}
We require that any QPT adversary $\adve$ wins the following game with negligible probability.
\begin{enumerate}
    \item The challenger samples the keys $sk, vk \samp \mathsf{Setup}(1)$.
    \item The adversary $\adve$ receives $vk$, and interacts with the signing oracle by sending classical messages $m \in \messpa$ and receiving the corresponding signatures that is computed by the challenger as $sig \samp \sign(sk, m)$.
    \item The adversary $\adve$ outputs a message $m^* \in \messpa$ that it has not queried the oracle with and a forged signature $sig^*$.
    \item The challenger outputs $1$ if and only if $\ver(vk, m^*, sig^*) = 1$.
\end{enumerate}
If the adversary $\adve$ is required to output the message $m^*$ before receiving $vk$, we call it \emph{selective EUF-CMA} security.
\end{definition}

\prelimlev{Indistinguishability Obfuscation}
In this section, we recall indistinguishability obfuscation (iO).
\begin{definition}[Indistinguishability Obfuscation]\label{predef:io}
    An indistinguishability obfuscation (iO) scheme for a class of circuits $\mathcal{C} = \{\mathcal{C}_\lambda\}_\lambda$ is an efficient algorithm $\io$ that satisfies the following.
    \paragraph{Correctness.} For all $\lambda \in \nat^+, C \in \mathcal{C}_\lambda$ and all inputs $x$ to $C$,
    $$\Pr[\Tilde{C}(x) = C(x): \Tilde{C} \samp \io(1^\lambda, C)] = 1.$$

    \paragraph{Security.} Let $\mathcal{B}$ be any QPT algorithm that outputs two circuits $C_0, C_1 \in \mathcal{C}$ of the same size, along with quantum auxiliary information $\regis{aux}$, such that $\Pr[\forall x ~ C_0(x)=C_1(x) : (C_0, C_1, \regis{aux}) \samp \mathcal{B}(1^\lambda)] \geq 1 - \negl(\lambda)$. Then, for any QPT adversary $\mathcal{A}$,
    \begin{align*}
      \bigg|&\Pr[\adve(\io(1^\lambda, C_0), \regis{aux}) = 1 :  (C_0, C_1, \regis{aux}) \samp \mathcal{B}(1^\lambda)] -\\ &\Pr[\adve(\io(1^\lambda, C_1), \regis{aux}) = 1 : (C_0, C_1, \regis{aux}) \samp \mathcal{B}(1^\lambda)]\bigg| \leq \negl(\lambda).  
    \end{align*}
\end{definition}

\subsection{Quantum Information}
In this section, we recall some quantum information results.

We first recall the gentle measurement lemma, which says that if some measurement has a high probability of giving a particular outcome on some state, then the measurement can be implemented in a way so that the state will not be disturbed too much when that outcome occurs.
\begin{lemma}[Gentle Measurement Lemma \cite{awinter}]\label{prelem:gentlemes}
    Let $\rho$ be a mixed state acting on $\C^{d}$. Let $U$ be a unitary and $(\Pi_0, \Pi_1 = I - \Pi_0)$ be projectors all acting on $\C^{d} \otimes \C^{d'}$. We interpret $(U, \Pi_0, \Pi_1)$ as a measurement performed by appending an ancillary system of dimension $d'$ in the state $\ketbra{0}{0}$, applying $U$ and then performing the projective measurement $\Pi_0, \Pi_1$ on the larger system. Assuming that the outcome corresponding to $\Pi_0$ has probability $1 - \epsilon$, we have
    \begin{equation*}
        \statdist{\rho}{\rho'} \leq \sqrt{\epsilon}
    \end{equation*}
    where $\rho'$ is the state after performing the measurement, undoing the unitary $U$ and tracing out the ancillary system.
\end{lemma}

We recall the \emph{hybrid lemma} for reprogramming quantumly-accessible classical oracles.
\begin{theorem}[Hybrid Lemma~\cite{bbbv97}]\label{prethm:bbbv97}
Let $\adve$ be a quantum algorithm making queries to an oracle $\mathcal{O}$. Let $\ket{\psi_t} = \sum_{w,x,y,t}\alpha_{w,x,y,t}\ket{w,x,y}$ denote the joint state of the working register, the query input register and the query output register of the algorithm right before execution of the $t$-th query. For a subset $S$ of the domain of $\mathcal{O}$, let $q_S(\ket{\psi_t}) = \sum_{x \in S} |\alpha_{w,x,y,t}|^2$ and $q_S = \sum_{t} q_s(\ket{\psi_t})$, and call $q_S$ the query weight of $S$.
Let $\mathcal{O}'$ be another oracle whose output differs from $\mathcal{O}$ only on points $x \in S$. Then, if $\adve$ makes $T$ queries to the oracle $\mathcal{O}$ and $S$ is a subset such that $q_S \leq \epsilon^2/T$, we have $\left|\ketbra{\psi}{\psi} - \ketbra{\psi'}{\psi'}\right|_1 \leq 2\epsilon$ where $\ket{\psi},\ket{\psi'}$ denote the final state of the algorithm $\adve$ when given access to the oracles $\mathcal{O}, \mathcal{O}'$ respectively. In the case where $\ora, \ora', S$ are distributional (along with some auxiliary information register $\regis{aux}$) and possibly correlated, we have $(\ketbra{\psi}{\psi}, \ora, \ora', S,\regis{aux}) \approx_{\epsilon}(\ketbra{\psi'}{\psi'}, \ora, \ora', S, \regis{aux})$ or equivalently $(\ketbra{\psi}{\psi}, \ora, \ora', S,\regis{aux}) \approx_{\sqrt{q_S\cdot T}}(\ketbra{\psi'}{\psi'}, \ora, \ora', S,\regis{aux})$. 
\end{theorem}

We recall the following lemma about compressed oracle simulation technique for random oracles. We will use this technique in \cref{sec:incomprrom}, thus, we refer the reader to \cite{C:Zhandry19} for details of this technique. As a reminder of the notation, we have $\mathsf{CStO} = \mathsf{StdDecomp}\circ\mathsf{CStO}'\circ \mathsf{StdDecomp}\circ \mathsf{Increase}$.
\begin{theorem}[Compressed Oracle Lemma~\cite{C:Zhandry19}]\label{prethm:compor}
Consider a quantum algorithm $\adve$ making queries to a random oracle $H: \zo^m \to \zo^n$ and outputting a tuple $(x_1, \dots, x_k, y_1, \dots, y_k, z)$. Let $R$ be a collection of such tuples. Suppose with probability $p$, $\adve$ outputs a tuple such that (1) the tuple is in $R$ and (2) $H(x_i) = y_i$ for all $i$. Now consider running the \emph{compressed oracle simulation} of $\adve$, suppose the database $D$ is measured after $\adve$ produces its output. Let $p'$ be the probability that (1) the tuple is in $R$, and (2) $D(x_i) = y_i$ for all $i$ (and in
particular $D(x_i) \neq \bot$). Then $\sqrt{p} \leq \sqrt{p'} + \sqrt{\frac{k}{2^n}}$.
\end{theorem}

We recall the small-range distributions lemma.
\begin{theorem}[Small Range Distributions \cite{FOCS:Zhandry12}]\label{thm:srd}
There is a universal constant $\csrd > 1$ such that, for any sets $\mathcal{X}$ and $\mathcal{Y}$, distribution $\mathcal{D}$ on $\mathcal{Y}$, any integer $\ell$, and any quantum algorithm $\adve$ that makes $k$ queries, we have
\begin{equation*}
    \left|\Pr_{\mathcal{O} \samp \mathfrak{O}_1}[\adve^{\mathcal{O}}() = 1] - \Pr_{\mathcal{O} \samp \mathfrak{O}_2}[\adve^{\mathcal{O}}() = 1] \right| \leq \csrd\frac{k^3}{\ell}
\end{equation*}
where the distributions $\mathfrak{O}_1, \mathfrak{O}_2$ are defined as follows.
\begin{itemize}
    \item For $\mathfrak{O}_1$, independently sample  $y_x \samp \mathcal{D}$ for each $x \in \mathcal{X}$ and set $\mathcal{O}(x) = {y}_x$.
    \item For $\mathfrak{O}_2$, sample a random function $P: \mathcal{X} \to [\ell]$, independently sample  $y_x \samp \mathcal{D}$ for each $x \in [\ell]$ and set $\mathcal{O}(x) = {y}_{P(x)}$.
\end{itemize}
\end{theorem}

\newpage\part{Definitions}
\section{Quantum Protection Definitions}\label{sec:prot}
In this section, we recall the definitions of various \emph{quantum protection }(\cite{cryptoeprint:2024/1876,cryptoeprint:2025/1197}) security notions, which is the question of encoding a key/functionality into a quantum state to obtain a classically-impossible security guarantee (such as copy-protection or unbounded leakage-resilience).

\subsection{Quantum Protection}
We recall the notions of classically unlearnable functionalities and quantum protection schemes.
\begin{definition}[Classically Unlearnable Functionalities \cite{cryptoeprint:2024/1876,cryptoeprint:2025/1197}]\label{predef:clunlearn}
A functionality $\mathcal{F}$ is a tuple $(F$, $\mathsf{SampFunc}$, $\mathsf{SampChal}$, $\mathsf{Pred})$ whose elements are defined as follows.
\begin{itemize}
    \item $F$: A sequence consisting of sets (for each $\lambda \in \nat^+$) of classical circuits of polynomial size.
    \item $\mathsf{SampFunc}(1^\lambda):$ An interactive efficient (quantum) algorithm  that outputs a value $f \in F$ at the end of the interaction.
    \item $\mathsf{SampChal}(f)$: An efficient (quantum) algorithm that takes as input a value $f \in F$ and outputs a challenge input $x$ and an answer key $ak$.
    \item $\mathsf{Pred}(f, ak, x, y)$: An efficient (quantum) algorithm that takes as input a value $f \in F$, an answer key $ak$, an input $x$ and a value $y$, and outputs $1$ or $0$, denoting if $y$ is a valid answer for input $x$ with respect to $f$ and $ak$ or not.
\end{itemize}

Consider the following between a challenger and an adversary $\adve$.
\paragraph{\underline{$\learninggame{\adve}(1^\lambda)$}}
\begin{enumerate}
\item The challenger and the adversary interact where the challenger simply simulates $\mathsf{SampFunc}$. Let $f^*$ denote the output of $\mathsf{SampFunc}$ at the end of the interaction.
\item For any polynomial number of rounds: The adversary $\adve$ submits the classical description of a oracular quantum circuit $C$ with classical output, the challenger executes $C$ with oracle access to $f^*$ and submits the outcome to the adversary.
\item The challenger samples a challenge input $x^*, ak^* \samp \mathsf{SampChal}(f^*)$ and submits $x^*$ to the adversary $\adve$.
\item $\adve$ outputs $y^*$.
\item The challenger outputs $1$ if and only if $\mathsf{Pred}(f^*, ak^*, x^*,y^*)$ outputs $1$.
\end{enumerate}
A functionality $\mathcal{F}$ is said to satisfy \emph{$\ptriv$-classical-unlearnability} if for any QPT adversary $\adve$,
\begin{equation*}
    \Pr[\learninggame{\adve}(1^\lambda) = 1] \leq \ptriv(\lambda) + \negl(\lambda).
\end{equation*}
\end{definition}
Note that for any non-trivial game (i.e. $\ptriv$ is not close to $1$), the size of the challenge $x^*$ must be superlogarithmic.

\begin{definition}[Quantum Protection Scheme \cite{cryptoeprint:2024/1876,cryptoeprint:2025/1197}]\label{predef:qprot}
  Let $\mathcal{F} = (F, \mathsf{SampFunc}, \mathsf{SampChal}, \mathsf{Pred})$ be a functionality.  A quantum protection scheme $\qprot$ for $\mathcal{F}$ consists of the following efficient algorithms.

  \begin{itemize}
      \item $\qprot.\prot(1^\lambda, f)$: On input a circuit $f \in {F}$, outputs a quantum register $\regis{key}$.
      \item $\qprot.\ceval(\regis{key}, x)$: On input a key register and a circuit input $x$, outputs an evaluation.
  \end{itemize}

  We require evaluation correctness: For each $f \in F$ and input $x$ to $f$, 
  \begin{equation*}
      \Pr[y = f(x) : \begin{array}{c}
           \regis{key} \samp \qprot.\prot(1^\lambda, f) \\
           y \samp \qprot.\ceval(\regis{key}, x)
      \end{array}] \geq 1 - \negl(\lambda).
  \end{equation*}
\end{definition}

\subsection{LOCC Leakage-Resilience}
 We recall LOCC leakage-resilience security for general functionalities.
\begin{definition}[LOCC Leakage-Resilience Security \cite{TCC:CGLR24,cryptoeprint:2024/1876,cryptoeprint:2025/1197}]\label{predef:ublrfunc}
Let $\mathcal{F} = (F$, $\mathsf{SampFunc}$, $\mathsf{SampChal}$, $\mathsf{Pred})$ be a functionality that is $\ptriv$-classical-unlearnable and let $\qprot$ be a quantum protection scheme for $\mathcal{F}$. Consider the following game between the challenger and an adversary $\adve$.

\paragraph{\underline{$\lrgame{\adve}(1^\lambda)$}}
\begin{enumerate}
\item The challenger and the adversary $\adve$ interact where the challenger simply simulates $\mathsf{SampFunc}$. Let $f^*$ denote the output of $\mathsf{SampFunc}$ at the end of the interaction.
\item The challenger samples $\regi_0 \samp \qprot.\prot(1^\lambda, f^*)$.
\item \textbf{Leakage Phase:} Then, for any number\footnote{Note that when $\adve$ is QPT, implicitly this will be any \emph{polynomial} number of rounds.} of rounds, the following is executed (starting with $i= 1$).
    \begin{enumerate}[label=\arabic*.]
        \item $\adve$ specifies a quantum leakage circuit $E_i$ that takes as input a quantum register and outputs a classical string and the updated quantum register.
        \item The challenger executes $L_i, \regi_i \samp E_i(\regi_{i - 1})$ and submits the classical string $L_i$ to $\adve$. Then $i$ is incremented.
    \end{enumerate}
    \item The challenger samples a challenge input $x^* \samp \mathsf{SampChal}(f^*)$ and submits $x^*$ to the adversary $\adve$.
    \item $\adve$ outputs $y^*$.
    \item The challenger outputs $1$ if and only if $\mathsf{Pred}(f^*, ak^*, x^*,y^*)$ outputs $1$.
\end{enumerate}
A quantum protection scheme $\qprot$ is said to satisfy \emph{LOCC leakage-resilience} for the functionality $\mathcal{F}$ if for any QPT adversary $\adve$,
\begin{equation*}
    \Pr[\lrgame{\adve}(1^\lambda) = 1] \leq \ptriv(\lambda) + \negl(\lambda).
\end{equation*}
\end{definition}

Note that previous works also define unbounded (i.e. 1-round) leakage-resilience. In this work, since we achieve the strongest notion, LOCC leakage-resilience, we will not define unbounded leakage-resilience.

In \cref{appn:alternativedefnslr}, for ease of exposition, we also explicitly write out the special case of this definition for digital signatures and public-key encryption explicitly, since these are the two simplest and most natural use cases.

\subsection{One-Shot Signatures}
We recall one-shot signatures.
\begin{definition}[One-Shot Signatures \cite{STOC:AGKZ20}]\label{predef:oss}
     A one-shot signature scheme $\oss$ consists of the following efficient algorithms.
    \begin{itemize}
        \item $\oss.\setup(1^\lambda)$: Takes in a security parameter, outputs a common reference string $crs$.
        \item $\oss.\genqkey(crs)$: Takes in the CRS value $crs$, outputs a verification key $vk$ and a quantum signing key $\regis{key}$.
        \item $\oss.\sign(crs, vk, \regis{key}, m)$: Takes in the CRS value,a verification key, a signing key $\regis{key}$, and a message $m$; outputs a classical signature $sig$.
        \item $\oss.\ver(crs, vk, m, sig)$: Takes in the CRS value $crs$, a verification key $vk$, a message $m$ and a signature $sig$,; outputs $0$ or $1$.
    \end{itemize}

    \paragraph{Correctness:} For any message $m \in \messpa$, \begin{equation*}
        \Pr[out_{\ver} = \outtrue  : \begin{array}{c}
        crs \samp \oss.\setup(1^\lambda) \\
        vk, \regis{key} \samp \oss.\genqkey(crs) \\
        sig \samp \oss.\sign(crs, vk, \regis{key}, m) \\
        out_{\ver} \samp \oss.\ver(crs, vk, m, sig)
    \end{array}] \geq 1 - \negl(\lambda).
    \end{equation*}

    \paragraph{One-Shot Security:} For any QPT adversary $\adve$,
    \begin{equation*}
        \Pr[  \begin{array}{c}m_1 \neq m_2 \\ \wedge \\ out_{1,\ver} = \outtrue \\ \wedge \\ out_{2,\ver} = \outtrue\end{array}  : \begin{array}{c}
        crs \samp \oss.\setup(1^\lambda) \\
        vk, sig_1, m_1, sig_2, m_2 \samp \adve(crs) \\
        out_{1,\ver} \samp \oss.\ver(crs, vk, m_1, sig_1) \\
        out_{2,\ver} \samp \oss.\ver(crs, vk, m_2, sig_2)
    \end{array}] \leq  \negl(\lambda).
    \end{equation*}
\end{definition}

In \cref{sec:incompross}, we introduce a new security notion called incompressibility for one-shot signature in the classical oracle model. Thus, we also define OSS in the oracle model below\footnote{This is slightly different from the standard oracle model where there is a single global oracle (rather than an oracle being output for each instance of the scheme) that all the algorithms and adversaries query. We use this model for simplicity, and they are equivalent to each other.}.

\begin{definition}[One-Shot Signatures in the Classical Oracle Model]\label{predef:ossora}
     A one-shot signature scheme $\oss$ in the classical oracle model consists of the following efficient algorithms.
    \begin{itemize}
        \item $\oss.\setuporacles(1^\lambda)$: Takes in a security parameter, outputs a classical oracle $\ora$.
        \item $\oss.\genqkey^{(\cdot)}(1^\lambda)$: An oracle algorithm that takes in a security parameter, outputs a verification key $vk$ and a quantum signing key $\regis{key}$.
        \item $\oss.\sign^{(\cdot)}(vk, \regis{key}, m)$:  An oracle algorithm that takes in a verification key, a signing key $\regis{key}$, and a message $m$; outputs a classical signature $sig$.
        \item $\oss.\ver^{(\cdot)}(vk, m, sig)$:  An oracle algorithm that takes in a verification key $vk$, a message $m$ and a signature $sig$,; outputs $0$ or $1$.
    \end{itemize}

    \paragraph{Correctness:} For any message $m \in \messpa$, \begin{equation*}
        \Pr[out_{\ver} = \outtrue  : \begin{array}{c}
        \ora \samp \oss.\setuporacles(1^\lambda) \\
        vk, \regis{key} \samp \oss^\ora.\genqkey(1^\lambda) \\
        sig \samp \oss.\sign^\ora(vk, \regis{key}, m) \\
        out_{\ver} \samp \oss.\ver^\ora(vk, m, sig)
    \end{array}] \geq 1 - \negl(\lambda).
    \end{equation*}

    \paragraph{One-Shot Security:} For any query-bounded adversary $\adve$,
    \begin{equation*}
        \Pr[  \begin{array}{c}m_1 \neq m_2 \\ \wedge \\ out_{1,\ver} = \outtrue \\ \wedge \\ out_{2,\ver} = \outtrue\end{array}  : \begin{array}{c}
        \ora \samp \oss.\setuporacles(1^\lambda) \\
        vk, sig_1, m_1, sig_2, m_2 \samp \adve^\ora(1^\lambda) \\
        out_{1,\ver} \samp \oss^\ora.\ver(vk, m_1, sig_1) \\
        out_{2,\ver} \samp \oss^\ora.\ver(vk, m_2, sig_2)
    \end{array}] \leq  \negl(\lambda).
    \end{equation*}
    We define strong one-shot security similarly, with the difference that it is only required to have $sig_1 \neq sig_2$ (and not necessarily $m_1 \neq m_2$).
\end{definition}
\newpage\section{Definitional Work on Quantum Fire}\label{sec:defnwork}
In this section, we recall some existing definitions related to quantum fire, and also introduce definitions for new primitives and security notions.
\subsection{Quantum Fire}
We first recall quantum fire (\cite{STOC:BosNehZha25}). Throughout the paper, we will only consider public-key quantum fire even when we ignore writing out \emph{public-key} explicitly.

\begin{definition}[Quantum Fire]\label{predef:qfire}
     A (public-key) quantum fire scheme $\qfire$ consists of the following efficient algorithms.
    \begin{itemize}
        \item $\qfire.\setup(1^\lambda)$: Takes in a security parameter, outputs a public key $pk$ and a secret key $sk$.
        \item $\qfire.\spark(sk)$: Takes in the secret key, outputs a serial number $sn$, and a \emph{flame state} in the register $\regis{flame}$.
        \item $\qfire.\clone(pk, sn, \regis{flame})$: Takes in the public key, a serial number, a flame state register $\regis{flame}$, outputs a new flame register $\regis{clone}$ (along with the input register $\regis{flame}$).
        \item $\qfire.\ver(pk, sn, \regi)$: Takes in the public key, a serial number, and an alleged flame state, outputs $\outfalse$ or $\outtrue$.
    \end{itemize}

We require the following correctness guarantees.
    \paragraph{Verification Correctness:} \begin{equation*}
        \Pr[out_{\ver} = \outtrue  : \begin{array}{c}
        pk, sk \samp \qfire.\setup(1^\lambda) \\
        sn, \regis{flame} \samp \qfire.\spark(sk) \\
        out_{\ver} \samp \qfire.\ver(pk, sn, \regis{flame}) \\
    \end{array}] \geq 1 - \negl(\lambda).
    \end{equation*}

    \paragraph{Cloning Correctness:} For any polynomial $p(\cdot)$, \begin{equation*}
        \Pr[\forall i\in[p(\lambda)]~out_{i,\ver} = \outtrue : \begin{array}{c}
        pk, sk \samp \qfire.\setup(1^\lambda) \\
        sn, \regi_1  \samp \qfire.\spark(sk) \\
        \text{For }i=2,\dots,p(\lambda)~\regi_i \samp \qfire.\clone(pk,sn,\regi_{i-1}) \\
        \text{For }i=1,\dots,p(\lambda)~out_{i,\ver} \samp \qfire.\ver(pk, sn, \regi_i)
    \end{array}] \geq 1 - \negl(\lambda).
    \end{equation*}
\end{definition}

We introduce a stronger correctness condition called \emph{strong cloning correctness}.
\begin{definition}[Strong Cloning Correctness]
 A quantum fire scheme $\qfire$ is said to satisfy \emph{strong cloning correctness} if the following are satisfied: There exists a collection of pure states $\ket{\psi_{sk,sn}}$ such that with probability $1$ the output of $\qfire.\spark(sk)$ is $sn, \ket{\psi_{sk, sn}}$ for some $sn$. Let $M_{sk, sn}$ denote the binary outcome projective measurement onto $\ket{\psi_{sk, sn}}$. \begin{equation*}
        \Pr[b_1 = b_2 = 1  : \begin{array}{c}
        pk, sk \samp \qfire.\setup(1^\lambda) \\
        sn, \regi_{flame} \samp \qfire.\spark(sk) \\
       \regi_{clone} \samp \qfire.\clone(pk, sn, \regi_{flame}) \\
       b_1 \samp M_{sk, sn}(\regi_{flame}) \\
       b_2 \samp M_{sk, sn}(\regi_{clone})
    \end{array}] \geq 1 - \negl(\lambda).
    \end{equation*}

    Note that in oracle-based constructions, the flame states also depend on the oracle itself\footnote{Equivalently, we can consider the whole oracle as part of the secret key for purposes of this definition.} in addition to $sk, sn$.
\end{definition}

We now recall untelegraphability security for quantum fire. We slightly extend the definition of \cite{STOC:BosNehZha25} to consider collusion-resistant quantum fire, where the adversary gets multiple independent flame states.
\begin{definition}[Untelegraphability Security]\label{predef:qfireunteleg}
    Consider the following game between an adversary $\adve = (\adve_1, \adve_2)$ and a challenger.

 \paragraph{\underline{$\qfiregame{\adve}(1^\lambda)$}}
    \begin{enumerate}
    \item The challenger samples $pk, sk \samp \qfire.\setup(1^\lambda)$ and submits $pk$ to $\adve_1$ and $\adve_2$.
    \item For polynomially many number of rounds: The adversary sends an empty query, and the challenger samples $sn, \regis{flame} \samp \qfire.\spark(sk)$ and submits $sn, \regis{flame}$ to the adversary $\adve_1$.
    \item The adversary $\adve_2$ outputs a message $T$.
    \item The adversary $\adve_1$ receives $T$ and  outputs a classical string $L$.
    \item The adversary $\adve_2$ receives $L$, outputs $sn^*, \regi'$.
    \item The challenger outputs $1$ if $b=1$ where $b \samp \qfire.\ver(pk, sn^*, \regi') = \outtrue$, and otherwise it outputs $0$.
    \end{enumerate}

    We say that the quantum fire scheme $\qfire$ satisfies \emph{untelegraphability security} if for any for any polynomial-time (or query-bounded, for unconditional oracle-based schemes)  adversary $\adve$, we have 
    \begin{equation*}
        \Pr[\qfiregame{\adve}(1^\lambda) = 1] \leq \negl(\lambda).
    \end{equation*}
\end{definition}
Note that the weaker notion of \emph{non-adaptive untelegraphability security} can also be defined, where $\adve_1$ produces its output without receiving a message from $\adve_2$.

Finally, we introduce the notion of mini-scheme for quantum fire, analogous to the notion of mini-scheme for quantum money. This definition makes proofs easier, and it is easy to see that any mini-scheme for quantum fire can generically be upgraded to a full scheme using digital signatures (\cref{predef:digsig}), simply by signing the serial number (signature is made part of the new serial number) and verifying the signature on the serial number as part of verification - just like in the case of upgrading quantum money mini-schemes to full schemes (\cite{STOC:AarChr12}).
\begin{definition}[Quantum Fire Mini-Scheme]\label{predef:qfiremini}
     A public-key quantum fire mini-scheme $\qfiremini$ consists of the following efficient algorithms.
    \begin{itemize}
        \item $\qfiremini.\spark(1^\lambda)$: Takes in the security parameter, outputs a serial number $sn$, and a \emph{flame state} $\regis{flame}$.
        \item $\qfiremini.\clone(sn, \regis{flame})$: Takes in a serial number, a flame register, outputs a new flame register $\regis{clone}$ (along with the input register $\regis{flame}$).
        \item $\qfiremini.\ver(sn, \regi)$: Takes in a serial number, and an alleged flame state, outputs $\outfalse$ or $\outtrue$.
    \end{itemize}

 \paragraph{Verification Correctness:} \begin{equation*}
        \Pr[out_{\ver} = \outtrue  : \begin{array}{c}
        sn, \regis{flame} \samp \qfiremini.\spark(1^\lambda) \\
        out_{\ver} \samp \qfiremini.\ver(sn, \regis{flame}) \\
    \end{array}] \geq 1 - \negl(\lambda).
    \end{equation*}

 \paragraph{Strong Cloning Correctness:} We require that there exists a collection of pure states $\ket{\psi_{k}}$ such that with probability $1$ the output of $\qfiremini.\spark(1^\lambda)$ is $sn, \ket{\psi_{k}}$ for some $k,sn$ and  \begin{equation*}
        \Pr[b_1 = b_2 = 1  : \begin{array}{c}
      sn, \regis{flame} \samp \qfiremini.\spark(1^\lambda) \\
       \regi_{clone} \samp \qfiremini.\clone(sn, \regi_{flame}) \\
       b_1 \samp M_{k}(\regi_{flame}) \\
       b_2 \samp M_{k}(\regi_{clone})
    \end{array}] \geq 1 - \negl(\lambda).
    \end{equation*}
    where $M_{k}$ denotes the binary outcome projective measurement onto $\ket{\psi_{k}}$.

    Note that in oracle-based constructions, the flame states also depend on the oracle itself\footnote{Equivalently, we can consider the whole oracle as part of the key $k$ for purposes of this definition.} in addition to $k$.
\end{definition}

\begin{definition}[Untelegraphability Security for Quantum Fire Mini-Scheme]\label{def:qfireminiunteleg}
    Consider the following game between an adversary $\adve = (\adve_1, \adve_2)$ and a challenger.

 \paragraph{\underline{$\qfireminigame{\adve}(1^\lambda)$}}
    \begin{enumerate}
    \item The challenger samples $sn, \regis{flame} \samp \qfiremini.\spark(1^\lambda)$ and submits $sn, \regis{flame}$ to the adversary $\adve_1$.
    \item The adversary $\adve_2$ receives $sn$ and outputs a message $T$.
    \item The adversary $\adve_1$ receives $T$ and  outputs a classical string $L$.
    \item The adversary $\adve_2$ receives $L$, outputs $\regi'$.
    \item The challenger executes $\qfiremini.\ver(sn, \regi')$ outputs if it outputs $1$, otherwise it outputs $0$.
    \end{enumerate}

    We say that the quantum fire scheme $\qfiremini$ satisfies \emph{untelegraphability security} if for any QPT (or query bounded, in case of unconditional oracle-based schemes) adversary $\adve$, we have 
    \begin{equation*}
        \Pr[\qfireminigame{\adve}(1^\lambda) = 1] \leq \negl(\lambda).
    \end{equation*}
\end{definition}

\subsection{Interactive Untelegraphability Security}
We introduce the notion of interactive telegraphing security, inspired by the notion of \emph{LOCC leakage-resilience} (\cref{predef:ublrfunc}) defined by \cite{TCC:CGLR24}. For simplicity, we will explicitly define it only for mini-schemes, but the collusion-resistant case is defined similarly  to \cref{predef:qfireunteleg} (where $\adve_1$ receives multiple independent flame states).

\begin{definition}[Interactive Untelegraphability Security]\label{defn:interactive}
    Consider the following game between an adversary $\adve = (\adve_1, \adve_2)$ and a challenger.

 \paragraph{\underline{$\qfireinteractivemini{\adve}(1^\lambda)$}}
    \begin{enumerate}
    \item The challenger samples $sn, \regis{flame} \samp \qfiremini.\spark(1^\lambda)$ and submits $sn, \regis{flame}$ to the adversary $\adve_1$.
    \item The adversaries $\adve_1$ and $\adve_2$ interact classically by exchanging classical messages for any number of rounds\footnote{Note that when $\adve_1, \adve_2$ are QPT, implicitly this will be any \emph{polynomial} number of rounds.}.
    \item The adversary $\adve_2$ receives $L$, outputs $\regi'$.
    \item The challenger executes $\qfiremini.\ver(sn, \regi')$ outputs if it outputs $1$, otherwise it outputs $0$.
    \end{enumerate}

    We say that the quantum fire scheme $\qfiremini$ satisfies \emph{interactive untelegraphability security} if for any QPT (or query bounded, in case of unconditional oracle-based schemes) adversary $\adve$, we have 
    \begin{equation*}
        \Pr[\qfireinteractivemini{\adve}(1^\lambda) = 1] \leq \negl(\lambda).
    \end{equation*}
\end{definition}

\subsection{Quantum Key-Fire}\label{sec:qkeyfire}
We now introduce a new notion called \emph{quantum key-fire}. Quantum key-fire is the functional version of quantum fire, where the flame state can be used to evaluate a functionality, and we require \emph{unbounded/LOCC leakage-resilience} (\cite{TCC:CGLR24}) as our security notion, which is a generalization of the notion of untelegraphability. Quantum key-fire is to quantum fire what copy-protection (\cite{C:ALLZZ21,C:CLLZ21}) is to public-key quantum money. Just like the case of copy-protection and quantum money, it is easy to see that quantum key-fire implies quantum fire.

A quantum key-fire scheme is an \emph{quantum protection scheme} (\cref{predef:qprot}) with unbounded/LOCC leakage-resilience, with the addition of a cloning algorithm. We write the full definition for reference.
\begin{definition}[Quantum Key-Fire]\label{def:qkeyfire}
       A public-key quantum key-fire scheme for a functionality $\mathcal{F}$ (\cref{predef:clunlearn}) consists of the following efficient algorithms.
    \begin{itemize}
        \item $\qkeyfire.\prot(1^\lambda, f)$: Takes in a function $f \in F$ and outputs a \emph{flame state} $\regis{key}$.
        \item $\qkeyfire.\clone(\regis{key})$: Takes in a flame register $\regis{key}$, outputs a new register $\regis{clone}$ (along with the input register $\regis{key}$).
        \item $\qkeyfire.\ceval(\regis{key}, x):$ Takes in a flame register $\regis{key}$, and an input $x$, outputs a value.
    \end{itemize}
    
    \paragraph{Evaluation Correctness:} For all $f \in F$, for all $x$,
    \begin{equation*}
        \Pr[y = f(x)  : \begin{array}{c}
        \regis{key} \samp \qkeyfire.\prot(1^\lambda, f) \\
       y \samp \qkeyfire.\ceval(\regis{key}, x)
    \end{array}] \geq 1 - \negl(\lambda).
    \end{equation*}

\paragraph{Strong Cloning Correctness:} We require that there exists a collection of pure states $\ket{\psi_{k,f}}$ such that for all $f \in F$, with probability $1$ the output of $\qkeyfire.\prot(1^\lambda, f)$ is $\ket{\psi_{k,f}}$ for some $k$ and \begin{equation*}
        \Pr[b_1 = b_2 = 1  : \begin{array}{c}
        k, sn, \regis{flame} \samp \qkeyfire.\prot(1^\lambda, f) \\
       \regi_{clone} \samp \qkeyfire.\clone(\regi_{flame}) \\
       b_1 \samp M_{k,f}(\regi_{flame}) \\
       b_2 \samp M_{k,f}(\regi_{clone})
    \end{array}] \geq 1 - \negl(\lambda).
    \end{equation*}
    where $M_{k,f}$ denotes the binary outcome projective measurement onto $\ket{\psi_{k,f}}$.

    Note that in oracle-based constructions, the flame states also depend on the oracle itself\footnote{Equivalently, we can consider the whole oracle as part of the key $k$ for purposes of this definition.} in addition to $f,k$.

\paragraph{Security:} When $\qkeyfire$ is a key-fire scheme for some (classically-unlearnable) functionality $\mathcal{F}$ (\cref{predef:clunlearn}), we will require that $\qkeyfire$ satisfies LOCC leakage-resilience (\cref{predef:ublrfunc}) for $\mathcal{F}$. 
\end{definition}

\subsubsection{Quantum Key-Fire for Signing}
In this section, we provide the explicit definition of the special case of quantum-key fire for signing, since this the main functionality we consider in this work. 

\begin{definition}[Quantum Key-Fire for Signing]\label{def:qkeyfiresign}
       A public-key quantum key-fire scheme for signing $\qkeyfiresign$ consists of the following efficient algorithms.
    \begin{itemize}
        \item $\qkeyfiresign.\setup(1^\lambda)$: Takes in a security parameter, outputs a verification key (also called serial number) $vk$, and a \emph{key state} $\regis{flame}$.
        \item $\qkeyfiresign.\clone(vk, \regis{key})$: Takes in a key register $\regis{key}$, outputs a new register $\regis{clone}$ along with the input register $\regis{key}$.
        \item $\qkeyfiresign.\sign(vk, \regis{key}, m):$ Takes in a key register and a message $m$, outputs a signature.
        \item $\qkeyfiresign.\ver(vk, m, sig)$: Takes in a verification key (also called serial number), a message and an alleged signature, outputs $\outtrue$ or $\outfalse$.
    \end{itemize}
   
    \paragraph{Evaluation Correctness:} For all $m \in \messpa$,
    \begin{equation*}
        \Pr[\qkeyfiresign.\ver(vk, m, sig) = 1  : \begin{array}{c}
        vk, \regis{key} \samp \qkeyfiresign.\setup(1^\lambda) \\
       sig \samp \qkeyfiresign.\sign(vk, \regis{key}, m)
    \end{array}] \geq 1 - \negl(\lambda).
    \end{equation*}

\paragraph{Strong cloning correctness:} This is defined the same as before: We require that output of $\qkeyfiresign.\setup(1^\lambda)$ is a pure state $\ket{\psi_{k}}$ and that clone on input $\ket{\psi_{k}}$ outputs two copies, $\ket{\psi_{k}}\otimes \ket{\psi_{k}}$.
\end{definition}

\begin{definition}[LOCC Leakage-Resilience Security for Quantum Key-Fire for Signing]\label{def:qkeyfiresignsec}
    Consider the following game between an adversary $\adve$ and a challenger.

 \paragraph{\underline{$\lrsiggame{\adve}(1^\lambda)$}}
    \begin{enumerate}
    \item The challenger samples $vk, \regis{key} \samp \qkeyfiresign.\setup(1^\lambda)$ and submits $vk$ to $\adve$.
    \item Set $\regi_0 = \regis{key}$.
    \item \textbf{Leakage Phase:} For any number\footnote{Note that when $\adve$ is QPT, implicitly this will be any \emph{polynomial} number of rounds.} of rounds, the following is executed (starting with $i= 1$).
    \begin{enumerate}[label=\arabic*.]
        \item $\adve$ specifies a quantum leakage circuit $E_i$ that takes as input a quantum register and outputs a classical string and the updated quantum register.
        \item The challenger executes $L_i, \regi_i \samp E_i(\regi_{i - 1})$ and submits the classical string $L_i$ to $\adve$. Then $i$ is incremented.
    \end{enumerate}
    \item The challenger samples a random message $m^* \samp \messpa$.
    \item The adversary $\adve$ receives $m^*$, outputs $sig^*$.
    \item The challenger checks if $\qkeyfiresign.\ver(sn, m^*, sig^*) = \outtrue$. If so, it outputs $1$, otherwise it outputs $0$.
    \end{enumerate}

    We say that the quantum key-fire scheme $\qkeyfiresign$ satisfies \emph{LOCC leakage-resilience security} if for any QPT (or query bounded, in case of unconditional oracle-based schemes) adversary $\adve$, we have 
    \begin{equation*}
        \Pr[\lrsiggame{\adve}(1^\lambda) = 1] \leq \negl(\lambda).
    \end{equation*}
\end{definition}

\newpage\part{Technical Tools}
\section{One-Shot Signatures with Incompressibility Security}\label{sec:incompross}
In this section, we introduce a new notion called \emph{incompressibility} security for one-shot signatures in the classical oracle model, and then construct a scheme that satisfies it. The section is organized as follows. First we introduce our definition of incompressibility and give our theorem statement about existence of an incompressible OSS scheme. Then in \cref{sec:oprincompr}, we introduce a technical tool called \emph{operational incompressibility} that simplifies the proof of our incompressibility notion. Then we give our construction in \cref{sec:ossincompexists} and prove its incompressibility property in \cref{sec:ossproof}. In \cref{sec:subspaces}, we show some technical results related to subspace oracles that will be used in our proof in \cref{sec:ossproof} and might be of independent interest.

For our incompressibility notion, we consider a one-shot signature scheme $\oss$ with \emph{split oracles.}

\begin{definition}[One-Shot Signature Scheme with Split Oracles]\label{defn:splitora}
    Let $\oss$ be a one-shot signature scheme with classical oracles (\cref{predef:ossora}). Write $\ora_{\oss}$ to denote its oracle. $\oss$ is said to have \emph{split oracles} if $\ora_{\oss}$ is a tuple of (correlated) oracles $(\ora_{\oss.\genqkey}$, $\ora_{\oss.\sign}$, $\ora_{\oss.\ver}) = \ora_{\oss}$, where each algorithm of $\oss$ uses only the corresponding oracle (e.g. $\oss.\genqkey$ only uses $\ora_{\oss.\genqkey}$). We also assume that the inputs to $\ora_{\oss.\sign}$ and $\ora_{\oss.\ver}$ are of the form $vk, x$ where $vk$ denotes a verification key.
\end{definition}
Note that any one-shot signature scheme can trivially be converted into a scheme with split oracles by setting all three oracles to be the full oracle, but this will not satisfy incompressibility.

Now we define incompressibility.
\begin{definition}[Incompressibility]\label{def:ossincomprnew}
$\oss$ is said to satisfy \emph{incompressibility security} for $\ora \in \{\ora_{\oss.\sign}, \ora_{\oss.\ver}\}$ if the following is satisfied. There exists some $c > 0$ such that for any query-bounded adversary $\adve_1$ with unbounded length classical output, there exists a mapping $\mathcal{M}$ (which outputs a list\footnote{We will refer to this list as the incompressibility list.} $\mathfrak{L}$ with at most $s(\lambda)$ elements where $s(\lambda) = (q(\lambda))^c$ and $q(\lambda)$ is the number of $\ora_{\oss.\genqkey}$ queries made by $\adve_1$) such that for all query-bounded $\adve_2$, we have

\begin{equation*}
    \E\left[\sum_{\substack{vk, x:\\\ora(vk, x) \neq 0 \\ vk\not\in  \mathcal{M}(\ora_{\oss}, L, \adve_2)}}\mathsf{QW}^{(vk, x)}_{\adve_2, L, \ora_{\oss.\sign}, \ora_{\oss.\ver}} : \begin{array}{c}
              \ora_{\oss} \samp \oss.\setuporacles(1^\lambda) \\
             L \samp \adve_1^{\ora_{\oss}}(1^\lambda)
        \end{array} \right] \leq \negl(\lambda).
\end{equation*}
where $\mathsf{QW}^{(vk, x)}_{\adve_2, L, \ora_{\oss.\sign}, \ora_{\oss.\ver}}$ denotes the total query weight of $\adve_2^{\ora_{\oss.\sign}, \ora_{\oss.\ver}}(L)$ to the oracle $\ora$ on oracle input $vk, x$, and $\adve_1, \adve_2$ are allowed to share an unbounded length classical randomness in advance (i.e. before oracles are sampled).

If incompressibility is satisfied for both $\ora \in \{\ora_{\oss.\sign}, \ora_{\oss.\ver}\}$, then we will simply say that OSS satisfies incompressibility.
\end{definition}
We define \emph{strong incompressibility} similarly, where we require the list $\mathcal{M}$ to output pairs $(vk, z)$ and the summation above is taken over $(vk, z) \not\in  \mathcal{M}(\ora_{\oss}, L, \adve_2)$. It is easy to see that strong incompressibility for $\ora_{\oss.\ver}$ follows from incompressibility for $\ora_{\oss.\ver}$ and strong one-shot security.

We claim that a one-shot signature scheme with incompressibility security exists.
\begin{theorem}\label{thm:ossincompexists}
Relative to a classical oracle, there exists a one-shot signature scheme for $1$-bit messages that satisfies one-shot security and incompressibility security.
\end{theorem}

We give our construction in \cref{sec:ossincompexists}  and prove its incompressibility in \cref{sec:ossproof}.

\subsection{Operational Incompressibility}\label{sec:oprincompr}
As a technical tool for proving incompressibility, we define the notion of operational incompressibility.
\begin{definition}[Operational Incompressibility]\label{def:oprincompr}
    $\oss$ is said to satisfy \emph{operational incompressibility security} for $\ora \in \{\ora_{\oss.\sign}, \ora_{\oss.\ver}\}$ if the following is satisfied. There exists some $c > 0$ such that for any query-bounded adversary $\adve_1$ with unbounded-length classical output, for all query-bounded $\adve_2$, we have
\begin{align*}
      \Pr[\begin{array}{c}
             \forall i\in [s(\lambda) + 1]~\ora(vk_i, x_i) \neq 0 \\
             \wedge \\
            (vk_i)_{i\in [s(\lambda) + 1]}~\text{distinct}
      \end{array}  : \begin{array}{c}
              \ora_{\oss} \samp \oss.\setuporacles(1^\lambda) \\
             L \samp \adve_1^{\ora_{\oss}}(1^\lambda) \\
             (vk_i, x_i)_{i \in [s(\lambda) + 1]} \samp \adve_2^{\ora_{\oss.\sign}, \ora_{\oss.\ver}}(L)
        \end{array}]
        \leq \negl(\lambda).
\end{align*}

where $s(\lambda) = (q(\lambda))^c$ and $q(\lambda)$ is the number of $\ora_{\oss.\genqkey}$ queries made by the adversary $\adve_1$ and $\adve_1, \adve_2$ are allowed to share an unbounded length classical randomness in advance (i.e. before oracles are sampled).
\end{definition}

\begin{theorem}\label{thm:incomprequiv}
If an OSS scheme satisfies operational incompressibility, then it also satisfies incompressibility.
\end{theorem}

Now we prove this theorem.
Let $\oss$ be a scheme that satisfies operational incompressibility. Fix $\ora \in \{\ora_{\oss.\sign}, \ora_{\oss.\ver}\}$. For any $vk$, define 
\begin{equation*}
    q_{vk} = \sum_{x: \ora(vk,x) \neq 0}\mathsf{QW}^{vk, x}_{\adve_2, L, \ora_{\oss.\sign}, \ora_{\oss.\ver}}
\end{equation*}
The dependence on $\adve_2, L, \ora_{\oss}$ are implicit.

Let $vk_1, \dots, vk_{s}, vk_{s+1}$ be the inputs such that 
\begin{itemize}
\item $q_{{vk}_i}\geq q_{{vk}_{i+1}}$ for $i \in [s]$
    \item $q_{{vk}_{s+1}} \geq q_{{vk}'}$  for any  ${vk}' \not\in \{{vk}_i\}_{i \in [s+1]}$
\end{itemize}
 That is, ${vk}_i$ are the inputs with the highest \emph{valid} query weights. Note that these all implicitly depend on $\ora_{\oss}, L, \adve_2$.
 
 We set $\mathcal{M}(\ora_{\oss}, L, \adve_2)$ to be $\{{vk}_1, \dots, {vk}_s\}$. Now suppose for a contradiction that incompressibility property is not satisfied, that is, there exists some $\adve_1, \adve_2$ and polynomial $t(\cdot)$ such that for infinitely many $\lambda > 0$,
\begin{equation*}
p_{out} =    \E\left[\sum_{\substack{{vk}:\\ {vk} \not\in  \mathcal{M}(\ora_{\oss}, L, \adve_2)}} q_{vk} : \begin{array}{c}
              \ora_{\oss} \samp \oss.\setuporacles(1^\lambda) \\
             L \samp \adve_1^{\ora_{\oss}}(1^\lambda)
        \end{array} \right] \geq \frac{1}{t(\lambda)}.
\end{equation*}

Let $u(\lambda)$ denote the number of queries to $\ora$ by $\adve_2$. Consider the following adversary $\adve_2'$ for the operational incompressibility game. \paragraph{\underline{$\adve_2'$}}
\begin{enumerate}
    \item Set $D = \emptyset$.
    \item For $n(\lambda) = u(\lambda)\cdot t(\lambda)\cdot s(\lambda) \cdot \lambda^2\cdot (s(\lambda)+1)$ times (in $s+1$ \emph{groups}):
    \begin{enumerate}[label=\arabic*.]
        \item Simulate $\adve_2(L)$, but measure a random query to $\ora$ to obtain $vk,x$.
        \item If $\ora({vk},x) \neq 0$ and $vk \not\in D$\footnote{Here we mean that there is no $x$ such that $({vk}, x)\in D$}, add $({vk}, x)$ to $D$ and skip to the next \emph{group}.
    \end{enumerate}    
    \item Output $D$.
\end{enumerate}

We will show that we  can violate operational incompressibility using $\adve'_2$. We continue our proof by considering two cases/events.  Define $E_1$ to be the event that $q_{{vk}_{s+1}} \geq \frac{1}{t(\lambda)\cdot s(\lambda) \cdot \lambda}$, and $E_2$ to be its complement. Thus we have either $\Pr[E_1] \geq \frac{1}{4\cdot t\cdot u - 2}$ or $\Pr[E_2] \geq 1 - \frac{1}{4\cdot t\cdot u - 2}$.

Now we consider the case $\Pr[E_1] \geq \frac{1}{4\cdot t\cdot u - 2}$.  We claim that $\adve_1, \adve_2'$ wins the operational incompressibility game with probability at least $ \frac{1}{8\cdot t\cdot u - 4}$. We condition on the event $E_1$, and we will show that the adversaries will win with probability $\geq 1 - \negl(\lambda)$. Observe that for each ${vk}_i$ with $i \in [s+1]$, we have that the probability of ${vk}_i \in D$ is $1 - \exp{-\frac{n}{u \cdot t\cdot s \cdot \lambda}} \geq 1 - \exp{-\lambda}$ since the weight of each ${vk}_i$ is at least $\frac{1}{t\cdot s \cdot \lambda}$. Thus, by the union bound, we get that ${vk}_1, \dots, {vk}_{s+1} \in D$ with overwhelming probability, which indeed violates incompressibility security.

Now we consider the case $\Pr[E_2] \geq 1 - \frac{1}{4\cdot t\cdot u - 2}$. Let $\mathsf{GOOD}$ be the event that 
\begin{equation*}
    \sum_{\substack{{vk}: {vk} \not\in  \mathcal{M}(\ora_{\oss}, L, \adve_2)}} q_{vk} \geq \frac{1}{2\cdot t}
\end{equation*}
Since expectation of this value ($p_{out}$) is greater than $\frac{1}{t(\lambda)}$ and since $\Pr[E_2] \geq 1 - \frac{1}{4\cdot t\cdot u - 2}$, we get $\Pr[E_2 \wedge \mathsf{GOOD}] \geq \frac{1}{4\cdot t\cdot u - 2}$ by an application of Markov's inequality and union bound. 

We claim that $\adve_1, \adve_2'$ wins the operational incompressibility game with overwhelming probability conditioned on $E_2 \wedge \mathsf{GOOD}$, which happens with non-negligible probability. Let $G$ denote the complement set of $\mathcal{M}(\adve_2, \ora_\oss, L)$. Let ${vk}^*_1, \dots, {vk}^*_{s+1}$ be the values added to the database $D$ during each group during execution of $\adve_2'$, in the order of addition. In particular, if no item was added during a group, we set ${vk}^*_i = \bot$. We now claim that with overwhelming probability, ${vk}^*_1, \dots, {vk}^*_{s+1} \neq\bot$. Note that once we establish this, we will have shown that we can violate operational incompressibility with non-negligible probability. Now to prove our claim, first observe that ${vk}^*_1 \in G$ with overwhelming probability since $q_{G} := \sum_{{vk} \in G}q_{vk} \geq \frac{1}{2t}$ (since we conditioned on $\mathsf{GOOD}$). Then, we claim that ${vk}^*_2 \neq \bot$ with overwhelming probability. To see this, observe that $q_{G} \geq \frac{1}{2t}$ and for any possible value of ${vk}_1^* \in G$, we have $q_{{vk}_1^*} \leq \frac{1}{t\cdot s\cdot \lambda}$ (since we conditioned on $E_2$) and thus $q_{G} - q_{{vk}_1^*} \geq \frac{1}{2t}$. Continuing like this, we can show that each ${vk}_i^*$ is $\in G$ with overwhelming probability, no matter what the previous values ${vk}_1^*, \dots, {vk}_{i-1}^*$ are. This completes case 2.

Observe that we showed violations of the  operational incompressibility game in both cases. This completes the proof.
\subsection{Construction and Proofs of Security}\label{sec:ossincompexists}
In this section, we prove \cref{thm:ossincompexists}: We show that there exists a one-shot signature scheme that satisfies incompressibility (\cref{def:ossincomprnew}) relative to a classical oracle. Our construction will be a delicately modified version of the construction of \cite{C:ShmZha25}. 

Now we move onto our construction. 

Define the following parameters: Let $q(\lambda) = 16\lambda, r(\lambda) = q(\lambda)\cdot(\lambda - 1), n(\lambda) = r(\lambda) + \frac{3}{2}q(\lambda), k(\lambda) = n(\lambda)$. 

\paragraph{\underline{$\oss.\setuporacles(1^\lambda)$}}
\begin{enumerate}
    \item Sample a random permutation $\pi$ on $\zo^n$. Let $H(x), J(x)$ denote the first $r$ and last $n - r$ bits of $\pi(x)$.
    \item For each $y \in \zo^r$, sample a random full column-rank matrix  $A_y \in \F_2^{k \times (n-r)}$ and a random vector $b_y$ in $\F_2^k$. 
    \item Set $\mathcal{P}(x) = (y, A_y \cdot J(x) + b_y)$ where $y = H(x)$.
    \item Set $\mathcal{P}^{-1}(y,v) = \pi^{-1}(y || z)$ if $v = A_y\cdot z + b_y$ for some $z \in \F_2^{n - r}$, otherwise $\bot$.
    \item Set $\mathcal{D}(y,v) = 1$ if $v^T A_y = 0$ and $v \neq 0$, otherwise $0$.
    \item Set $\mathcal{D}_0(y,m,v) = 1$ if $v = A_y\cdot z + b_y$ for some  $z \in \F_2^{n - r}$ and the first bit of $v$ is $m$; otherwise $0$.
    \item \textbf{The oracles:}   Set $\ora_\ver = \mathcal{D}_0$, $\ora_\sign = \mathcal{D}$, $\ora_\genqkey = (\mathcal{P},\mathcal{P}^{-1}, \mathcal{D}, \mathcal{D}_0)$. Set the oracle $\ora_\oss$ to be $(\ora_\ver, \ora_\sign, \ora_\genqkey)$.
\end{enumerate}

\paragraph{\underline{$\oss.\genqkey^{\ora_\genqkey}(1^\lambda)$}}
\begin{enumerate}
    \item Parse $\ora_\genqkey = (\mathcal{P},\mathcal{P}^{-1}, \mathcal{D}, \mathcal{D}_0)$.
    \item Prepare the states $\ket{0^n},\ket{0^r},\ket{0^k}$ over the registers $\regi_1, \regi_2, \regi_3$.
    \item Apply $H^{\otimes n}$ to $\regi_1$.
    \item Execute the oracle $\mathcal{P}$ with input register $\regi_1$ and output register $(\regi_2, \regi_3)$.
     \item Execute the oracle $\mathcal{P}^{-1}$ with input register $(\regi_2, \regi_3)$ and output register $\regi_1$.
     \item Measure the register $\regi_2$, let $y$ be the measurement outcome.
     \item Output $y, \regi_3$.
\end{enumerate}

\paragraph{\underline{$\oss.\ver^{\ora_\ver}(vk, sig, m)$}}
\begin{enumerate}
\item Parse $\ora_\ver = \mathcal{D}_0$.
    \item Output $\mathcal{D}_0(vk, m, sig) = 1$.
\end{enumerate}

\paragraph{}
Our signing algorithm is the same as the signing algorithm of \cite{C:ShmZha25} (which in turn is the same as the signing mechanism of \cite{STOC:AGKZ20,C:Shmueli22}). For completeness, we recall it here.
\paragraph{\underline{$\oss.\sign^{\ora_\sign}(\regis{key}, m)$}}
\begin{enumerate}
    \item Parse $\ora_\sign = \mathcal{D}$.
    \item Parse $(vk, \regis{key}') = \regis{key}$
    \item For $j = 1, \dots, \lambda$:
    \begin{enumerate}[label=\arabic*.]
        \item Measure the first qubit of $\regis{key}'$. In case the outcome is $m$, measure the rest of $\regis{key}'$ to obtain $sig'$, and output $m || sig'$ and terminate.

        \item Apply $H^{\otimes k}$ to $\regis{key}'$, then apply the following mapping coherently (writing the outcome to a new output register) to $\regis{key}'$ and measure the output bit: $(vk, z) \to 1$ if $\mathcal{D}(vk, z) = 1$ or if $z = 0^k$, otherwise $(vk, z) \to 0$. Then, apply $H^{\otimes k}$ again to $\regis{key}'$. Proceed to the next iteration $j \gets j + 1$.
    \end{enumerate}

    \item If all $\lambda$ iterations failed, output $\bot$.
  \end{enumerate}
  
\begin{theorem}
$\oss$ satisfies correctness.
\end{theorem}
\begin{proof}
While the correctness of the scheme follows from previous work (\cite{C:ShmZha25}, which in turn uses the correctness proof of \cite{STOC:AGKZ20,C:Shmueli22}), for ease of reference, we recall it below. Fix any $m \in \zo$ for the rest of the proof.

First, we assume that for all but (at most) $2\cdot 2^{2r(\lambda) - n(\lambda)}$ many values of $y$, we have that $A_{y}$ has at least a single column whose first entry is $1$. Since the probability of this happening for any individual $y$ value is $1 - (1/2)^{n-r}$, by a simple Chernoff bound it is easy to see that our assumption holds with probability at least $1 - \exp{-\lambda}$.

It is easy to see that $\genqkey$ outputs $y, \sum_{u \in \mathsf{ColSpan}(A_{y})} \ket{v + b_y}$ for a uniformly random $y$, thus we start with this state in the register $\regis{key}$. We assume that $A_{y}$ has at least a single column whose first entry is $1$. Note that we had already assumed that this is already true for all but (at most) $2\cdot 2^{2r(\lambda) - n(\lambda)}$ many verification keys, and since $y$ is chosen uniformly at random from $2^r$ verification keys, our assumption holds with probability at least $1 - \exp{-\lambda}$.

 Note that by definition signatures for $m$ correspond to vectors in $\mathsf{ColSpan}(A_{y}) + b_{y}$ that start with the bit $m$. Since there is at least one column whose first entry is $1$, it is easy to see that there is an equal number of vectors that start with either bit in $\mathsf{Colspan}(A_y) + b_y$. Thus, the probability of signing in the first iteration of the algorithm is accordingly $1/2$. The algorithm makes $\lambda$ tries and we would like to argue it succeeds with probability $1 - 2^{\lambda}$. We will prove that each try indeed succeeds with at least probability $1/2$ (so far it is clear only for the first iteration), which will complete the proof. 
 
 Now let $S := \mathsf{ColSpan}(A_{y})$ and let $S_{0}$ the set of vectors in $S$ that start with $0$, which is a subspace of $S$ with one less dimension. Observe that after the failed first try, the state we have in $\regis{key}'$ is the state $\sum_{u \in S_{0}} \ket{b_y + x + u}$, for some value $x \in S$ which will not matter for our proof (for the interested reader, $x$ will be the zero vector if $b_y$ starts with the bit $\overline{m}$, and otherwise it will be a vector that \emph{completes} $S_0$ to $S$). Now we claim that that if we have a state of the form $\sum_{u \in S_{0}} (-1)^{\langle z, u \rangle} \ket{b_y + x + u}$, for any $z \in S_{0}^{\bot}$, after performing the correction part of the signing algorithm, we obtain a state of the form $\sum_{u \in S} (-1)^{\langle z', u \rangle} \ket{b_y + x + u}$ for some $z' \in S_{0}^{\bot}$. To see this, note that after executing $H^{\otimes k}$ the state we get is $\sum_{v \in S^{\bot}_{0}} (-1)^{\langle b_y + x, v \rangle} \ket{z + v}$, and since $S_{0}$ is a subspace of $S$ with one less dimension, the subspace $S^{\bot} := \mathsf{ColSpan}(A_{y})^{\bot}$ is a subspace of $S^{\bot}_{0}$ with one less dimension. Thus, when we measure the output qubit of $\mathcal{D}$ (note that we are not using the output of $\mathcal{D}$ and effectively discarding its answer) the state collapses to a state of the form $\sum_{v \in S^{\bot}} (-1)^{\langle b_y + x, v \rangle} \ket{z' + v}$ for some value $z'  \in S_{0}^{\bot}$. When we execute the second Hadamard, we get a state of the form $\sum_{u \in S} (-1)^{\langle z', u \rangle} \ket{b_y + x + u}$, which is equal to $\sum_{u \in S} (-1)^{\langle z', u \rangle} \ket{b_y + u}$ since $x \in S$. Thus, in summary, as long as we start an iteration with a state of this form, we end the iteration with a state of this form (even the signing attempt fails). With this, it is easy to see that each iteration succeeds with probability $1/2$ independently. This finishes the proof.
\end{proof}

\begin{theorem}
$\oss$ satisfies strong one-shot security.
\end{theorem}
\begin{proof}
Observe that the oracle $\mathcal{D}_0$ in our construction can be implemented using $\mathcal{P}^{-1}$. Since strong one-shot security relative to $\mathcal{P}, \mathcal{P}^{-1}, \mathcal{D}$ has already been proven by \cite{C:ShmZha25}, the security of our scheme follows.
\end{proof}

\begin{theorem}\label{thm:ossincomprthmstatement}
$\oss$ satisfies incompressibility.
\end{theorem}
We prove this theorem in \cref{sec:ossproof}.
\subsection{Properties of Subspace Oracles}\label{sec:subspaces}
In this section, we recall some results about subspace oracles and also prove some new technical results. Throughout the section, we will write $\ora_{S'}$ to denote the oracle that checks membership in $S'$ (outputs $1$ if and only if the input is inside $S'$) where $S'$ is a subspace.

First we recall the subspace-hiding oracles.
\begin{lemma}[Subspace-Hiding Oracles \cite{C:ShmZha25}]\label{thm:sho}
Let $k, r, s \in \mathds{N}$ be such that $r + s \leq k$ and let $S$ be a subspace of $ \F_{2}^{k}$ of dimension $r$. Let $\mathcal{U}$ be the uniform distribution over superspaces $T$ of $S$ of dimension $r + s$. 

Then, for any adversary $\adve$ making at most $q$ quantum queries, we have the following.
$$
\{
\adve^{\ora_{S}}()
\}
\approx_{O\left( \frac{ q \cdot s }{ \sqrt{ 2^{k - r - s} } } \right)}
\{
\adve^{\ora_{T}}()
\;
:
\;
T \gets \mathcal{U}
\} \enspace .
$$
\end{lemma}

Now we prove the following theorem, which significantly improves upon a similar result of \cite{C:ShmZha25}.

\begin{theorem}\label{thm:subspacehardness}
 Let $d_2 \leq d_1 \leq n$. Let $S$ be a subspace of dimension $d_1$ of $\F_2^n$, and let   $\mathcal{U}$ denote the uniform distribution of subspaces of $S$ of dimension $d_2$. Then, for any adversary $\adve$ making at most $q$ quantum queries, we have
    \begin{equation*}
        \Pr[v \in T \setminus \{0\} : \begin{array}{c}
             T \samp \mathcal{U} \\
             v \samp \adve^{\ora_{T^\perp}} 
        \end{array}] \leq  \frac{2^{d_1-d_2} - 1}{2^{d_1} - 1} + q\cdot \sqrt{ \frac{2^{d_1-d_2} - 1}{2^{d_1} - 1}}
    \end{equation*}
\end{theorem}
\begin{proof}
Let $p_{\adve} = \Pr[v \in T \setminus \{0\} : \begin{array}{c}
             T \samp \mathcal{U} \\
             v \samp \adve^{\ora_{T^\perp}} 
        \end{array}]$. We observe that for any adversary $\mathcal{B}$ that is not given query access to $\ora_{T^\perp}$ (call such an adversary an \emph{oblivious adversary}), the probability of outputting a vector $v \in T^\perp \setminus S^\perp$ is $p = \frac{2^{d_1-d_2} - 1}{2^{d_1} - 1}$. Thus, by \cref{prethm:bbbv97}, if we modify the first oracle call of $\adve$ to instead use $\mathcal{O}_{S^\perp}$, the probability of winning decreases by at most $\sqrt{p}$, since the part of the adversary before its first call to $\ora_{T^\perp}$ can be considered an oblivious adversary. Further, once we modify the first oracle call of the adversary to $\ora_{S^\perp}$, now the part until the second oracle call is also an oblivious adversary. Continuing this argument for all $q$ oracle calls, we can conclude that when $\adve$ is run with the oracle $\ora_{S^\perp}$, its probability of winning is at least $p_{\adve} - q\cdot \sqrt{p}$. However, by the same oblivious algorithm argument as above, we have $p_{\adve} - q\cdot \sqrt{p} \leq p$. Thus, $p_{\adve} \leq q\cdot \sqrt{p} + p$.

\end{proof}
\subsection{Proof of Incompressibility}\label{sec:ossproof}
In this section, we prove \cref{thm:ossincomprthmstatement}. In the proof, instead of setting $s(\lambda) = (q(\lambda))^c$ for some $c > 0$, we actually set $s(\lambda) = 2q(\lambda)$, which is even stronger.

Now, we move onto our proof. We will explicitly prove only the case of $\ora = \ora_{\oss.\ver}$, and the case of $\ora = \ora_{\oss.\sign}$ follows essentially the same way (note the symmetry between $A_y$ and $A_y^{\perp}$ in our proof). 

Suppose for a contradiction that incompressibility is not satisfied: There exists query-bounded $\adhocadv_1, \adhocadv_2$ and a polynomial $p(\cdot)$ such that $p_{win}  \geq \frac{1}{p(\lambda)}$, where we define \begin{align*}
       p_{win} = \Pr[
        \begin{array}{c}
            \forall i\in [s(\lambda) + 1]~\ora_{\ver}(vk_i, m_i, sig_i) \neq 0 \\ 
             \wedge \\
            {(vk_i)}_{i\in [s(\lambda) + 1]}~\text{distinct}
      \end{array}
        : \begin{array}{c}
              (\ora_{\oss.\genqkey}, \ora_{\oss.\sign}, \ora_{\oss.\ver}) \samp \oss.\setuporacles(1^\lambda) \\
             L \samp \adhocadv_1^{(\ora_{\oss.\genqkey}, \ora_{\oss.\sign}, \ora_{\oss.\ver})}(1^\lambda) \\
             (vk_i, m_i, sig_i)_{i \in [s(\lambda) + 1]} \samp \adhocadv_2^{\ora_{\oss.\sign}, \ora_{\oss.\ver}}(L)
        \end{array}]
    \end{align*}
    where $s(\lambda) = 2q(\lambda)$ and $q(\lambda)$ is the number of queries to $\ora_{\oss.\genqkey}$ made by $\adhocadv_1$.

 Now define a sequence of hybrids for some adversaries $\adve_1, \adve_2$.
    \paragraph{$\hyb_0$:} We define this hybrid as follows.
    \begin{mdframed}
        \underline{$\hyb_0$}

        \begin{enumerate}
        \item Sample a random permutation $\pi$ on $\zo^n$.
    \item For each $y \in \zo^r$, sample a random full column-rank matrix  $A_y \in \F_2^{k \times (n-r)}$ and a random vector $b_y$ in $\F_2^k$. 
    \item Set $\mathcal{P}(x) = (y, {A_{{y}}} \cdot J(x) + b_y)$ where $y = H(x)$.
    \item Set $\mathcal{P}^{-1}(y,v) = \pi^{-1}(y || z)$ if $v = {A_{{y}}}\cdot z + b_y$ for some $z \in \F_2^{n - r}$, otherwise $\bot$.
    \item Set $\mathcal{D}(y,v) = 1$ if $v^T {A_{{y}}} = 0$ and $v\neq 0$, otherwise $0$.
    \item Set $\mathcal{D}_0(y,m,v) = 1$ if $v = {A_{{y}}}\cdot z + b_y$ for some  $z \in \F_2^{n - r}$ and the first bit of $v$ is $m$; otherwise $0$.
    
            \item  Set $\ora_\ver = \mathcal{D}_0$, $\ora_\sign = \mathcal{D}$, $\ora_\genqkey = (\mathcal{P},\mathcal{P}^{-1}, \mathcal{D}, \mathcal{D}_0)$.

            \item Sample $L \samp \adve_1^{(\ora_{\oss.\genqkey}, \ora_{\oss.\sign}, \ora_{\oss.\ver})}(1^\lambda)$.
            \item Sample $(vk_i, m_i, v_i)_{i \in [q(\lambda) + 1]} \samp \adve_2^{\ora_{\oss.\sign}, \ora_{\oss.\ver}}(L, (b_y)_{y \in \zo^r})$.
            \item Check that $vk_i$ are all distinct, and that $v_i + b_{vk_i}$ starts with the bit $m_i$ and that $v_i \in \mathsf{Colspan}(A_{vk_i}) \setminus \{0\}$ for all $i \in [q + 1]$. If so, output $1$, otherwise output $0$.
        \end{enumerate}
    \end{mdframed}
    We set $\adve_1$ to $\adhocadv_1$ and $\adve_2$ to be the following adversary. It first runs $\adve_1$ to obtain $(vk_i, m_i, sig_i)_{i \in [s(\lambda) + 1]}$, then it checks if there are at least $q+1$ values of ${i \in [s(\lambda) + 1]}$ such that $sig_i \neq b_{vk_i}$. If so, let $I \subset [s(\lambda) + 1]$ (with $|I| = q + 1$) denote those values and $\adve_2$ outputs $(vk_i, m_i, sig_i - b_{vk_i})_{i \in I}$. Otherwise, it outputs $\bot$.

Then, we claim that we have $\Pr[\hyb_0 = 1] \geq \frac{1}{2p}$. To see this, first define the event $E_1$ to be the event in the definition of $p_{win}$. Define $E_2$ to be the event that for at least $q+1$ values of $i \in [s+1]$, we have $sig_i \neq b_{vk_i}$ (where the same sampling procedure is the same as in the definition of $p_{win}$). Now, observe that we have $\Pr[E_1 \wedge E_2] + \Pr[E_1 \wedge \overline{E_2}] = p_{win} \geq \frac{1}{p}$. Now we claim that $ \Pr[E_1 \wedge \overline{E_2}] \leq\negl(\lambda)$. Observe that once we prove this, $\Pr[\hyb_0 = 1] \geq \frac{1}{2p}$ follows easily for $\adve_1, \adve_2$ we chose above.

Now we prove $\Pr[E_1 \wedge \overline{E_2}] \leq \negl(\lambda)$. Observe that once $E_1 \wedge \overline{E_2}$ occurs, it means that we have recovered the value $b_{vk}$ as part of the output $\adhocadv_2$ for at least $q+1$ different $vk$ values. Also note that each query to $\ora_{\oss.\ver}$ made by $\adhocadv_2$ can be simulated by making $2^{n-r}$ queries to an oracle that verifies equality for $b_y$ values (i.e. the oracle that outputs $1$ on $b_y, y$ and $0$ otherwise). Thus, $\adhocadv_1$ makes $q$ queries to the oracle $\ora(y) = b_y$ and $\adhocadv_2$ makes $\poly(\lambda)\cdot 2^{n-r}$ queries to the verification/equality check oracle for $b_y$ values. Thus, by \cref{thm:randoraincomp} and by the values of the parameters $n,r,k$, we get that (except with negligibly small probability) $\adhocadv_2$ cannot output $b_{vk}$ values for $q+1$ different $vk$ values. Hence, $\Pr[E_1 \wedge \overline{E_2}] \leq \negl(\lambda)$ as desired.

    \paragraph{$\hyb_1$:} Let $t(\lambda)$ denote the total number of queries in the experiment. Set $\ell(\lambda) = 1024\cdot\csrd\cdot p(\lambda)\cdot (t(\lambda))^3$ where $\csrd$ is the constant from \cref{thm:srd}. 
    \begin{mdframed}
        \underline{$\hyb_1$}

        \begin{enumerate}
        \item Sample a random permutation $\pi$ on $\zo^n$.
    \item For each $y \in \zo^r$, sample a random full column-rank matrix  $A_y \in \F_2^{k \times (n-r)}$ and a random vector $b_y$ in $\F_2^k$. 
    \item \textcolor{red}{Sample a random function $SRD: \zo^{r} \to [\ell]$}.
    \item Set $\mathcal{P}(x) = (y, \textcolor{red}{A_{{SRD(y)}}} \cdot J(x) + b_y)$ where $y = H(x)$.
    \item Set $\mathcal{P}^{-1}(y,v) = \pi^{-1}(y || z)$ if $v = \textcolor{red}{A_{{SRD(y)}}}\cdot z + b_y$ for some $z \in \F_2^{n - r}$, otherwise $\bot$.
    \item Set $\mathcal{D}(y,v) = 1$ if $v^T \textcolor{red}{A_{{SRD(y)}}} = 0$ and $v\neq 0$, otherwise $0$.
    \item Set $\mathcal{D}_0(y,m,v) = 1$ if $v = \textcolor{red}{A_{{SRD(y)}}}\cdot z + b_y$ for some  $z \in \F_2^{n - r}$ and the first bit of $v$ is $m$; otherwise $0$.
    
            \item  Set $\ora_\ver = \mathcal{D}_0$, $\ora_\sign = \mathcal{D}$, $\ora_\genqkey = (\mathcal{P},\mathcal{P}^{-1}, \mathcal{D}, \mathcal{D}_0)$.

            \item Sample $L \samp \adve_1^{(\ora_{\oss.\genqkey}, \ora_{\oss.\sign}, \ora_{\oss.\ver})}(1^\lambda)$.
            \item Sample $(vk_i, m_i, v_i)_{i \in [q(\lambda) + 1]} \samp \adve_2^{\ora_{\oss.\sign}, \ora_{\oss.\ver}}(L, (b_y)_{y \in \zo^r})$.
            \item Check that $vk_i$ are all distinct, and that $v_i + b_{vk_i}$ starts with the bit $m_i$ and that $v_i \in\mathsf{Colspan}(\textcolor{red}{A_{SRD(vk_i)}}) \setminus \{0\}$ for all $i \in [q + 1]$. If so, output $1$, otherwise output $0$.

        \end{enumerate}
    \end{mdframed}
We have $\hyb_0 \approx_{1/16p} \hyb_1$ by \cref{thm:srd}, since the whole experiment can be implemented with oracle access to the oracle $\ora_{A}(y) = A_y$ (where each $\ora_A(y)$ is sampled independently from the full-column-rank matrix distribution as above)

   \paragraph{$\hyb_2$:} We define this hybrid as follows.
    \begin{mdframed}
        \underline{$\hyb_2$}

        \begin{enumerate}
        \item Sample a random permutation $\pi$ on $\zo^n$.
    \item For each $y \in \zo^r$, sample a random full column-rank matrix  $A_y \in \F_2^{k \times (n-r)}$ and a random vector $b_y$ in $\F_2^k$. 
    \item \textcolor{red}{For each $y \in \zo^r$, sample $A'_y \in \F_2^{k \times (n-\frac{r}{2})}$ and $A''_y \in \F_2^{k \times (n-\frac{r}{2})}$ so that $\mathsf{Colspan}(A'_y)$ is a random superspace of $\mathsf{Colspan}(A_y)$ of dimension $n-\frac{r}{2}$ and $\mathsf{Colspan}(A''_y)^\perp$ is a random superspace of $\mathsf{Colspan}(A_y)^\perp$ of dimension $\frac{3r}{2}$.}
    
    \item {Sample a random function $SRD: \zo^{r} \to [\ell]$}.
    \item Set $\mathcal{P}(x) = (y, {A_{{SRD(y)}}} \cdot J(x) + b_y)$ where $y = H(x)$.
    \item Set $\mathcal{P}^{-1}(y,v) = \pi^{-1}(y || z)$ if $v = {A_{{SRD(y)}}}\cdot z + b_y$ for some $z \in \F_2^{n - r}$, otherwise $\bot$.
    \item Set $\mathcal{D}(y,v) = 1$ if $v^T \textcolor{red}{A''_{{SRD(y)}}} = 0$ and $v \neq 0$, otherwise $0$.
    \item Set $\mathcal{D}_0(y,m,v) = 1$ if $v = \textcolor{red}{A'_{{SRD(y)}}}\cdot z + b_y$ for some  $z \in \F_2^{n - r}$ and the first bit of $v$ is $m$; otherwise $0$.
    
            \item  Set $\ora_\ver = \mathcal{D}_0$, $\ora_\sign = \mathcal{D}$, $\ora_\genqkey = (\mathcal{P},\mathcal{P}^{-1}, \mathcal{D}, \mathcal{D}_0)$.
            \item Sample $L \samp \adve_1^{(\ora_{\oss.\genqkey}, \ora_{\oss.\sign}, \ora_{\oss.\ver})}(1^\lambda)$.

      \item Sample $(vk_i, m_i, v_i)_{i \in [q(\lambda) + 1]} \samp \adve_2^{\ora_{\oss.\sign}, \ora_{\oss.\ver}}(L, (b_y)_{y \in \zo^r})$.
            \item Check that $vk_i$ are all distinct, and that $v_i + b_{vk_i}$ starts with the bit $m_i$ and that $v_i \in \mathsf{Colspan}({A_{SRD(vk_i)}}) \setminus \textcolor{red}{\mathsf{Colspan}({A''_{SRD(vk_i)}})}$ for all $i \in [q + 1]$. If so, output $1$, otherwise output $0$.
        \end{enumerate}
    \end{mdframed}
We have $\hyb_1 \approx \hyb_2$ by the following argument. First, note that the changes to the oracles $\mathcal{D}, \mathcal{D}_0$ make negligible difference by \cref{thm:sho} (since we only deal with $\ell$, which is polynomial, many subspaces). Also note that \cref{thm:sho} allows the adversary to choose the base subspace, thus, the oracles $\mathcal{P},\mathcal{P}^{-1}$ and so on do not preclude us from invoking the theorem. After we modify the oracles, the change to the winning condition also makes negligible difference by \cref{thm:subspacehardness}.

   \paragraph{$\hyb_3$:} We define this hybrid as follows.
    \begin{mdframed}
        \underline{$\hyb_3$}

        \begin{enumerate}
        \item Sample a random permutation $\pi$ on $\zo^n$.
    \item For each $y \in \zo^r$, sample a random full column-rank matrix  $A_y \in \F_2^{k \times (n-r)}$ and a random vector $b_y$ in $\F_2^k$. 
    \item{For each $y \in \zo^r$, sample $A'_y \in \F_2^{k \times (n-\frac{r}{2})}$ and $A''_y \in \F_2^{k \times (n-\frac{r}{2})}$ so that $\mathsf{Colspan}(A'_y)$ is a random superspace of $\mathsf{Colspan}(A_y)$ of dimension $n-\frac{r}{2}$ and $\mathsf{Colspan}(A''_y)^\perp$ is a random superspace of $\mathsf{Colspan}(A_y)^\perp$ of dimension $\frac{3r}{2}$ (or, equivalently, $\mathsf{Colspan}(A''_y)$  is a random subspace of $\mathsf{Colspan}(A_y)$ of dimension $n - \frac{3r}{2}$.)}
    
    \item Set $\mathcal{P}(x) = (y,  \textcolor{red}{A_y} \cdot J(x) + b_y)$ where $y = H(x)$.
    \item Set $\mathcal{P}^{-1}(y,v) = \pi^{-1}(y || z)$ if $v = { \textcolor{red}{A_y}}\cdot z + b_y$ for some $z \in \F_2^{n - r}$, otherwise $\bot$.
    \item Set $\mathcal{D}(y,v) = 1$ if $v^T \textcolor{red}{A''_{y}} = 0$ and $v \neq 0$, otherwise $0$.
    \item Set $\mathcal{D}_0(y,m,v) = 1$ if $v = \textcolor{red}{A'_{y}}\cdot z + b_y$ for some  $z \in \F_2^{n - r}$ and the first bit of $v$ is $m$; otherwise $0$.
    
            \item  Set $\ora_\ver = \mathcal{D}_0$, $\ora_\sign = \mathcal{D}$, $\ora_\genqkey = (\mathcal{P},\mathcal{P}^{-1}, \mathcal{D}, \mathcal{D}_0)$.

  \item Sample $L \samp \adve_1^{(\ora_{\oss.\genqkey}, \ora_{\oss.\sign}, \ora_{\oss.\ver})}(1^\lambda)$.

      \item Sample $(vk_i, m_i, v_i)_{i \in [q(\lambda) + 1]} \samp \adve_2^{\ora_{\oss.\sign}, \ora_{\oss.\ver}}(L, (b_y)_{y \in \zo^r})$.
            \item Check that $vk_i$ are all distinct, and that $v_i + b_{vk_i}$ starts with the bit $m_i$ and that $v_i \in \textcolor{red}{\mathsf{Colspan}({A_{vk_i}}) \setminus \mathsf{Colspan}({A''_{vk_i}})}$ for all $i \in [q + 1]$. If so, output $1$, otherwise output $0$.
        \end{enumerate}
    \end{mdframed}
We have $\hyb_2 \approx_{1/16p} \hyb_3$ by \cref{thm:srd}, since the whole experiment can be implemented with oracle access to the oracle $\ora_{A}(y) = A_y'' ||  A_y || A_y'$ (where each $\ora_A(y)$ is sampled independently as in the experiment).

     \paragraph{$\hyb_{final}$:} We now define our final hybrid, for some adversaries $\adve_1', \adve_2'$.
  \begin{mdframed}
        \underline{$\hyb_{final}$}

        \begin{enumerate}
        \item Sample a random permutation $\pi$ on $\zo^n$.
        \item For each $y \in \zo^r$, sample a random vector $b_y$ in $\F_2^k$. 
        
         \item For each $y \in \zo^r$, sample $A'_y \in \F_2^{k \times (n-\frac{r}{2})}$ and $A''_y \in \F_2^{k \times (n-\frac{r}{2})}$ so that  $\mathsf{Colspan}(A'_y)$ is a random subspace of $\F_2^k$ of dimension $n-\frac{r}{2}$ and so that $\mathsf{Colspan}(A''_y)$  is a random subspace of   $\mathsf{Colspan}(A'_y)$ of dimension $n - \frac{3r}{2}$.

         \item For each $y \in \zo^r$, sample $A_y \in \F_2^{k \times (n-r)}$ so that $\mathsf{Colspan}(A_y)$ is a random subspace from the set $\mathcal{S}$ of subspaces of dimension $n-r$ where each subspace in the set is a superspace of $\mathsf{Colspan}(A''_y)$ and subspace of $\mathsf{Colspan}(A'_y)$.

         \item Define the oracle $\ora(y)$ that outputs the index of $\mathsf{Colspan}(A_y)$ in the set $\mathcal{S}$ when it is ordered in a canonical way. That is, each $\ora(y)$ is independently and uniformly sampled from $[|\mathcal{S}|]$ (we will refer to the elements of $[|\mathcal{S}|]$ and the subspaces $A_y$ in an interchangeable way).

  \item Sample $L \samp \adve_1^{\ora}(1^\lambda, \pi, (b_y, A'_y, A''_y)_{y \in \zo^r})$.
\item Sample $(vk_i, m_i, v_i)_{i \in [q(\lambda) + 1]} \samp \adve_2(L, \pi, (b_y, A'_y, A''_y)_{y \in \zo^r})$.
            \item \textbf{Winning Condition Check:} Check that $vk_i$ are all distinct, and that $v_i + b_{vk_i}$ starts with the bit $m_i$ and that $v_i \in {\mathsf{Colspan}({A_{vk_i}}) \setminus \mathsf{Colspan}({A''_{vk_i}})}$ for all $i \in [q + 1]$. If so, output $1$, otherwise output $0$.
        \end{enumerate}
    \end{mdframed}

    Since $\Pr[\hyb_3 = 1] \geq\frac{1}{2p}$, it is easy to see that there exists $\adve_1', \adve_2'$ such that $\adve_1'$ makes $q$ queries to $\ora$ and $\Pr[\hyb_{final} = 1] \geq \frac{1}{2p}$. Now we will show that this is a contradiction. Moving forward, without loss of generality, we will assume that $\adve_2$ never outputs (i) $v_i \not\in {\mathsf{Colspan}({A'_{vk_i}}) \setminus \mathsf{Colspan}({A''_{vk_i}})}$, (ii) it never outputs non-distinct $vk_i$ values and (iii) it never outputs $\bot$ (e.g. at worst, it can output random vectors in ${\mathsf{Colspan}({A'_{vk_i}}) \setminus \mathsf{Colspan}({A''_{vk_i}})}$). Now, first, consider the compressed oracle simulation, $\mathsf{Exp}$ of the hybrid $\hyb_{final}$. In particular, we also execute the final part of the experiment (i.e. winning condition check) using queries to $\ora$: For $i \in [q+1]$, we first query the oracle $\ora$ on $vk_i$ and write the output to an empty register, and then we perform the membership check for $v_i$ by reading from this register. We assume that the output (win or lose bit) is also written to a new register (call it the experiment outcome register). Since compressed oracle simulation is a perfect simulation, we have $\Pr[\mathsf{Exp} = 1] = \Pr[\hyb_{final} = 1]$. Now we define the experiment $\mathsf{Exp}'$ where we modify the execution of the compressed oracle simulation during the winning condition check as follows. Instead of applying $\mathsf{StdDecomp}\circ \mathsf{CStO}' \circ \mathsf{StdDecomp}$ (we assume that $\mathsf{Increase}$ has been applied for sufficiently many times in advance), we instead apply  $\mathsf{StdDecomp}\circ \mathsf{CStO}' \circ \mathsf{StdDecomp}'$ where $\mathsf{StdDecomp}'$ is a linear mapping defined as $\mathsf{StdDecomp}': \ket{vk}\ket{v}\ket{D} \to \ket{vk}\ket{v}\sum_{A \in [|S|]: v \not\in {\mathsf{Colspan}(A) \setminus \mathsf{Colspan}({A''_{vk}})} } \frac{1}{\sqrt{a}} \ket{D \cup (vk, A)}$ for $D,vk$ such that $D(vk) = \bot$ (and defined the same as $\mathsf{StdDecomp}$ for other basis elements), where $a = \begin{bmatrix} r \\ r/2 \end{bmatrix}_2 - \begin{bmatrix} r-1 \\ r/2 - 1 \end{bmatrix}_2$. We will call the case of  $D, vk$ such that $D(vk) = \bot$ a \emph{good case}. Since the original mapping $\mathsf{StdDecomp}$ maps $\ket{vk}\ket{v}\ket{D} \to \ket{vk}\ket{v}\sum_{A \in [|S|]} \frac{1}{\sqrt{|S|}} \ket{D \cup (vk, A)}$ for $D$ such that $D(vk) = \bot$, and since $|S|= \begin{bmatrix} r \\ r/2 \end{bmatrix}_2$, it is easy to see that $\mathsf{Exp}' \approx \mathsf{Exp}$. Finally, observe that $\Pr[\mathsf{Exp}' = 1] = 0$: For any output $(vk_i, m_i, v_i)_{i \in [q(\lambda) + 1]}$ (of $\adve_2$) and any compressed oracle database $D$ that is in the support of the system when $\adve_2$ yields its output, it will be the case that there is at least one value  $i \in [q+1]$ such that $vk_i \not\in D$ since $D$ contains $q$ valid pairs (also recall that when the good case does not apply, the number of valid pairs in the database does not increase). Thus, while applying the winning condition check, there will be a point where the good case of $\mathsf{StdDecomp}'$ applies, and after that it is easy to see that verification for that index fails. Thus, any  $(vk_i, m_i, v_i)_{i \in [q(\lambda) + 1]}$ and any compressed oracle database $D$ in the support of the system will map to a state where the experiment outcome register will have the state $\ket{0}$. Thus,  $\Pr[\mathsf{Exp}' = 1] = 0$. However, we had $\mathsf{Exp}' \approx \mathsf{Exp}$ and  $\Pr[\mathsf{Exp} = 1] = \Pr[\hyb_{final} = 1]$ and $\Pr[\hyb_{final} = 1] \geq \frac{1}{2p}$. Thus, this is a contradiction, as desired.

\newpage
\section{Incompressibility of Random Oracles}\label{sec:incomprrom}
In this section, we show that random oracles satisfy a natural notion of incompressibility. First we give our theorem statement regarding this property, then we prove a technical lemma that will help us in our proof of the incompressibility theorem, and finally we prove our incompressibility theorem.

\begin{theorem}[Incompressibility of Random Oracles]\label{thm:randoraincomp}
Let $H: \zo^{g_1(\lambda)} \to \zo^{g_2(\lambda)}$ be a random oracle and define $\oraimgver$ to be the following oracle (called \emph{(image) verification oracle}).
\begin{mdframed}
    {\bf \underline{$\oraimgver(x,y)$}}
    
    {\bf Hardcoded: $H$}
    \begin{enumerate}[label=\arabic*.]
    \item Check if $y = H(x)$. If so, output $1$. Otherwise, output $0$.
    \end{enumerate}
    \end{mdframed}

   For a predicate $P: \zo^{g_1(\lambda)} \to \zo$, define $\ora_P$ as follows.
 \begin{mdframed}
    {\bf \underline{$\ora(x)$}}
    
    {\bf Hardcoded: $H, P$}
    \begin{enumerate}[label=\arabic*.]
    \item Check if $P(x) = 0$. If so, output $\bot$ and terminate.
      \item Output $H(x)$.
    \end{enumerate}
    \end{mdframed}

Define the following experiment $\mathsf{Exp}_H$.
 \begin{mdframed}
 {\underline{$\mathsf{Exp}_H$}}
   \begin{enumerate}
        \item Set $i = 1$.
       \item For any polynomial number of rounds (denote this by $u(\lambda)$):
       \begin{enumerate}[label=\arabic*.]
           \item $\adve_1$ makes polynomial number of queries (denote the number by $s_i(\lambda)$) to $H$ and unbounded number of queries to $\ora_{\cup^{i-1}_{j=1} P_j}$ where ${\cup^{i-1}_{j=1} P_j}$ denotes the union of the predicates $P_1, \dots, P_{i-1}$ (i.e. ${\cup^{i-1}_{j=1} P_j}(x) = 1$ if $P_j(x) = 1$ for some $j=1,\dots,i-1$).

        \item $\adve_1$ outputs a predicate $P_i$.
           \item Increase $i$ by one.
       \end{enumerate}
       \item $\adve$ outputs an unbounded length message $L$.
   \end{enumerate}
    \end{mdframed}

For any $g_1(\lambda), g_2(\lambda)$ such that $g_2(\lambda)$ is superlogarithmic, for an adversary $\adve_1$ as described above,
 there exists a mapping $\mathcal{M}$ (which outputs a list\footnote{We will refer to this as the incompressibility list.} $\mathfrak{L}$ of size at most $\sum_{j \in [u(\lambda)]} s_j(\lambda)$) such that for all $\adve_2$ that makes at most $q(\lambda)$ queries to $\oraimgver$ such that $\frac{q(\lambda)}{\sqrt{2^{g_2(\lambda)}}}$ is negligible (but can make any number of queries to $\ora_{P^*}$), we have

\begin{equation*}
    \E\left[\sum_{\substack{x,y:\\ \oraimgver(x,y) \neq 0 \\ x\not\in  \mathcal{M}(H, L, (P_j)_j, \adve_2) \\ P^*(x) = 0 }}\mathsf{QW}^{x,y}_{\adve_2, L, H, (P_j)_j} : \begin{array}{c}
           H \samp \zo^{g_1(\lambda)} \to \zo^{g_2(\lambda)} \\
        L, (P_j)_{j \in [u(\lambda)]} \samp \adve_1^{H(\cdot)}(1^\lambda)
        \end{array} \right] \leq \negl(\lambda).
\end{equation*}

where $P^* = {\cup_{j\in[u(\lambda)]} P_j}$ and $\mathsf{QW}^{x,y}_{\adve_2, L, H, (P_j)_j}$ denotes the total query weight of $\adve_2^{\ora_{P^*}, \oraimgver}(L)$ to the oracle $\oraimgver$ on oracle input $x,y$.
\end{theorem}
We will refer to the queries to $\ora_{P^*}$ (for $\adve_2$) and $\ora_{\cup^{i-1}_{j=1} P_j}$ (for $\adve_1$) as \emph{section queries} (and to these oracles as \emph{section oracles}), and queries to $\oraimgver$ as verification queries. A normal query will mean a query to $H$.

Before proving our theorem, we will prove a different technical lemma that will be useful in our proof. This lemma roughly says that, for a compressed random oracle adversary with additional verification oracle access to the random oracle; if at any point, the weight of databases that contain at most $i$ inputs that satisfy a certain predicate is negligible, then this still holds after a verification query.

\begin{lemma}\label{lem:predqrom}
    Let $\adve$ be an adversary with access to a random oracle $H: \zo^{m(\lambda)}\to \zo^{n(\lambda)}$ and the verification oracle for the random oracle that outputs $1$ on input $x, y$ if $H(x) = y$, and outputs $0$ otherwise. Define the registers
    \begin{itemize}
        \item $\mathsf{X}, \mathsf{Y}$, the input registers to the verification oracle,
        \item $\mathsf{Q}$, the 1-qubit phase register for the verification oracle output,
        \item $\mathsf{W}$, the internal workspace register of the adversary,
        \item $\mathsf{P}$, the register containing a predicate\footnote{We will write $P$ to both mean the predicate function, and to mean the set of inputs $x$ such that $P(x) = 1$. Similarly, $\overline{P}$ means the complement of the set $P$}, chosen by the adversary,
        \item $\mathsf{D}$ compressed oracle database register,
        \item $\mathsf{T}$, the register used for implementing the verification oracle queries (initialized at the beginning to zeroes and is not accessed by the adversary).
    \end{itemize}
    
    Consider the following simulation of execution of $\adve$:
    \begin{itemize}
    \item The random oracle queries are implemented as a standard oracle (i.e. the output XORed into the output register) using a compressed oracle simulation,
    \item The verification oracle queries are implemented as a phase oracle as follows: 
        \begin{enumerate}
        \item Query the random oracle with input and output registers $X$ and the output register $T$.
        \item Apply a controlled phase gate that maps $\ket{x}\ket{y}\ket{q}\ket{t}$ to $(-1)^{b\cdot q}\ket{x}\ket{y}\ket{q}\ket{t}$ where $b = 1$ if and only if $y=t$ (and otherwise $b = 0$).
        \item Query the random oracle with input and output registers $X$ and the output register $T$.
        \end{enumerate}
    \end{itemize}
Then,
\begin{itemize}
    \item This is a perfect simulation of execution of $\adve$,
    \item Let $\ket{\phi} = \sum_{x,y,q,w,P,D} \alpha_{x,y,q,w,P,D} \ket{x}\ket{y}\ket{q}\ket{w}\ket{P}\ket{D}\ket{0^{n(\lambda)}}$ be the state of the system right before executing one of the verification oracle queries, respectively over the registers $\mathsf{X}, \mathsf{Y}, \mathsf{Q}, \mathsf{W}, \mathsf{P}, \mathsf{D}, \mathsf{T}$. Let $i \in \nat$ and let $\mathsf{Inp}(D)$ denote the set consisting of all $a$ such that $(a, b_a) \in D$ for some $b_a \neq \perp$. Let  let $\ket{\psi} = \sum_{x,y,q,w,P,D} \beta_{x,y,q,w,P,D} \ket{x}\allowbreak\ket{y}\ket{q}\allowbreak\ket{w}\allowbreak\ket{P}\ket{D}\ket{0^{n(\lambda)}}$  be the state right after executing the verification oracle query on $\ket{\phi}$.
    
    Then, $\sum_{\substack{x,y,q,w,P,D: \\ |\mathsf{Inp}(D) \cap \overline{P}| > i}} |\beta_{x,y,q,w,P,D}|^2 \leq \sum_{\substack{x,y,q,w,P,D: \\ |\mathsf{Inp}(D) \cap \overline{P}| > i}} |\alpha_{x,y,q,w,P,D}|^2 + \frac{6}{\sqrt{2^{n(\lambda)}}}$.
\end{itemize}
    
\end{lemma}
\begin{proof}
The fact this simulation is a perfect simulation follows easily from the fact that compressed oracle simulation is a perfect simulation for the random oracle (\cite{C:Zhandry19}). Now we prove the second part.
Let $\mathsf{CtPh}$ denote the unitary for the controlled phase described above that is used in the implementation of the verification queries. Let $U = \mathsf{CStO}\circ \mathsf{CtPh} \circ \mathsf{CStO}$ (where $\mathsf{CStO}$ is the compressed oracle query unitary with input register $X$ and output register $T$). Note that $U$ is an isometry. Thus, $\ket{\psi} = U\ket{\phi}$ (here our notation means applying the isometry $U$ to the state $\ket{\phi}$).  We also let $\pi$ denote\footnote{By abuse of notation, we will also write $\pi$ to denote the measurement $\pi\otimes I$ where the identity $I$ is over the remaining registers.} the projection onto $\ket{P}\ket{D}$ such that $|\mathsf{Inp}(D) \cap \overline{P}| > i$. Thus, we have $\sum_{\substack{x,y,q,w,P,D: \\ |\mathsf{Inp}(D) \cap \overline{P}| > i}} |\beta_{x,y,q,w,P,D}|^2 = \norm{\pi U\ket{\phi}}^2$.

We write \begin{align*} 
        \ket{\phi} =  &\sum_{\substack{x,y,q,w,P,D:\\ |\mathsf{Inp}(D) \cap \overline{P}| > i }}  \alpha_{x,y,q,w,P,D}\ket{x}\ket{y}\ket{q}\ket{w}\ket{P}\ket{D}\ket{0^{n(\lambda)}}  + \sum_{\substack{x,y,q,w,P,D:\\ |\mathsf{Inp}(D) \cap \overline{P}| \leq i \\ x \in D}} \alpha_{x,y,w,P,D}\ket{x}\ket{y}\ket{q}\ket{w}\ket{P}\ket{D}\ket{0^{n(\lambda)}}  +\\
        &\sum_{\substack{x,y,q,w,P,D:\\ |\mathsf{Inp}(D) \cap \overline{P}| \leq i\\ x \not\in D \\ P(x) = 1}} \alpha_{x,y,w,P,D} \ket{x}\ket{y}\ket{q}\ket{w}\ket{P}\ket{D}\ket{0^{n(\lambda)}} + 
        \sum_{\substack{x,y,q,w,P,D:\\ |\mathsf{Inp}(D) \cap \overline{P}| \leq i\\ x \not\in D \\ P(x) = 0}} \alpha_{x,y,q,w,P,D} \ket{x}\ket{y}\ket{q}\ket{w}\ket{P}\ket{D}\ket{0^{n(\lambda)}} 
    \end{align*}

    Let $\ket{\phi_1}, \ket{\phi_2}, \ket{\phi_3}, \ket{\phi_4}$ denote each unnormalized state in the summation above in the order they are written. Observe that when executing a verification query on $\ket{x}\ket{y}\ket{q}\ket{w}\ket{P}\ket{D}$, only $x$ can be added to the database (if not already in the database), and the other entries in the database do not change. Thus, we get $\pi U \ket{\phi_2} = 0$. By the same argument, we get $\pi U \ket{\phi_3} = 0$.  We also have $\norm{\pi U \ket{\phi_1}}^2 \leq \norm{\ket{\phi_1}}^2 =  \sum_{\substack{x,y,q,w,P,D: \\ |\mathsf{Inp}(D) \cap \overline{P}| > i}} |\alpha_{x,y,q,w,P,D}|^2$. 
    
    Now we claim $\norm{\pi U \ket{\phi_4}}^2 \leq \frac{4}{{2^n}}$ will prove our claim. Observe that once we prove this, the theorem follows. To see this, first observe that  $\sum_{\substack{x,y,q,w,P,D: \\ |\mathsf{Inp}(D) \cap \overline{P}| > i}} |\beta_{x,y,q,w,P,D}|^2 = \norm{\pi U\ket{\phi}}^2 \leq \norm{\pi U \ket{\phi_1}}^2 + \norm{\pi U \ket{\phi_4}}^2 + 2\cdot \norm{\pi U \ket{\phi_1}}\cdot \norm{\pi U \ket{\phi_4}}$ by triangle inequality and since $\pi U \ket{\phi_2} = \pi U \ket{\phi_3} = 0$. Finally, note that $\norm{\pi U \ket{\phi_1}}^2 + \norm{\pi U \ket{\phi_4}}^2 + 2\cdot \norm{\pi U \ket{\phi_1}}\cdot \norm{\pi U \ket{\phi_4}} = \norm{\pi U \ket{\phi_1}}^2 + \norm{\pi U \ket{\phi_4}}(\norm{\pi U \ket{\phi_4}} + 2\norm{\pi U \ket{\phi_1}})$ and $(\norm{\pi U \ket{\phi_4}} + 2\norm{\pi U \ket{\phi_1}}) \leq 3$.

Now we move onto proving  $\norm{\pi U \ket{\phi_4}}^2 \leq \frac{4}{{2^n}}$.
    For all valid states (i.e. states where compressed oracle invariants are kept) $\ket{\tau}$,  we have $U\ket{\tau} =\mathsf{StdDecomp}\circ  \mathsf{CStO}' \circ \mathsf{CPh}\circ \mathsf{CStO}' \circ\mathsf{StdDecomp}\circ (\mathsf{Inc})^2 \ket{\tau}$. This follows from the fact that unitaries on different registers commute, that $\mathsf{Increase}$ can be moved to the beginning, and that $\mathsf{StdDecomp}$ is an involution. Write $U_1 = \mathsf{CStO}' \circ\mathsf{StdDecomp}\circ(\mathsf{Inc})^2$ and $U_3 =\mathsf{StdDecomp}\circ  \mathsf{CStO}'$; and thus $U = U_3\circ \mathsf{CPh}\circ U_1$.

    After applying $U_1$ to $\ket{\phi_4}$, the state becomes
    \begin{equation*}
          \sum_{\substack{x,y,q,w,P,D, z:\\ z \in \zo^{n(\lambda)} \\ |\mathsf{Inp}(D) \cap \overline{P}| \leq i\\ x \not\in D \\ P(x) = 0}} \alpha_{x,y,q,w,P,D} \frac{1}{\sqrt{2^{n(\lambda)}}} \ket{x}\ket{y}\ket{q}\ket{w}\ket{P}\ket{D \cup (x, z) \cup (\bot, 0^n)}\ket{z}
    \end{equation*}
    Once we apply the controlled phase based on the verification, the state becomes
    \begin{align*}
          &\sum_{\substack{x,y,q,w,P,D: \\ |\mathsf{Inp}(D) \cap \overline{P}| \leq i\\ x \not\in D \\ P(x) = 0}} (-1)^q\alpha_{x,y,q,w,P,D} \frac{1}{\sqrt{2^{n(\lambda)}}} \ket{x}\ket{y}\ket{q}\ket{w}\ket{P}\ket{D \cup (x, y) \cup (\bot, 0^n)}\ket{y}+ \\  &\sum_{\substack{x,y,q,w,P,D: \\ |\mathsf{Inp}(D) \cap \overline{P}| \leq i\\ x \not\in D \\ P(x) = 0}} \sum_{z \neq y} \alpha_{x,y,q,w,P,D} \frac{1}{\sqrt{2^{n(\lambda)}}} \ket{x}\ket{y}\ket{q}\ket{w}\ket{P}\ket{D \cup (x, z) \cup (\bot, 0^n)}\ket{z}
    \end{align*}
    which is equal to
     \begin{align*}
         &\sum_{\substack{x,y,q,w,P,D: \\ |\mathsf{Inp}(D) \cap \overline{P}| \leq i\\ x \not\in D \\ P(x) = 0}} ((-1)^q - 1)\alpha_{x,y,q,w,P,D} \frac{1}{\sqrt{2^{n(\lambda)}}} \ket{x}\ket{y}\ket{q}\ket{w}\ket{P}\ket{D \cup (x, y) \cup (\bot, 0^n)}\ket{y} + \\  &\sum_{\substack{x,y,q,w,P,D: \\ |\mathsf{Inp}(D) \cap \overline{P}| \leq i\\ x \not\in D \\ P(x) = 0}} \sum_{z } \alpha_{x,y,q,w,P,D} \frac{1}{\sqrt{2^{n(\lambda)}}} \ket{x}\ket{y}\ket{q}\ket{w}\ket{P}\ket{D \cup (x, z) \cup (\bot, 0^n)}\ket{z}
    \end{align*}

Denote the two unnormalized states in the summation as $\ket{\eta_1}, \ket{\eta_2}$ respectively. Since $\sum_{x,y,q,w,P,D}\allowbreak|\alpha_{x,y,q,w,P,D} |^2 = 1$, we get $\norm{\ket{\eta_1}}^2 \leq \frac{4}{2^n}$ and thus $\norm{\pi\circ U_3\ket{\eta_1}}^2 \leq \frac{4}{2^n}$. Now, we show ${\pi\circ U_3\ket{\eta_2}} = 0$. Observe that once we apply $U_3$ to  $\ket{\eta_2}$, we get 
    \begin{equation*}
      \sum_{\substack{x,y,w,P,D: \\ |\mathsf{Inp}(D) \cap \overline{P}| \leq i\\ x \not\in D \\ P(x) = 0}} \alpha_{x,y,q,w,P,D}  \ket{x}\ket{y}\ket{q}\ket{w}\ket{P}\ket{D\cup \{(\bot, 0^n), (\bot, 0^n)\}}\ket{0^{n(\lambda)}}
    \end{equation*}
    Hence, ${\pi\circ U_3\ket{\eta_2}} = 0$, which completes the proof.
\end{proof}

Now we prove \cref{thm:randoraincomp}.
\begin{proof}
 We will prove the operational version of the lemma (similar to \cref{def:oprincompr}) and the actual incompressibility theorem directly follows (as in \cref{thm:incomprequiv}).

 We will consider the compressed oracle simulation of the operational game, as in \cref{lem:predqrom}. Note that this simulation is purified: The output of $\adve_1$ is measured at the very end. Let $\mathsf{P}_1, \dots, \mathsf{P}_{u(\lambda)}$ denote the registers on which the predicates will appear - without loss of generality, we can assume that these registers exist from the beginning, and are initialized to zeroes (which will be interpreted as a predicate that rejects all inputs).%

 For any $j \in \nat$, let $\pi_j$ denote the projection onto database states (over the database register $\mathsf{D}$) and predicate states (over the registers $\mathsf{P}_1, \dots, \mathsf{P}_{u(\lambda)}$) such that $|\mathsf{Inp}(D) \cap \overline{(\cup_{j=1}^{u(\lambda)} P_j)}| > j$. Let $\ket{\tau}$ be the state right before some query to the verification oracle, or the random oracle, or the section oracle, or right before updating a predicate register; and let $\ket{\tau'}$ denote the state right after the execution of the query. We observe the following.
 \begin{itemize}
     \item If the query is a verification query, then $\norm{\pi_j \ket{\tau'}}^2 \leq  \norm{\pi_j \ket{\tau}}^2 + \frac{6}{\sqrt{2^{g_2(\lambda)}}}$. This follows directly from \cref{lem:predqrom}.
     \item If the query is a section query, then $\norm{\pi_j \ket{\tau'}}^2 \leq  \norm{\pi_j \ket{\tau}}^2$. This follows since when a compressed oracle query is executed on the state $\ket{x}\ket{D}$, only $x$ can be added to the database, no other values.
     \item If the query is a normal query, then $\norm{\pi_{j+1} \ket{\tau'}}^2 \leq  \norm{\pi_j \ket{\tau}}^2$. This is because a single query can add at most one input to the database, and cannot change the other entries (other than deleting an entry).
     \item If we are considering a point where a predicate register is being updated, then  $\norm{\pi_j \ket{\tau'}}^2 \leq  \norm{\pi_j \ket{\tau}}^2$. This is because we are considering the union of the predicates in the definition of $\pi_j$, and adding more predicates makes the \emph{good} set  ($\mathsf{Inp}(D) \cap \overline{(\cup P_j)}$) smaller.
 \end{itemize}
 
 Applying this to each query of the adversaries $\adve_1, \adve_2$, it is easy to see that $\pi_{u(\lambda)}\ket{\eta} \leq \negl(\lambda)$ where $\ket{\eta}$ is the state of the system right before $\adve_2$ produces its output in the operational incompressibility game, since there is a total of $u(\lambda)$ normal queries. By \cref{prethm:compor}, this means that except with negligibly small probability, $\adve_2$ can output at most $u(\lambda)$ different valid input - output pairs such that the input satisfies $P^*(x) = 0$, which completes the proof of operational incompressibility.
\end{proof}


\newpage\part{Quantum-Key Fire}
\section{Quantum Key-Fire for Signing in the Classical Oracle Model}
In this section, we define quantum key-fire for signing in the classical oracle model.

For simplicity, we work with an oracle model where the oracle is sampled by $\qkeyfiresign.\setup$ (i.e. there is an independent oracle for each instance) and output as the serial number. However, we note that, through elementary tricks, constructions in this model can also be easily converted to the standard classical oracle model where there is a single global classical oracle $\ora$ (that is sampled from some distribution) that everyone (adversaries and the construction) has quantum-query access to, and $\qkeyfiresign.\setup^\ora$ simply outputs a classical string as the verification key/serial number instead of outputting oracles. Again for simplicity, we consider the leakage phase as two adversaries communicating over a classical channel: note that this is equivalent to the leakage phase from the original game.

\begin{definition}[Quantum Key-Fire for Signing in the Classical Oracle Model]\label{def:qfireora}
       A public-key quantum key-fire scheme for signing $\qkeyfiresign$ in the classical oracle model consists of the following efficient algorithms.
    \begin{itemize}
        \item $\qkeyfiresign.\setup(1^\lambda)$: Takes in a security parameter, outputs an oracle $\ora$ as the verification key/serial number and a \emph{key state} $\regis{flame}$.
        \item $\qkeyfiresign.\clone^\ora(\regis{key})$: Takes in a key register $\regis{key}$, outputs a new register $\regis{clone}$ along with the input register $\regis{key}$.
        \item $\qkeyfiresign.\sign^\ora(\regis{key}, m):$ Takes in a key register and a message $m$, outputs a signature.
        \item $\qkeyfiresign.\ver^\ora(m, sig)$: Takes in a message and an alleged signature, outputs $\outtrue$ or $\outfalse$.
    \end{itemize}
   
    \paragraph{Evaluation Correctness:} For all $m \in \messpa$,
    \begin{equation*}
        \Pr[\qkeyfiresign.\ver^{\ora^*}(m, sig) = 1  : \begin{array}{c}
        {\ora^*}, \regis{key} \samp \qkeyfiresign.\setup(1^\lambda) \\
       sig \samp \qkeyfiresign.\sign^{\ora^*}(\regis{key}, m)
    \end{array}] \geq 1 - \negl(\lambda).
    \end{equation*}

\paragraph{Strong cloning correctness:} This is defined the same as before: We require that output of $\qkeyfiresign.\setup(1^\lambda)$ is a pure state $\ket{\psi_{\ora^*}}$ and that clone on input $\ket{\psi_{\ora^*}}$ outputs two copies, $\ket{\psi_{\ora^*}}\otimes \ket{\psi_{\ora^*}}$.
\end{definition}

\begin{definition}[LOCC Leakage-Resilience Security for Quantum Key-Fire for Signing in the Classical Oracle Model]\label{def:qfireorasec}
Let $\qkeyfiresign$ be a quantum key-fire scheme with classical oracles.
Consider the following game between an adversary $\adve = (\adve_1, \adve_2)$ and a challenger.

 \paragraph{\underline{$\lrsiggame{\adve}(1^\lambda)$}}
    \begin{enumerate}
    \item The challenger samples $sn, \regis{key} \samp \qkeyfiresign.\setup(1^\lambda)$ and submits $\regis{key}$ to the adversary $\adve_1$.
    \item Parse $\ora = sn$.
    \item \textbf{Leakage Phase:}The adversaries $\adve_1^\ora$ and $\adve_2^\ora$ interact classically by exchanging classical messages for any number of rounds.
    \item The challenger samples a random message $m^* \samp \messpa$.
    \item The adversary $\adve_2^\ora$ receives $m^*$, outputs $sig^*$.
    \item The challenger checks if $\qkeyfiresign.\ver^\ora(m^*, sig^*) = \outtrue$. If so, it outputs $1$, otherwise it outputs $0$.
    \end{enumerate}

    We say that the quantum key-fire scheme $\qkeyfiresign$ satisfies \emph{LOCC leakage-resilience security} if for any QPT (or query bounded, in case of unconditional oracle-based schemes) adversary $\adve$, we have 
    \begin{equation*}
        \Pr[\lrsiggame{\adve}(1^\lambda) = 1] \leq \negl(\lambda).
    \end{equation*}
\end{definition}

Note that this model also gives us the following conclusion. If the oracles of the scheme can be efficiently implemented (which is the case for our construction in \cref{sec:cons},  assuming one-way functions), our oracle security proof will also imply that our scheme is secure assuming virtual black-box obfuscation in the plain model and thus heuristically secure assuming indistinguishability obfuscation. Separately, as discussed, through elementary tricks our model also can be converted to the standard classical oracle model with unconditional security.

\newpage \newcommand{\firemeslen}{\nu(\lambda)}
\newcommand{\ossvklen}{p_1(\lambda)}
\newcommand{\ossgenqkeyinplen}{p_2(\lambda)}

\newcommand{\hshare}{H_\mathsf{share}}
\newcommand{\hsig}{H_{\mathsf{sig}}}

\newcommand{\fireora}{{\ora^*_{\qkeyfiresign}}} 

\newcommand{\numstepsosssign}{t_1}
\newcommand{\numstepsosskey}{t_2}
\newcommand{\auxsizeossign}{t_3}
\newcommand{\auxsizeosskey}{t_4}

\section{Construction}\label{sec:cons}
In this section, we give a quantum key-fire  construction  relative to a quantumly-accessible classical oracle (\cref{def:qfireora}) for signing $\firemeslen$-bit messages (where $\firemeslen$ is superlogarithmic), then prove its signing and cloning correctness. The security is proven in \cref{sec:proofsec}.

In \cref{sec:qfireblocks}, we describe our building block: incompressible one-shot signatures (OSS). Then, in \cref{sec:constructionitself}, we give our construction.

Before moving onto the building block and construction, we introduce some notation.
\paragraph{Oracle Algorithms} For a classical oracle $\ora$, let $U_\ora$ denote its \emph{query execution unitary}, that is, the unitary that maps $\ket{x}\ket{w}\ket{z}$ to $\ket{x}\ket{w}\ket{z \oplus \ora(x)}$ where the registers respectively are the query input, workspace and query output registers. For a quantum query algorithm $\mathcal{B}$, there is an associated \emph{algorithm unitary} $U_{\mathcal{B}}$, and the execution of  $\mathcal{B}$ with input $\regi$ and oracle access to $\ora$ is simply applying $(\mathcal{U}_{\ora}\cdot U_{\mathcal{B}})^{p(\lambda)}$ to $(\regi, \regis{aux})$ (where $p(\lambda)$ is some polynomial and $\regis{aux}$ is initialized to zeroes), and outputting (the whole or part of) the resulting register (after possibly measuring some parts of it in the computational basis)
 
\subsection{Building Block: Incompressible One-Shot Signature Scheme}\label{sec:qfireblocks}
We will use an incompressible one-shot signature scheme $\oss$ for $1$-bit messages relative to a classical oracle (\cref{def:ossincomprnew}) as a building block.

We make  the following assumptions about $\oss$ for presentation clarity, and all of these are without loss of generality. 

Let $\numstepsosssign(\cdot), \numstepsosskey(\cdot), \auxsizeossign(\cdot), \auxsizeosskey(\cdot)$ be some polynomials.
\begin{itemize}

\item Since $\oss$ is an incompressible OSS, it has split oracles (\cref{defn:splitora}): write $(\ora_{\oss.\genqkey}$, $\ora_{\oss.\sign}$, $\ora_{\oss.\ver}) = \ora_\mathsf{OSS}$ to denote its oracles. Let $U_{\oss.\genqkey}$ be the algorithm unitary and let $U_{\ora_{\oss.\genqkey}}$ be the query execution unitary for $\oss.\genqkey$, and similarly define $U_{\oss.\sign}$, $U_{\ora_{\oss.\sign}}$ for $\oss.\sign$.

    \item $\oss.\genqkey^{\ora_{\oss.\genqkey}}(1^\lambda)$: Define $U^{'}_{\ora_{\oss.\genqkey}, \oss.\genqkey} = (U_{\ora_{\oss.\genqkey}}\cdot U_{\oss.\genqkey})^{\numstepsosskey(\lambda)}$. Then, this algorithm proceeds as follows:
    \begin{enumerate}
        \item Introduce the auxiliary input register $\regis{inp-aux}$ initialized to $\ket{0^{\auxsizeosskey(\lambda)}}$.
        \item Apply $U^{'}_{\ora_{\oss.\genqkey}, \oss.\genqkey}$ to  $\regis{inp-aux}$ and parse the output into the registers: $\regis{vk}, \regis{key}, \regis{out-aux}$.
        \item  Measure $\regis{vk}$ to obtain the verification key $vk$.
        \item Output $\regis{key}$ as the signing key and $vk$ as the verification key.
    \end{enumerate}

    \item $\oss.\sign^{\ora_{\oss.\sign}}(vk, \regis{sk}, m)$: Define $U^{'}_{\ora_{\oss.\sign}, \oss.\sign} = (U_{\ora_{\oss.\sign}}\cdot U_{\oss.\sign})^{\numstepsosssign(\lambda)}$. Then, this algorithm proceeds as follows:
    \begin{enumerate}
        \item Introduce the auxiliary input register $\regis{inp-aux}$ initialized to $\ket{0^{\auxsizeossign(\lambda)}}$. 
        \item Introduce the message register $\regis{msg}$ initialized to $\ket{m}$.
        \item Introduce the verification key register $\regis{vk}$ initialized to $\ket{vk}$.
        \item Apply $U^{'}_{\ora_{\oss.\sign}, \oss.\sign}$ to $(\regis{vk}, \regis{msg}, \regis{sk}, \regis{inp-aux})$ and parse the output into the registers: $\regis{vk}$,  $\regis{msg}$, $\regis{sig}$, $\regis{out-aux}$ (registers $\regis{vk}$,  $\regis{msg}$ are unmodified).
        \item  Measure $\regis{sig}$ to obtain the signature $sig$.
        \item Output $sig$.
    \end{enumerate}

    \item $\oss.\ver^{\ora_{\oss.\ver}}(vk,m,sig)$: Applies the oracle $\ora_{\oss.\ver}$ to obtain the output $\ora_{\oss.\ver}(vk, m, sig)$. If the output is $1$, the algorithm outputs $\outtrue$, otherwise it outputs $\outfalse$\footnote{The assumption about $\oss.\ver$ is essentially that the signature verification procedure is classical and deterministic. It is easy to see that deterministic verification assumption is without loss of generality due to correctness with overwhelming probability. When we do not require the oracles to be efficiently implementable, classical verification assumption is also without loss of generality since any \emph{classical input - classical output} quantum algorithm can be simulated by a classical oracle. However, in the efficient oracle case, the assumption about $\oss.\ver$ might not be without loss of generality, but it is nevertheless true for our OSS construction in \cref{sec:incompross}. The assumptions about $\oss.\genqkey$ and $\oss.\sign$ are indeed without loss of generality in all cases.}.
\end{itemize}

\subsection{Construction}\label{sec:constructionitself}
\newcommand{\funcspace}[2]{\mathcal{F}_{#1,#2}}
Let $\firemeslen$ be any polynomially bounded function. We now give our public-key quantum key-fire construction for signing $\firemeslen$-bit messages, with the classical oracle $\ora_{\qkeyfiresign} = (\ora_0$, $\ora_1$, $\ora_2$, $\ora_{\oss.\sign}$, $\ora_3, \ora_4)$.

Let $\oss$ be a one-shot signature scheme for $1$-bit messages as described in \cref{sec:qfireblocks}. Let $\ossvklen$ denote the length of verification keys of $\oss$, $\ossgenqkeyinplen$ denote the input length of the oracle $\ora_{\oss.\genqkey}$. 

\paragraph{\underline{$\qkeyfiresign.\setup(1^\lambda)$}}
\begin{enumerate}
\item Sample random functions $H_0, H_1, \hsig$ where $H_0: \zo^{\ossvklen+\ossgenqkeyinplen} \to\zo^{\lambda}$, $H_1:\zo^{\ossvklen+\ossgenqkeyinplen} \to\zo^{\lambda}$.
    \item Sample random function $\hsig: \zo^{\firemeslen} \to \zo^\lambda$.
    \item Sample $(\ora_{\oss.\genqkey}$, $\ora_{\oss.\sign}$, $\ora_{\oss.\ver}) \samp \oss.\setuporacles(1^\lambda)$.
    \item Construct the oracle $\mathcal{O}_0$.
    \begin{mdframed}
    {\bf \underline{$\ora_0(ivk, isig, z)$}}
    
     {\bf Hardcoded: $\ora_{\oss.\ver}, H_0$}
    \begin{enumerate}[label=\arabic*.]
        \item Check if $\ora_{\oss.\ver}(ivk, 0, isig) = 1$. If not, output $\bot$ and terminate.

        \item Output $H_0({ivk} || z)$.
    \end{enumerate}
    \end{mdframed}

    \item Construct the oracle $\mathcal{O}_1$.
    \begin{mdframed}
    {\bf \underline{$\ora_1(ivk, isig, z)$}}
    
    {\bf Hardcoded: $\ora_{\oss.\ver}, H_1$}
    \begin{enumerate}[label=\arabic*.]
         \item Check if $\ora_{\oss.\ver}(ivk, 1, isig) = 1$. If not, output $\bot$ and terminate.

        \item Output $H_1({ivk} || z)$.
    \end{enumerate}
    \end{mdframed}

    \item Construct the oracle $\mathcal{O}_2$.
    \begin{mdframed}
    {\bf \underline{$\ora_2(ivk, y_0, y_1, z)$}}
    
    {\bf Hardcoded: $H_0, H_1, \ora_{\oss.\genqkey}$}
    \begin{enumerate}[label=\arabic*.]
    \item Check if $y_0 = H_0(ivk || z)$ and if $y_1 = H_1(ivk || z)$. If not, output $\bot$ and terminate.
      \item Output $\ora_{\oss.\genqkey}(z)$.
    \end{enumerate}
    \end{mdframed}

  \item Construct the oracle $\mathcal{O}_3$.
    \begin{mdframed}
    {\bf \underline{$\ora_3(ivk, y_0, y_1, m)$}}

    {\bf Hardcoded: $H_0, H_1, \hsig$}
    \begin{enumerate}[label=\arabic*.]
    \item Check if $y_0 = H_0(ivk || m)$ and if $y_1 = H_1(ivk || m)$. If not, output $\bot$ and terminate.
      \item Output $\hsig(m)$.
    \end{enumerate}
    \end{mdframed}

 \item Construct the oracle $\mathcal{O}_4$.
    \begin{mdframed}
    {\bf \underline{$\ora_4(m, y)$}}
    
    {\bf Hardcoded: $\hsig$}
    \begin{enumerate}[label=\arabic*.]
    \item Output $1$ if $\hsig(m) = y$, and output $0$ otherwise.
    \end{enumerate}
    \end{mdframed}

    \item Set $sn = (\ora_0, \ora_1, \ora_2, \ora_{\oss.\sign}, \ora_3, \ora_4)$.

    \item Simulate the purified version of $\oss.\genqkey^{\ora_{\oss.\genqkey}}(1^\lambda)$ and let $(\regis{vk}, \regis{key}, \regis{out-aux})$ be the output registers.
    
    \item Output $sn, (\regis{vk}, \regis{key}, \regis{out-aux})$.
\end{enumerate}

\paragraph{\underline{$\qkeyfiresign.\sign^{\ora}(\regis{flame}, m)$}}
\begin{enumerate}
    \item Parse $(\ora_0, \ora_1, \ora_2, \ora_{\oss.\sign}, \ora_{\oss.\ver}, \ora_3, \ora_4) = \ora$.
 \item Parse $(\regis{vk}, \regis{key}, \regis{out-aux}) = \regis{flame}$.
\item Measure $\regis{vk}$ to obtain $ivk$.
\item Coherently run the following to obtain $y_0$ and then rewind (as in \cref{prelem:gentlemes}): 
\begin{enumerate}[label=\arabic*.]
    \item Simulate $\oss.\sign^{\ora_{\oss.\sign}}(ivk, \regis{key}, 0)$, let $isig$ be the output. Run $\ora_0(ivk, isig, m)$ and let $y_0$ be the output.
\end{enumerate}
\item Coherently run the following to obtain $y_1$ and then rewind (as in \cref{prelem:gentlemes}): 
\begin{enumerate}[label=\arabic*.]
    \item Simulate $\oss.\sign^{\ora_{\oss.\sign}}(ivk, \regis{key}, 0)$, let $isig$ be the output. Run $\ora_1(ivk, isig, m)$ and let $y_1$ be the output.
\end{enumerate}

\item Output $\ora_3(ivk, y_0, y_1, m)$.
\end{enumerate}

\paragraph{\underline{$\qkeyfiresign.\ver^{\ora}(m, sig)$}}
\begin{enumerate}
\item Parse $(\ora_0, \ora_1, \ora_2, \ora_{\oss.\sign}, \ora_3, \ora_4) = \ora$.
    \item Output $\ora_4(m, sig)$.
\end{enumerate}

\paragraph{\underline{$\qkeyfiresign.\clone^{\ora}(\regis{flame})$}}
\begin{enumerate}
    \item Parse $(\ora_0, \ora_1, \ora_2, \ora_{\oss.\sign}, \ora_3, \ora_4) = \ora$.
 \item Parse $(\regis{vk}, \regis{key}, \regis{out-aux,1}) = \regis{flame}$.
    \item Create the registers $\regis{msg}, \regis{aux,2}, \regis{aux,3}, \regis{ora-out,4}, \regis{ora-out,5}$. Initialize  $\regis{aux,2}$ and $\regis{aux,3}$ to $\ket{0^{\auxsizeossign(\lambda)}}$ and $\ket{0^{\auxsizeosskey(\lambda)}}$. Initialize $\regis{ora-out,4}$ and $\regis{ora-out,5}$ to $\ket{0^\lambda}$. Initialize $\regis{msg}$ to $\ket{0}$. At some steps, we will parse $\regis{aux,3}$ as $\regis{ora-inp,3}, \regis{ora-out,3}, \regi'_{\mathsf{aux,3}}$, as in the oracle query step of $\oss.\genqkey$, where $U_{\ora_{\oss.\genqkey}}$ reads the oracle input input from $\regis{ora-inp,3}$ and writes the oracle output onto $\regis{ora-out,3}$.
    \item\label{item:cloneloop} For $\numstepsosskey(\lambda)$ steps, execute the following.
    \begin{enumerate}[label=\arabic*.]
    \item Apply $U_{\oss.\genqkey}$ to $\regis{aux,3}$. 
    \item Execute the following (which essentially simulates the oracle call to $\ora_{\oss.\genqkey}$):
    \begin{enumerate}[label=\arabic*.]
        \item Apply $U'_{\ora_{\oss.\sign}, \oss.\sign}$ to $\regis{vk}$, $\regis{msg}$,$\regis{key}$, $\regis{aux,2}$. Parse the output as $\regis{vk}$,$\regis{msg}$,$\regis{sig}$, $\regis{out-aux,2}$ as in $\oss.\sign$.
    \item Apply $U_{\ora_0}$ with input registers $\regis{vk}, \regis{sig}, \regis{ora-inp,3}$ and the output register $\regis{ora-out,4}$.
    \item Apply $(U'_{\ora_{\oss.\sign}, \oss.\sign})^{-1}$ to $\regis{vk}, \regis{msg}, \regis{sig}, \regis{aux,2}$, parse the output as $\regis{vk}$, $\regis{msg}$, $\regis{key}$, $\regis{aux,2}$.
    \item Set $\regis{msg}$ to $\ket{1}$.
    \item Apply $U'_{\ora_{\oss.\sign}, \oss.\sign}$ to $\regis{vk}, \regis{msg}, \regis{key}, \regis{aux,2}$. Parse the output as $\regis{vk}$,$\regis{msg}$, $\regis{sig}$, $\regis{out-aux,2}$ as in $\oss.\sign$.
    \item Apply $U_{\ora_1}$ with input registers $\regis{vk}, \regis{sig}, \regis{ora-inp,3}$ and the output register $\regis{ora-out,5}$.
    \item Apply $U_{\ora_2}$ with input registers $\regis{vk}, \regis{ora-out,4}, \regis{ora-out,5}$ and the output register $\regis{ora-out,3}$.
    \item Apply $U_{\ora_1}$ with input registers $\regis{vk}, \regis{sig}, \regis{ora-inp,3}$ and the output register $\regis{ora-out,5}$.
    \item Apply $(U'_{\ora_{\oss.\sign}, \oss.\sign})^{-1}$ to $, \regis{vk}, \regis{msg}, \regis{sig}, \regis{aux,2}$, parse the output as $\regis{vk}$, 
 $\regis{msg}$, $\regis{key}$, $\regis{aux,2}$.
        \item Set $\regis{msg}$ to $\ket{0}$.
    \item Apply $U'_{\ora_{\oss.\sign}, \oss.\sign}$ to $\regis{vk},\regis{msg}$,$\regis{key}$,  $\regis{aux,2}$.
    \item Apply $U_{\ora_0}$ with input registers $\regis{vk}, \regis{sig}, \regis{ora-inp,3}$ and the output register $\regis{ora-out,4}$.
    \item Apply $(U'_{\ora_{\oss.\sign}, \oss.\sign})^{-1}$ to $\regis{vk}, \regis{msg}, \regis{sig}, \regis{aux,2}$, parse the output as $\regis{vk}$, 
 $\regis{msg}$, $\regis{key}$, $\regis{aux,2}$.
    
    \end{enumerate}
    \end{enumerate}

    \item Parse $\regis{aux,3}$ as $\regi_{\mathsf{vk}}^*, \regi_{\mathsf{key}}^*, \regi_{\mathsf{out-aux}}^*$ as in $\oss.\genqkey$.
    \item Set $\regis{clone} = (\regi_{\mathsf{vk}}^*, \regi_{\mathsf{key}}^*, \regi_{\mathsf{out-aux}}^*)$.
    \item Output $\regis{flame}, \regis{clone}$.
\end{enumerate}

\begin{remark}
    As described above, the signing algorithm $\qkeyfiresign.\sign$ of our scheme consumes the flame state given to it. Note that thanks to the strong clonability correctness of the scheme, this is not a issue since one can clone the flame state before signing a message. However, if one insists on not consuming/altering the flame state while signing, this is also possible: Using the gentle measurement technique (\cref{prelem:gentlemes}) it is also easy to see that the signing algorithm can be made so that it does not alter the state of the flame register (except up to negligible trace distance), since the output is pseudodeterministic (it is equal to $\hsig(m)$ with overwhelming probability).
\end{remark}
\begin{remark}
    Aside from the conversion from quantum protection schemes for signing to quantum protection schemes for general unlearnable functionalities shown by the previous work, we can also slightly modify our construction above to achieve key-fire for unlearnable functionalities directly: Instead of $\ora_3, \ora_4$, we will just have a new $\ora_3$ which outputs $f^*(x)$ after verifying $y_0, y_1$. The proof also follows the same, with the difference that at the final hybrid, instead of invoking \cref{lem:qromunlearn}, we simply invoke the unlearnability of the functionality.
\end{remark}

We claim that the construction satisfies correctness and security.
\begin{theorem}
$\qkeyfiresign$ satisfies quantum key-fire signing correctness, strong cloning correctness (\cref{def:qkeyfiresign}). Let $\nu(\lambda)$ be superlogarithmic. Suppose $\oss$ satisfies $\ora_{\oss.\ver}$ incompressibility and one-shot security, then $\qkeyfiresign$ satisfies unbounded (i.e. non-interactive) leakage-resilience for signing unconditionally.  Suppose $\oss$ satisfies strong incompressibility and one-shot security, then $\qkeyfiresign$ satisfies LOCC leakage-resilience for signing (\cref{def:qfireorasec}) unconditionally. Further, the oracles of $\qkeyfiresign$ can be implemented efficiently using a one-way function while preserving security (in which case, the security becomes computational rather than information-theoretic).
\end{theorem}
We prove the signing correctness and cloning correctness in \cref{sec:signcorrectfire,sec:clonecorrect}, and we prove security in \cref{sec:proofsec}.

By implementing our construction using the incompressible one-shot signature we construct in \cref{thm:ossincompexists}, we get the following corollary.
\begin{theorem}
     Relative to a classical oracle, there exists a quantum key-fire scheme that satisfies LOCC leakage-resilience for signing functionality.
 \end{theorem}

\begin{theorem}
     Relative to a classical oracle, there exists a quantum key-fire scheme that satisfies LOCC leakage-resilience for any unlearnable functionality.
 \end{theorem}

\newpage\subsection{Proof of Signing Correctness}\label{sec:signcorrectfire}
In this section, we prove the signing correctness of our scheme.

For simplicity, we will assume that $\oss$ satisfies perfect correctness, and we will show that $\qkeyfiresign$ satisfies perfect signing correctness. Through elementary lemmas, this implies that if $\oss$ satisfies correctness with overwhelming probability, then $\qkeyfiresign$ also satisfies signing correctness with overwhelming probability.

We first observe that the flame states of our scheme ($sn, \ket{\psi}_{\regis{vk}, \regis{key}, \regis{out-aux}}$) output by $\qkeyfiresign.\setup(1^\lambda)$ are the same as the purified output of $\oss.\genqkey$ (whose original version traces out $\regis{out-aux}$ and measures $\regis{vk}$, but $\qkeyfiresign.\setup$ does neither since it is purified). Then, $\qkeyfiresign.\sign$ first measures $\regis{vk}$ to obtain some value $ivk$ and (effectively) traces out $\regis{out-aux}$. This collapses the state to the one-shot signature key with the serial number $ivk$. By signing correctness of $\oss$ and definition of $\ora_0$, we get $y_0 = H_0(ivk || m)$ with probability $1$. Thus, by \cref{prelem:gentlemes}, the signing key is indeed not altered since we are rewinding. Then again by signing correctness of $\oss$ and definition of $\ora_1$, we get $y_1 = H_1(ivk || m)$ with probability $1$. Hence, the output of the signing algorithm is $\ora_3(ivk, H_0(ivk || m), H_1(ivk || m), m)$ which is equal to  $\hsig(m)$. Due to the definition of the verification algorithm, this completes the proof of signing correctness.

\subsection{Proof of Strong Cloning Correctness}\label{sec:clonecorrect}
In this section, we prove the strong cloning correctness of our scheme.

First, we note that the flame states of our scheme are indeed pure states as required for strong cloning correctness. Observe that the state output by $\qkeyfiresign.\setup$ is simply $\ket{\psi_{flame}} = (U_{\ora_{\oss.\genqkey}}\cdot U_{\oss.\genqkey})^{\numstepsosskey(\lambda)}\ket{0^{\auxsizeosskey(\lambda)}}$. This is also the same as the purified output of $\oss.\genqkey$ (whose not-purified version traces out $\regis{out-aux}$ and measures $\regis{vk}$, but $\qkeyfiresign.\setup$ does neither).

For simplicity, we will assume that $\oss$ satisfies perfect correctness, and we will show that $\qkeyfiresign$ satisfies perfect strong cloning correctness. Through elementary lemmas, this implies that if $\oss$ satisfies correctness with overwhelming probability, then $\qkeyfiresign$ also satisfies strong cloning correctness with overwhelming probability.

We first prove the following lemma.
\begin{lemma}
    During the loop in $\qkeyfiresign.\clone$, if the state over the registers $\regis{vk}$, $\regis{key}$, $\regis{out-aux,1}$, $\regis{msg}$, $\regis{aux,2}$, $\regis{aux,3}$, $\regis{ora-out,4}$, $\regis{ora-out,5}$ is $$ \ket{\psi_{flame}}_{\regis{vk}, \regis{key}, \regis{out-aux,1}} \otimes \ket{0}_{\regis{msg}} \otimes \ket{0^{p_2(\lambda)}}_{\regis{aux,2}} \otimes  \ket{\zeta}_{\regis{aux,3}} \otimes \ket{0^{\lambda}}_{{\regis{ora-out,4}}} \otimes \ket{0^{\lambda}}_{{\regis{ora-out,5}}}$$ at the beginning of an iteration, then the state at the end of that iteration is $$\ket{\psi_{flame}}_{\regis{vk}, \regis{key}, \regis{out-aux,1}} \otimes \ket{0}_{\regis{msg}} \otimes \ket{0^{p_2(\lambda)}}_{\regis{aux,2}} \otimes  (U_{\ora_{\oss.\genqkey}}\cdot U_{\oss.\genqkey} \ket{\zeta}) \otimes \ket{0^{\lambda}}_{{\regis{ora-out,4}}} \otimes \ket{0^{\lambda}}_{{\regis{ora-out,5}}}.$$
\end{lemma}
\begin{proof}
    We will show the result by writing the state of the registers after each step of the iteration. The actual order of the registers will stay the same throughout, however, when writing the state we will reorder some registers for convenience of notation. 
    
    We also introduce the following notation for conciseness.
    \begin{itemize}
        \item $\ket{\psi_{flame}} = \sum_{vk, w} \beta_{vk,w}\ket{vk}_{\regis{vk}} \otimes \ket{\phi_{vk,w}}_{\regis{key}} \otimes \ket{w}_{\regis{out-aux}}$,
        \item $U_1 = U_{\oss.\genqkey}$,
        \item $U_2 = U_{\ora_{\oss.\genqkey}}$,
        \item $U = U_2\cdot U_1$,
        \item $U_1\ket{\zeta} = \sum_{z,o,w'} \gamma_{z,o,w'} \ket{z}_{\regis{ora-inp,3}} \otimes \ket{o}_{\regis{ora-out,3}} \otimes \ket{w'}_{\regi'_{\mathsf{aux,3}}}$.
    \end{itemize}

    Now we show the evolution of the state during the loop.

     \begin{enumerate}[label=\arabic*.]
    \item After applying $U_{\oss.\genqkey}$ to $\regis{aux,3}$, we get
    \begin{equation*}
\ket{\psi_{flame}}_{\regis{vk}, \regis{key}, \regis{out-aux,1}} \otimes \ket{0}_{\regis{msg}} \otimes \ket{0^{p_2(\lambda)}}_{\regis{aux,2}} \otimes  (U_1\ket{\zeta})_{\regis{aux,3}} \otimes \ket{0^{\lambda}}_{{\regis{ora-out,4}}} \otimes \ket{0^{\lambda}}_{{\regis{ora-out,5}}}
    \end{equation*}
    \item After applying $U'_{\ora_{\oss.\sign}, \oss.\sign}$ to $\regis{vk}, \regis{key}, \regis{msg}, \regis{aux,2}$, we get
 \begin{equation*}
        \ket{0}_{\regis{msg}} \otimes 
 \left(\sum_{vk, w} \beta_{vk,w}\ket{vk}_{\regis{vk}} \otimes \sum_{sig}\alpha_{vk,w,sig} \ket{sig} \right) \otimes  (U_1\ket{\zeta})_{\regis{aux,3}} \otimes \ket{0^{\lambda}}_{{\regis{ora-out,4}}} \otimes \ket{0^{\lambda}}_{{\regis{ora-out,5}}}
    \end{equation*}

    where for each $vk, w, sig$ with $\beta_{vk,w} \neq 0$ and $\alpha_{vk, w, sig} \neq 0$ we have that $\ora^*_{\oss.\ver}(vk, sig, 0) = 1$ by correctness of $\oss.\sign$.
    
    \item After applying $U_{\ora_0}$ with input registers $\regis{vk}, \regis{sig}, \regis{ora-inp,3}$ and the output register $\regis{ora-out,4}$, we get
    
\begin{align*}
\ket{0}_{\regis{msg}} \otimes 
 \bigg(&\sum_{\substack{vk,w \\ z,o,w'}}  \beta_{vk,w}\ket{vk}_{\regis{vk}} \otimes (\sum_{sig}\alpha_{vk,w,sig} ) \gamma_{z,o,w'} \ket{z}_{\regis{ora-inp,3}} \otimes \\ &\ket{o}_{\regis{ora-out,3}}\otimes\ket{\ora_0(vk, sig_{vk,w}, z)}_{\regis{ora-out,4}} \otimes \ket{w'}_{\regi'_{\mathsf{aux,3}}} \bigg) \otimes  \ket{0^{\lambda}}_{{\regis{ora-out,5}}}
\end{align*}
which is equal to
   
\begin{align*}
&\ket{0}_{\regis{msg}} \otimes \\
 &\Bigg(\sum_{\substack{vk,w}}  \beta_{vk,w} \left(\ket{vk}_{\regis{vk}}\otimes \big( \sum_{z,o,w'}\gamma_{z,o,w'} \ket{z}_{\regis{ora-inp,3}} \ket{o}_{\regis{ora-out,3}} \ket{H_0(vk, z)}_{\regis{ora-out,4}} \otimes \ket{w'}_{\regi'_{\mathsf{aux,3}}} \big)\right) \otimes \\ &\ket{sig_{vk,w}} \Bigg)  \otimes \ket{0^{\lambda}}_{{\regis{ora-out,5}}}
\end{align*}

    \item After applying $(U'_{\ora_{\oss.\sign}, \oss.\sign})^{-1}$ to $\regis{crs}, \regis{vk}, \regis{msg}, \regis{sig}, \regis{aux,2}$, we get
 
\begin{align*}
&\ket{0}_{\regis{msg}} \otimes \\
 &\bigg(\sum_{\substack{vk,w}}  \beta_{vk,w} \left(\ket{vk}_{\regis{vk}}\otimes \big( \sum_{z,o,w'}\gamma_{z,o,w'} \ket{z}_{\regis{ora-inp,3}} \ket{o}_{\regis{ora-out,3}} \ket{H_0(vk, z)}_{\regis{ora-out,4}} \otimes \ket{w'}_{\regi'_{\mathsf{aux,3}}} \big)\right) \otimes \\ &\ket{\phi_{vk,w}}_{\regis{key}} \otimes \ket{w}_{\regis{out-aux}} \otimes \ket{0^{p_2(\lambda)}}_{\regis{aux,2}} \bigg) \otimes \ket{0^{\lambda}}_{{\regis{ora-out,5}}}
\end{align*}

    \item After applying $X^{}$ to $\regis{msg}$, this register changes from $\ket{0}$ to $\ket{1}$, and the others are unaffected.
    
    \item After applying $U'_{\ora_{\oss.\sign}, \oss.\sign}$ to $\regis{vk}, \regis{key}, \regis{msg}, \regis{aux,2}$ we get

\begin{align*}
&\ket{1}_{\regis{msg}} \otimes \\
 &\bigg(\sum_{\substack{vk,w}}  \beta_{vk,w} \left(\ket{vk}_{\regis{vk}}\otimes \big( \sum_{z,o,w'}\gamma_{z,o,w'} \ket{z}_{\regis{ora-inp,3}} \ket{o}_{\regis{ora-out,3}} \ket{H_0(vk, z)}_{\regis{ora-out,4}} \otimes \ket{w'}_{\regi'_{\mathsf{aux,3}}} \big)\right) \otimes \\ &\sum_{sig}\alpha'_{vk,w,sig}\ket{sig} \bigg) \otimes \ket{0^{\lambda}}_{{\regis{ora-out,5}}}
\end{align*}
    where for each $vk, w, sig$ with $\beta_{vk,w} \neq 0$ and $\alpha'_{vk, w, sig} \neq 0$ we have that $\ora^*_{\oss.\ver}(vk, sig, 1) = 1$ by correctness of $\oss.\sign$.

    \item After applying $U_{\ora_1}$ with input registers $\regis{vk}, \regis{sig}, \regis{ora-inp,3}$ and the output register $\regis{ora-out,5}$, we get
\begin{align*}
&\ket{1}_{\regis{msg}} \otimes \\
 &\Bigg(\sum_{\substack{vk,w}}  \beta_{vk,w} \bigg(\ket{vk}_{\regis{vk}}\otimes \big( \sum_{z,o,w'}\gamma_{z,o,w'} \ket{z}_{\regis{ora-inp,3}} \ket{o}_{\regis{ora-out,3}} \ket{H_0(vk, z)}_{\regis{ora-out,4}} \ket{H_1(vk,z)}_{{\regis{ora-out,5}}} \otimes \\ &\ket{w'}_{\regi'_{\mathsf{aux,3}}} \big)\bigg) \otimes  \sum_{sig}\alpha'_{vk,w,sig}\ket{sig} \Bigg)  
\end{align*}

    \item After applying $U_{\ora_2}$ with input registers $\regis{vk}, \regis{ora-out,4}, \regis{ora-out,5}$ and the output register $\regis{ora-out,3}$, we get
    \begin{align*}
&\ket{1}_{\regis{msg}} \otimes \\
 &\bigg(\sum_{\substack{vk,w}}  \beta_{vk,w} \Bigg(\ket{vk}_{\regis{vk}}\otimes \big( \sum_{z,o,w'}\gamma_{z,o,w'} \ket{z}_{\regis{ora-inp,3}} \ket{o \oplus \ora^*_{\oss.\genqkey(z)}}_{\regis{ora-out,3}} \ket{H_0(vk, z)}_{\regis{ora-out,4}} \\&\ket{H_1(vk,z)}_{{\regis{ora-out,5}}} \otimes \ket{w'}_{\regi'_{\mathsf{aux,3}}} \big)\Bigg) \otimes \sum_{sig}\alpha'_{vk,w,sig}\ket{sig}\bigg)  
\end{align*}
    \item After applying $U_{\ora_1}$ with input registers $\regis{vk}, \regis{sig}, \regis{ora-inp,3}$ and the output register $\regis{ora-out,5}$, we get
    \begin{align*}
&\ket{1}_{\regis{msg}} \otimes \\
 &\bigg(\sum_{\substack{vk,w}}  \beta_{vk,w} \Bigg(\ket{vk}_{\regis{vk}}\otimes \big( \sum_{z,o,w'}\gamma_{z,o,w'} \ket{z}_{\regis{ora-inp,3}} \ket{o \oplus \ora^*_{\oss.\genqkey(z)}}_{\regis{ora-out,3}} \ket{H_0(vk, z)}_{\regis{ora-out,4}} \\ &\ket{0^{\lambda}}_{{\regis{ora-out,5}}} \otimes \ket{w'}_{\regi'_{\mathsf{aux,3}}} \big)\Bigg) \otimes \sum_{sig}\alpha'_{vk,w,sig}\ket{sig} \bigg)  
\end{align*}
    \item After applying $(U'_{\ora_{\oss.\sign}, \oss.\sign})^{-1}$ to $\regis{vk}, \regis{msg}, \regis{sig}, \regis{aux,2}$, we get
\begin{align*}
&\ket{1}_{\regis{msg}} \otimes \\
 &\bigg(\sum_{\substack{vk,w}}  \beta_{vk,w} \Bigg(\ket{vk}_{\regis{vk}}\otimes \big( \sum_{z,o,w'}\gamma_{z,o,w'} \ket{z}_{\regis{ora-inp,3}} \ket{o \oplus \ora^*_{\oss.\genqkey(z)}}_{\regis{ora-out,3}} \ket{H_0(vk, z)}_{\regis{ora-out,4}} \\ &\ket{0^{\lambda}}_{{\regis{ora-out,5}}} \otimes \ket{w'}_{\regi'_{\mathsf{aux,3}}} \big)\Bigg) \otimes  \ket{\phi_{vk,w}}_{\regis{key}} \otimes \ket{w}_{\regis{out-aux,1}}  \otimes \ket{0^{p_2(\lambda)}}_{\regis{aux,2}}\bigg)  
\end{align*}

    \item After applying $X^{}$ to $\regis{msg}$, this register changes from $\ket{1}$ to $\ket{0}$, and the others are unaffected.
    \item After applying $U'_{\ora_{\oss.\sign}, \oss.\sign}$ to $\regis{crs}, \regis{vk}, \regis{key}, \regis{msg}, \regis{aux,2}$, we get
    \begin{align*}
&\ket{0}_{\regis{msg}} \otimes \\
 &\bigg(\sum_{\substack{vk,w}}  \beta_{vk,w} \Bigg(\ket{vk}_{\regis{vk}}\otimes \big( \sum_{z,o,w'}\gamma_{z,o,w'} \ket{z}_{\regis{ora-inp,3}} \ket{o \oplus \ora^*_{\oss.\genqkey(z)}}_{\regis{ora-out,3}} \ket{H_0(vk, z)}_{\regis{ora-out,4}} \\ &\ket{0^{\lambda}}_{{\regis{ora-out,5}}} \otimes \ket{w'}_{\regi'_{\mathsf{aux,3}}} \big)\Bigg) \otimes \sum_{sig}\alpha_{vk,w,sig}\ket{sig} \bigg)  
\end{align*}
    where for each $vk, w, sig$ with $\beta_{vk,w} \neq 0$ and $\alpha_{vk, w, sig} \neq 0$ we have that $\ora^*_{\oss.\ver}(crs^*, vk, sig, 0) = 1$ by correctness of $\oss.\sign$.
    
    \item After applying $U_{\ora_0}$ with input registers $\regis{vk}, \regis{sig}, \regis{ora-inp,3}$ and the output register $\regis{ora-out,4}$, we get
     \begin{align*}
&\ket{0}_{\regis{msg}} \otimes \\
 &\bigg(\sum_{\substack{vk,w}}  \beta_{vk,w} \Bigg(\ket{vk}_{\regis{vk}}\otimes \big( \sum_{z,o,w'}\gamma_{z,o,w'} \ket{z}_{\regis{ora-inp,3}} \ket{o \oplus \ora^*_{\oss.\genqkey(z)}}_{\regis{ora-out,3}} \ket{0^\lambda}_{\regis{ora-out,4}} \\ &\ket{0^{\lambda}}_{{\regis{ora-out,5}}} \otimes \ket{w'}_{\regi'_{\mathsf{aux,3}}} \big)\Bigg) \otimes \sum_{sig}\alpha_{vk,w,sig}\ket{sig} \bigg)  
\end{align*}
    \item After applying $(U'_{\ora_{\oss.\sign}, \oss.\sign})^{-1}$ to $\regis{vk}, \regis{msg}, \regis{sig}, \regis{aux,2}$, we get
    \begin{align*}
&\ket{0}_{\regis{msg}} \otimes \\
 &\bigg(\sum_{\substack{vk,w}}  \beta_{vk,w} \Bigg(\ket{vk}_{\regis{vk}}\otimes \big( \sum_{z,o,w'}\gamma_{z,o,w'} \ket{z}_{\regis{ora-inp,3}} \ket{o \oplus \ora^*_{\oss.\genqkey(z)}}_{\regis{ora-out,3}} \ket{0^\lambda}_{\regis{ora-out,4}} \\ &\ket{0^{\lambda}}_{{\regis{ora-out,5}}} \otimes \ket{w'}_{\regi'_{\mathsf{aux,3}}} \big)\Bigg) \otimes \ket{\phi_{vk,w}}_{\regis{key}} \otimes \ket{w}_{\regis{out-aux,1}}  \otimes \ket{0^{p_2(\lambda)}}_{\regis{aux,2}} \bigg)  
\end{align*}
which is equal to
\begin{align*}
    &\ket{\psi_{flame}} \otimes  \ket{0}_{\regis{msg}} \otimes  \ket{0^{p_2(\lambda)}}_{\regis{aux,2}} \otimes U_2 (U_1\ket{\zeta}) \otimes \ket{0^{\lambda}}_{{\regis{ora-out,4}}} \otimes \ket{0^{\lambda}}_{{\regis{ora-out,5}}} = \\ &\ket{\psi_{flame}} \otimes  \ket{0}_{\regis{msg}} \otimes   \ket{0^{p_2(\lambda)}}_{\regis{aux,2}} \otimes U\ket{\zeta} \otimes \ket{0^{\lambda}}_{{\regis{ora-out,4}}} \otimes \ket{0^{\lambda}}_{{\regis{ora-out,5}}} 
\end{align*}
    as claimed. 
    \end{enumerate}
\end{proof}

 It is easy to see that the strong cloning correctness follows easily with this lemma: The loop starts with the state $$\ket{\psi}_{\regis{vk}, \regis{key}, \regis{out-aux,1}} \otimes \ket{0}_{\regis{msg}} \otimes \ket{0^{p_2(\lambda)}}_{\regis{aux,2}} \otimes  \ket{0^{p_2(\lambda)}}_{\regis{aux,3}} \otimes \ket{0^{\lambda}}_{{\regis{ora-out,4}}} \otimes \ket{0^{\lambda}}_{{\regis{ora-out,5}}},$$ thus, at the end, we get the state 
\begin{align*}
  & \ket{\psi_{flame}}_{\regis{vk}, \regis{key}, \regis{out-aux,1}} \otimes \ket{0}_{\regis{msg}} \otimes \ket{0^{p_2(\lambda)}}_{\regis{aux,2}} \otimes  (U^{p_1(\lambda)}\ket{0^{p_2(\lambda)}})_{\regis{aux,3}} \otimes \\ &\ket{0^{\lambda}}_{{\regis{ora-out,4}}} \otimes \ket{0^{\lambda}}_{{\regis{ora-out,5}}} = \\ & \ket{\psi_{flame}}_{\regis{vk}, \regis{key}, \regis{out-aux,1}} \otimes \ket{0}_{\regis{msg}} \otimes \ket{0^{p_2(\lambda)}}_{\regis{aux,2}} \otimes  (\ket{\psi})_{\regis{aux,3}} \otimes \ket{0^{\lambda}}_{{\regis{ora-out,4}}} \otimes \ket{0^{\lambda}}_{{\regis{ora-out,5}}}.  
\end{align*}
Hence, the algorithm outputs $\ket{\psi_{flame}} \otimes \ket{\psi_{flame}}$.

\newpage\newcommand{\origwinpoly}{q^*}
\newcommand{\numqueries}{t(\lambda)}
\newcommand{\numrounds}{u(\lambda)}
\newcommand{\finaladvsuffix}{\mathsf{final}}
\newcommand{\roundadve}[2]{\adve_{#1,#2}}

\newcommand{\advetworeply}[1]{S^{#1}}
\newcommand{\advetworeplynoind}{S}
\newcommand{\advintstatenosub}{\regis{internal}}
\newcommand{\advintstate}[2]{\regis{#1, internal}^{#2}} 
\newcommand{\leakmes}[1]{L^{#1}}
\newcommand{\hybindex}{\ell}
\newcommand{\newadv}[2]{\mathcal{B}_{{#1},{#2}}} 
\newcommand{\newadvnoind}{\mathcal{B}}
\newcommand{\newleak}[1]{T^{#1}}
\newcommand{\estadv}[2]{\mathcal{C}_{#1,#2}}
\newcommand{\estadvnoind}{\mathcal{C}}
\newcommand{\estadvsub}[4]{\mathcal{C}_{#1,#2,#3,#4}}
\newcommand{\qhelpstr}[1]{Y^{#1}}

\newcommand{\numest}{q(\lambda)} 
\newcommand{\numestbyind}[2]{r_{#1,#2}(\lambda)} 
\newcommand{\promisedlistsize}[2]{s_{#1,#2}(\lambda)}
\newcommand{\lastadv}[2]{\mathcal{Q}_{#1, #2}}
\newcommand{\lastadvout}[1]{V^{#1}}
\newcommand{\lastadvnoind}{\mathcal{Q}}

\newcommand{\plastadv}[2]{\mathcal{R}_{#1, #2}}
\newcommand{\plastadvnoind}{\mathcal{R}}
\newcommand{\plastadvout}[1]{G^{#1}}

\newcommand{\helplist}[3]{X^{#2}_{#1,#3}}
\newcommand{\helplistnew}[2]{X^{#2}_{#1}}
\newcommand{\helplistone}[1]{X_{#1}}
\newcommand{\helplistnosub}{X}

\newcommand{\dbadv}[3]{D^{#2}_{#1,#3}}
\newcommand{\dbadvone}[1]{D_{#1}}
\newcommand{\dbadvnosub}{D}

\newcommand{\numfourqueries}{t(\lambda)}
\newcommand{\numthreequeries}{t(\lambda)}

\section{Proof of Security of the Main Construction}\label{sec:proofsec}

In this section, we prove the security of our scheme. Throughout the proof, we will refer to the adversary that gets the flame state ($\adve_1$) as the sender adversary and to the adversary that outputs the forged signature as the receiver adversary ($\adve_2$).

Suppose for a contradiction that there exists a query-bounded adversary $\adve = (\adve_1, \adve_2)$ such that $\Pr[\lrsiggame{\adve}(1^\lambda) = 1] > \frac{1}{\origwinpoly(\lambda)}$ for some polynomial $\origwinpoly(\cdot)$ and infinitely many values $\lambda > 0$. Note that $\adve_1, \adve_2$ will have query access to the oracles $\ora_0, \ora_1, \ora_2, \ora_{\oss.\sign}, \ora_{\oss.\ver}, \ora_3, \ora_4$ of the construction.

    \paragraph{Adversary Assumptions} Without loss of generality, we make the following assumptions about the adversaries $\adve_1, \adve_2$ (where $\numrounds, \numqueries$ are some polynomials), and we also define some notation.

\begin{itemize}
    \item $(\adve_1, \adve_2)$ runs for exactly $\numrounds$ rounds with probability $1$, where each round consists of a message from $\adve_1$ to $\adve_2$ and then a message from $\adve_2$ to $\adve_1$ (except for the last round where $\adve_2$ outputs $\bot$ as its message).
    \item We will denote the $i$-round of $\adve_1$ as $\roundadve{1}{i}$, for $i \in [\numrounds]$. That is, at round $i$, $\adve_1$ simply executes $\roundadve{1}{i}$ on its internal state register to obtain its message and the updated internal state register. $\roundadve{1}{i}$ makes a total of $\numqueries$ queries to the oracles.
    \item Similar to above, we will denote the $i$-th round of $\adve_2$ as $\roundadve{2}{i}$ for $i \in [\numrounds]$. Note that for $\roundadve{2}{\numrounds}$, the message output by it is $\bot$ since there is no message from $\adve_2$ to $\adve_1$. Finally, we will write $\mathcal{A}_{2,\finaladvsuffix}$ to denote the challenge phase of the adversary $\adve_2$: It receives the internal state register produced by $\roundadve{2}{\numrounds}$ and the challenge message to be signed from the challenger and outputs a signature.
    \item Every round (i.e. for all $\hybindex \in [\numrounds]$), queries of $\roundadve{2}{i}$ to the oracles $\ora_0, \ora_1, \ora_2$ proceeds in triples: It first makes one query to $\ora_0$, then to $\ora_1$, and then to $\ora_2$. We will call these groups of three queries a \emph{query-triplet}. Further, we assume that each round, $\roundadve{2}{i}$ has exactly $\numqueries$ query-triplets, with any (polynomial) number of queries to the other oracles in between the triplets but with a total of exactly $\numthreequeries$ queries to each of the oracles $\ora_3,\ora_4,\ora_{\oss.\sign}, \ora_{\oss.\ver}$ during the round. Note that this means that $\adve_2$ has a total of $\numrounds\cdot \numqueries$ query-triplets and a total of $2\numrounds\cdot \numqueries$ over the leakage protocol.
\end{itemize}
Note that without loss of generality, we can take $t(\lambda) = 1$ by modifying our adversaries, since we do not impose an a-priori bound on the number of rounds. Similarly, without loss of generality, we can assume that each round, queries of $\roundadve{2}{i}$ to $\ora_{\oss.\sign}, \ora_3,\ora_4$ occur in this order, after its queries to $\ora_0,\ora_1,\ora_2$.

\paragraph{Organization of the proof} Our proof will be done through a sequence of hybrids, $\hyb_0$ (which is the original game $\lrsiggame{\adve}(1^\lambda)$) through $\hyb_{3}$ (with some additional subhybrids in between). In the remainder of this section, we simply state our main theorems: We state our theorem about the indistinguishability of the first and the last hybrid (\cref{thm:hybridindist}), and about the security in the final hybrid $\hyb_{3}$ (\cref{thm:finalhybridsecure}). Then, combining these main theorems simply proves the security of our scheme. The main theorems then are then proven in the subsequent subsections. In more detail, our proof is organized as follows.

In \cref{sec:subroutines}, we define various modified adversaries that we use in our hybrids. Then, in \cref{sec:hybdef}, we give the definitions of our hybrids. Then in \cref{sec:hybrid-indist},  we prove \cref{thm:hybridindist}, that is, we show the indistinguishability of our hybrids (except for $\hyb_{2,\hybindex,0}$,$\hyb_{2,\hybindex,1}$ which we prove in \cref{sec:bighybrid}). Finally, in  \cref{sec:finalhybridsecure}, we prove \cref{thm:finalhybridsecure}, that is, we show the security in the final hybrid.

\paragraph{}
We now give our main theorem statements about the hybrids (which are defined in \cref{sec:hybdef}).
\begin{theorem}\label{thm:hybridindist}
    $\left|\hyb_0 - \hyb_{3}\right| \leq \frac{1}{2\cdot \origwinpoly(\lambda)}$
\end{theorem}
We prove this theorem in \cref{sec:hybrid-indist}.

\begin{theorem}\label{thm:finalhybridsecure}
    $\Pr[\hyb_{3} = 1] \leq \negl(\lambda)$.
\end{theorem}
We prove this theorem in \cref{sec:finalhybridsecure}.
\paragraph{}
Now we complete the proof of security assuming these two theorems. We had assumed for a contradiction that $\Pr[\lrsiggame{\adve}(1^\lambda) = 1] = \Pr[\hyb_0 = 1] > \frac{1}{\origwinpoly(\lambda)}$ for some polynomial $\origwinpoly(\cdot)$ and infinitely many values $\lambda > 0$. Combining this with \cref{thm:hybridindist}, we get that $\Pr[\hyb_{3} = 1] > \frac{1}{2\origwinpoly(\lambda)}$ for infinitely many $\lambda > 0$. However, this is a contradiction by \cref{thm:finalhybridsecure}. Therefore, our contradiction hypothesis is wrong, and thus we get $\Pr[\lrsiggame{\adve}(1^\lambda) = 1] \leq \negl(\lambda)$, completing the proof of security.

\subsection{Defining Modified Adversaries}\label{sec:subroutines}
In this section, we define the modified adversaries we will be using in the hybrids. We will explicitly denote the section queries to the oracles $H_0, H_1$.

We define $\roundadve{1}{\numrounds+1}$ and $\roundadve{2}{\numrounds+1}$ as follows.
\begin{mdframed}
 {\bf \underline{$\roundadve{1}{\numrounds+1}^\ora(\advintstatenosub, \advetworeply{})$}}
 
    \begin{enumerate}
    \item Output $\bot$.
    \end{enumerate}

\end{mdframed}
\begin{mdframed}
 {\bf \underline{$\roundadve{2}{\numrounds+1}^\ora(\advintstatenosub, \leakmes{})$}}

    \begin{enumerate}
    \item Sample $m' \samp \messpa$.
    \item $sig \samp \adve_{2,\finaladvsuffix}^{\ora}(\advintstatenosub, m')$.
    \item Output $\bot, (m', sig)$.
    \end{enumerate}
\end{mdframed}

We define $\newadv{1}{0},\newadv{2}{0}$.
\begin{mdframed}
 {\bf \underline{$\newadv{1}{0}^{(\ora_0, \ora_1, \ora_2, \ora_{\oss.\sign}, \ora_3, \ora_4)}(\regi)$}}

 \begin{enumerate}
     \item Output $\regi, \bot$.
 \end{enumerate}
 \end{mdframed}
 \begin{mdframed}
 {\bf \underline{$\newadv{2}{0}^{(\ora_0, \ora_1, \ora_2, \ora_{\oss.\sign}, \ora_3, \ora_4)}(\newleak{})$}}

 \begin{enumerate}
     \item Output $\bot$.
 \end{enumerate}
 \end{mdframed}

For $\hybindex \in [\numrounds+2]$, we define the adversaries $\newadv{1}{\hybindex}$,$\newadv{2}{\hybindex}$.
\begin{mdframed}
 {\bf \underline{$\newadv{1}{\hybindex}^{(\ora_0, \ora_1, \ora_2, \ora_{\oss.\sign}, \ora_3, \ora_4)}(\regi)$}}

    \begin{enumerate}
        \item $\advintstatenosub, (T, S) \samp \newadv{1}{\hybindex - 1}^{(\ora_0, \ora_1, \ora_2, \ora_{\oss.\sign}, \ora_3, \ora_4)}(\regi)$.
        \item $\advintstatenosub', \leakmes{} \samp \roundadve{1}{\hybindex}^{(\ora_0, \ora_1, \ora_2, \ora_{\oss.\sign},   \ora_3, \ora_4)}(\advintstatenosub, S)$.
            \item $\lastadvout{} \samp \lastadv{1}{\hybindex}^{(\ora_0, \ora_1, \ora_2, \ora_{\oss.\sign},   \ora_3, \ora_4)}(\newleak{}, \leakmes{})$.
            \item Set $\helplistone{4}$ to be the empty set.
        \item For $j \in [512^2\cdot(t(\lambda))^6\cdot \lambda^8 \cdot (u(\lambda))^4 \cdot (q^*(\lambda))^4 \cdot (\ell)^2]$:
        \begin{enumerate}[label=\arabic*.]
            \item Sample a random index $ind$ between $1$ and the number of oracle queries made by $\lastadv{2}{\hybindex}$.
            \item Simulate $\lastadv{2}{\hybindex}^{\ora_4}(\lastadvout{})$ until right before its $ind$-th oracle query is executed and measure the query input register instead of executing the query. Let $m, y$ be the measurement outcome.
            \item Check if $\ora_4(m,y)$ is $1$. If so, add $(m,y)$ to $\helplistone{4}$.
        \end{enumerate}
        \item {Simulate $\lastadv{2}{\hybindex}(\lastadvout{})$ with the following oracle $\ora'$: On input $m,y$, output $1$ if $(m,y) \in \helplistone{4}$, otherwise output $0$.  Let $(\advintstatenosub, \advetworeply{})$ be the output.}
\item Set $\newleak{}_{new} = (\newleak{}, (\lastadvout{}, \helplistone{4}, \advetworeplynoind))$.
\item Output $\advintstatenosub', (\newleak{}_{new}, \advetworeplynoind)$.
    \end{enumerate}
\end{mdframed}

\begin{mdframed}
 {\bf \underline{$\newadv{2}{\hybindex}(\newleak{})$}}

   \begin{enumerate}
   \item Parse $(\newleak{}_{prev}, (\lastadvout{}_{}, \helplistone{4},\advetworeplynoind)) = \newleak{}$.
    \item Until $\advetworeplynoind' = \advetworeplynoind$:
    \begin{enumerate}[label=\arabic*.]
    \item {Simulate $\lastadv{2}{\hybindex}(\lastadvout{})$ with the following oracle $\ora'$: On input $m,y$, output $1$ if $(m,y) \in \helplistone{4}$, otherwise output $0$.  Let $(\advintstatenosub, \advetworeplynoind')$ be the output.}
    \end{enumerate}
    \item Output $\advintstatenosub, \advetworeplynoind$.
    \end{enumerate}
\end{mdframed}

\paragraph{$\estadvnoind$-Adversaries}  
We define $\estadvsub{1}{\hybindex}{0}{2}$ to be the following algorithm: On input $(\newleak{}, \leakmes{})$, output $(\newleak{}$, $\leakmes{}$,$(\helplistone{0,0}{}$,$\helplistone{1,0}{}$,$\helplistone{2,0}{}$,$\dbadvone{0,0}{}$,$\dbadvone{1,0}{},\dbadvone{2,0}{}))$ where $\helplistone{0,0}{},\helplistone{1,0}{},\helplistone{2,0}{}$ are empty sets and $\dbadvone{0,0}{},\dbadvone{1,0}{},\dbadvone{2,0}{}$ are empty databases.  We define $\estadvsub{2}{\hybindex}{0}{2}(\qhelpstr{})$ to be the algorithm that runs $\adve_{2,\hybindex}(\regi, L)$ after parsing $(\newleak{}, \leakmes{}, \qhelpstr{'}) = \qhelpstr{}$ and executing $\regi \samp \newadv{2}{\ell}(\newleak{})$. 

For $\hybindex \in [\numrounds+1]$, $j \in [\numqueries]$ we define $\estadvsub{1}{\hybindex}{j}{0}$,$\estadvsub{2}{\hybindex}{j}{0}$. 

\begin{mdframed}
 {\bf \underline{$\estadvsub{1}{\hybindex}{j}{0}^{(\ora_0, \ora_1, \ora_2, \ora_{\oss.\sign},   \ora_3, \ora_4)}(\newleak{},\leakmes{})$}}

    \begin{enumerate}
        \item $\qhelpstr{} \samp \estadvsub{1}{\hybindex}{j-1}{2}^{(\ora_0, \ora_1, \ora_2, \ora_{\oss.\sign},   \ora_3, \ora_4)}(\newleak{},\leakmes{})$.
        \item Parse $(\newleak{}$, $\leakmes{}$, $((\helplistone{0,j'})_{j' \in [j-1]}$,  $(\helplistone{1,j'})_{j' \in [j-1]}$, $(\helplistone{2,j'})_{j' \in [j-1]}$,   $(\dbadvone{0,j'}{})_{j' \in [j-1]}$, $(\dbadvone{1,j'}{})_{j' \in [j-1]}$,  $(\dbadvone{2,j'}{})_{j' \in [j-1]})) = \qhelpstr{}$.
        \item Initialize $\helplistone{0,j}$ to be the empty set and $\dbadvone{0,j}$ to be an empty set.
      \item For $i \in [ 4194304\cdot \lambda^{10} \cdot t^2(\lambda)\cdot (\hybindex)^5\cdot (\numrounds\cdot \origwinpoly(\lambda)\cdot\numqueries)^4]$:
        \begin{enumerate}[label=\arabic*.]
        \item Run $\estadvsub{2}{\hybindex}{j-1}{2}$ on input $\qhelpstr{}$ until its first query to $\ora_0$ but do not execute the query. Instead, measure the query input register and let $ivk,isig,z$ be the measurement result. During this procedure\footnote{Note that $\estadvsub{1}{\hybindex}{j-1}{2}$ does not make any queries to $\ora_1, \ora_2$ before making its first query to $\ora_0$.},  $\estadvsub{2}{\hybindex}{j-1}{2}$ makes queries to $\ora_{\oss.\sign},   \ora_3, \ora_4$ which are executed by forwarding them to the corresponding oracles.
        \item If $\ora_0(ivk, isig,z) \neq \bot$, add $(ivk, isig)$ to $\helplistone{0,j}$ and set $\dbadvone{0,j}(ivk || z) = \ora_0(ivk, isig, z)$ for all $z$ (\emph{section query}).
        \end{enumerate}
        \item Output $(\newleak{}$, $\leakmes{}$, $((\helplistone{0,j'})_{j' \in [j]}$,  $(\helplistone{1,j'})_{j' \in [j-1]}$, $(\helplistone{2,j'})_{j' \in [j-1]}$, $(\dbadvone{0,j'}{})_{j' \in [j]}$, $(\dbadvone{1,j'}{})_{j' \in [j-1]}$,  $(\dbadvone{2,j'}{})_{j' \in [j-1]}))$.
    \end{enumerate}
\end{mdframed}

\begin{mdframed}
 {\bf \underline{$\estadvsub{2}{\hybindex}{j}{0}^{(\ora_0, \ora_1, \ora_2, \ora_{\oss.\sign},   \ora_3, \ora_4)}(\qhelpstr{}{})$}}

  \begin{enumerate}[label=\arabic*.]
\item  Parse  $(\newleak{}$, $\leakmes{}$, $((\helplistone{0,j'})_{j' \in [j]}$,  $(\helplistone{1,j'})_{j' \in [j-1]}$, $(\helplistone{2,j'})_{j' \in [j-1]}$,   $(\dbadvone{0,j'}{})_{j' \in [j]}$, $(\dbadvone{1,j'}{})_{j' \in [j-1]}$, $(\dbadvone{2,j'}{})_{j' \in [j-1]})) = \qhelpstr{}$.
\item Set $\qhelpstr{'} = (\newleak{}$, $\leakmes{}$,$((\helplistone{0,j'})_{j' \in [j-1]}$,  $(\helplistone{1,j'})_{j' \in [j-1]}$,$(\helplistone{2,j'})_{j' \in [j-1]}$,  $(\dbadvone{0,j'}{})_{j' \in [j-1]}$,$(\dbadvone{1,j'}{})_{j' \in [j-1]}$, $(\dbadvone{2,j'}{})_{j' \in [j-1]}))$.
            \item Run $\estadvsub{2}{\hybindex}{j-1}{2}$ on input $\qhelpstr{'}$ with the oracles $(\ora_0, \ora_1, \ora_2, \ora_{\oss.\sign},   \ora_3, \ora_4)$, but only the first query to $\ora_0$ is instead simulated as follows: On input $ivk, isig, z$, output $\dbadvone{0,j}{}(ivk || z)$ if $(ivk,isig)\in\helplistone{0,j}$ and otherwise output $\bot$. Let  $\advintstatenosub, \advetworeplynoind$ be the output.
            \item Output  $\advintstatenosub, \advetworeplynoind$.
    \end{enumerate}
\end{mdframed}

 For $\hybindex \in [\numrounds+1]$, $j \in [\numqueries]$ we define $\estadvsub{1}{\hybindex}{j}{1}$,$\estadvsub{2}{\hybindex}{j}{1}$.
\begin{mdframed}
 {\bf \underline{$\estadvsub{1}{\hybindex}{j}{1}^{(\ora_0, \ora_1, \ora_2, \ora_{\oss.\sign},   \ora_3, \ora_4)}(\newleak{},\leakmes{})$}}

    \begin{enumerate}
        \item $\qhelpstr{} \samp \estadvsub{1}{\hybindex}{j}{0}^{(\ora_0, \ora_1, \ora_2, \ora_{\oss.\sign},   \ora_3, \ora_4)}(\newleak{},\leakmes{})$.
        \item Parse  $(\newleak{}$, $\leakmes{}$, $((\helplistone{0,j'})_{j' \in [j]}$, $(\helplistone{1,j'})_{j' \in [j-1]}$, $(\helplistone{2,j'})_{j' \in [j-1]}$,  $(\dbadvone{0,j'}{})_{j' \in [j]}$, $(\dbadvone{1,j'}{})_{j' \in [j-1]}$, $(\dbadvone{2,j'}{})_{j' \in [j-1]})) = \qhelpstr{}$.
                \item Initialize $\helplistone{1,j}$ to be the empty set and $\dbadvone{1,j}$ to be an empty set.
\item For $i \in [ 4194304\cdot \lambda^{10} \cdot t^2(\lambda)\cdot (\hybindex)^5\cdot (\numrounds\cdot \origwinpoly(\lambda)\cdot\numqueries)^4]$:
        \begin{enumerate}[label=\arabic*.]
        \item Run $\estadvsub{2}{\hybindex}{j}{0}$ on input $\qhelpstr{}$ until its first query to $\ora_1$ but do not execute the query. Instead, measure the query input register and let $ivk,isig$ be the measurement result. During this procedure\footnote{Note that $\estadvsub{1}{\hybindex}{j}{0}$ does not make any queries to $\ora_0, \ora_2$ before making its first query to $\ora_1$.},  $\estadvsub{2}{\hybindex}{j}{0}$ makes queries to $\ora_{\oss.\sign},   \ora_3, \ora_4$ which are executed by forwarding them to the corresponding oracles.
        \item If $\ora_1(ivk, isig, z) \neq \bot$, add $(ivk, isig)$ to $\helplistone{1,j}$ and set $\dbadvone{1,j}(ivk || z) = \ora_1(ivk, isig, z)$ for all $z$ (\emph{section query}).
        \end{enumerate}
        \item Output  $(\newleak{}$, $\leakmes{}$,$(\helplistone{0,j'})_{j' \in [j]}$,$(\helplistone{1,j'})_{j' \in [j]}$,$(\helplistone{2,j'})_{j' \in [j-1]}$,$(\dbadvone{0,j'}{})_{j' \in [j]}$,$(\dbadvone{1,j'}{})_{j' \in [j]}$,$(\dbadvone{2,j'}{})_{j' \in [j-1]})$.
    \end{enumerate}
\end{mdframed}

\begin{mdframed}
 {\bf \underline{$\estadvsub{2}{\hybindex}{j}{1}^{(\ora_0, \ora_1, \ora_2, \ora_{\oss.\sign},   \ora_3, \ora_4)}(\qhelpstr{}{})$}}

  \begin{enumerate}[label=\arabic*.]
\item  Parse $(\newleak{}$, $\leakmes{}$, $((\helplistone{0,j'})_{j' \in [j]}$, $(\helplistone{1,j'})_{j' \in [j]}$, $(\helplistone{2,j'})_{j' \in [j-1]}$, $(\dbadvone{0,j'}{})_{j' \in [j]}$, $(\dbadvone{1,j'}{})_{j' \in [j]}$, $(\dbadvone{2,j'}{})_{j' \in [j-1]})) = \qhelpstr{}$.
\item $\qhelpstr{'} = (\newleak{}$, $\leakmes{}$, $((\helplistone{0,j'})_{j' \in [j]}$, $(\helplistone{1,j'})_{j' \in [j-1]}$, $(\helplistone{2,j'})_{j' \in [j-1]}$, $(\dbadvone{0,j'}{})_{j' \in [j]}$, $(\dbadvone{1,j'}{})_{j' \in [j-1]}$,$(\dbadvone{2,j'}{})_{j' \in [j-1]}))$.
            \item Run $\estadvsub{2}{\hybindex}{j}{0}$ on input $\qhelpstr{'}$ with the oracles $(\ora_0, \ora_1, \ora_2, \ora_{\oss.\sign},   \ora_3, \ora_4)$, but only the first query to $\ora_1$ is instead simulated as follows: On input $ivk, isig, z$, output $\dbadvone{1,j}{}(ivk || z)$ if $(ivk,isig)\in\helplistone{1,j}$ and otherwise output $\bot$. Let  $\advintstatenosub, \advetworeplynoind$ be the output.
            \item Output  $\advintstatenosub, \advetworeplynoind$.
    \end{enumerate}
\end{mdframed}

For $\hybindex \in [\numrounds+1]$, $j \in [\numqueries]$ we define $\estadvsub{1}{\hybindex}{j}{2}$,$\estadvsub{2}{\hybindex}{j}{2}$. 
\begin{mdframed}
 {\bf \underline{$\estadvsub{1}{\hybindex}{j}{2}^{(\ora_0, \ora_1, \ora_2, \ora_{\oss.\sign},   \ora_3, \ora_4)}(\newleak{},\leakmes{})$}}

    \begin{enumerate}
        \item $\qhelpstr{} \samp \estadvsub{1}{\hybindex}{j}{1}^{(\ora_0, \ora_1, \ora_2, \ora_{\oss.\sign},   \ora_3, \ora_4)}(\newleak{},\leakmes{})$.
        \item Parse $(\newleak{}$, $\leakmes{}$,$((\helplistone{0,j'})_{j' \in [j]}$,$(\helplistone{1,j'})_{j' \in [j]}$,$(\helplistone{2,j'})_{j' \in [j-1]}$,$(\dbadvone{0,j'}{})_{j' \in [j]}$,$(\dbadvone{1,j'}{})_{j' \in [j]}$,$(\dbadvone{2,j'}{})_{j' \in [j-1]})) = \qhelpstr{}$.
                \item Initialize $\helplistone{2,j}$ to be the empty set and $\dbadvone{2,j}$ to be an empty set.
      \item For $i \in [512\cdot\lambda\cdot (t(\lambda)\cdot \ell)\cdot\lambda^4\cdot (\numrounds\cdot \origwinpoly(\lambda)\cdot\numqueries)^2]$:
        \begin{enumerate}[label=\arabic*.]
        \item Run $\estadvsub{2}{\hybindex}{j}{1}$ on input $\qhelpstr{}$ until its first query to $\ora_2$ but do not execute the query. Instead, measure the query input register and let $ivk,y_0,y_1,z$ be the measurement result. During this procedure\footnote{Note that $\estadvsub{1}{\hybindex}{j}{1}$ does not make any queries to $\ora_0, \ora_1$ before making its first query to $\ora_2$.},  $\estadvsub{2}{\hybindex}{j}{1}$ makes queries to $\ora_{\oss.\sign},   \ora_3, \ora_4$ which are executed by forwarding them to the corresponding oracles.
        \item If $\ora_2(ivk, y_0, y_1, z) \neq \bot$, add $(ivk, y_0, y_1, z)$ to $\helplistone{2,j}$ and set $\dbadvone{2,j}(z) = \ora_2(ivk, y_0, y_1, z)$.
        \end{enumerate}
        \item Output $(\newleak{}$, $\leakmes{}$,$(\helplistone{0,j'})_{j' \in [j]}$,$(\helplistone{1,j'})_{j' \in [j]}$,$(\helplistone{2,j'})_{j' \in [j]}$,$(\dbadvone{0,j'}{})_{j' \in [j]}$,$(\dbadvone{1,j'}{})_{j' \in [j]}$,$(\dbadvone{2,j'}{})_{j' \in [j]})$.
    \end{enumerate}
\end{mdframed}

\begin{mdframed}
 {\bf \underline{$\estadvsub{2}{\hybindex}{j}{2}^{(\ora_0, \ora_1, \ora_2, \ora_{\oss.\sign},   \ora_3, \ora_4)}(\qhelpstr{}{})$}}

  \begin{enumerate}[label=\arabic*.]
\item  Parse $(\newleak{}$, $\leakmes{}$,$((\helplistone{0,j'})_{j' \in [j]}$,$(\helplistone{1,j'})_{j' \in [j]}$,$(\helplistone{2,j'})_{j' \in [j]}$,$(\dbadvone{0,j'}{})_{j' \in [j]}$,$(\dbadvone{1,j'}{})_{j' \in [j]}$,$(\dbadvone{2,j'}{})_{j' \in [j]})) = \qhelpstr{}$.
\item $\qhelpstr{'} = (\newleak{}$, $\leakmes{}$,$((\helplistone{0,j'})_{j' \in [j]}$,$(\helplistone{1,j'})_{j' \in [j]}$,$(\helplistone{2,j'})_{j' \in [j-1]}$,$(\dbadvone{0,j'}{})_{j' \in [j]}$,$(\dbadvone{1,j'}{})_{j' \in [j]}$,$(\dbadvone{2,j'}{})_{j' \in [j-1]}))$.
            \item Run $\estadvsub{2}{\hybindex}{j}{1}$ on input $\qhelpstr{'}$ with the oracles $(\ora_0, \ora_1, \ora_2, \ora_{\oss.\sign},   \ora_3, \ora_4)$, but only the first query to $\ora_2$ is instead simulated as follows: On input $ivk, y_0, y_1, z$, output $\dbadvone{2,j}{}(z)$ if $(ivk,y_0,y_1,z)\in\helplistone{2,j}$ and otherwise output $\bot$. Let  $\advintstatenosub, \advetworeplynoind$ be the output.
            \item Output  $\advintstatenosub, \advetworeplynoind$.
    \end{enumerate}
\end{mdframed}

 We define $\estadv{1}{\hybindex}$,$\estadv{2}{\hybindex}$ to be $\estadvsub{1}{\hybindex}{\numqueries}{2}$ and $\estadvsub{2}{\hybindex}{\numqueries}{2}$, respectively.
For ease of reference, we write the full description of $\estadv{2}{\hybindex}$ with the internal calls unwound.

\begin{mdframed}
 {\bf \underline{$\estadv{2}{\hybindex}^{\ora_{\oss.\sign},   \ora_3, \ora_4}(\qhelpstr{}{})$}}
 
    \begin{enumerate}[label=\arabic*.]
        \item Parse $(\newleak{}$, $\leakmes{}$, $((\helplistone{0,j'})_{j' \in [j]}$, $(\helplistone{1,j'})_{j' \in [\numqueries]}$, $(\helplistone{2,j'})_{j' \in [j]}$, $(\dbadvone{0,j'}{})_{j' \in [\numqueries]}$, $(\dbadvone{1,j'}{})_{j' \in [\numqueries]}$, $(\dbadvone{2,j'}{})_{j' \in [\numqueries]})) = \qhelpstr{}$.
        
            \item Run $\newadv{2}{\hybindex-1}(\newleak{})$ to get $\advintstatenosub$.
            
        \item Simulate $\adve_{2,\hybindex}^{(\ora_0', \ora_1', \ora_2', {\ora_{\oss.\sign}^*, \ora_{\oss.\ver}^*, \ora_3^*, \ora_4^*)}}(\advintstatenosub, \leakmes{})$ where
            \begin{itemize}
                \item The $i$-th (for $i \in [\numqueries]$) query-triplet to the oracles $\ora_0', \ora_1', \ora_2'$ is executed with the oracles $\ora^i_0, \ora^i_1, \ora^i_2$ defined as follows.
        \begin{itemize}
            \item $\ora^i_0(ivk, isig, z)$: If $(ivk, isig) \not\in \helplistone{0,i}$, output $\bot$. Otherwise, output $(\dbadvone{0,i}(ivk || z))$.
            \item $\ora^i_1(ivk, isig, z)$: If $(ivk, isig) \not\in \helplistone{1,i}$, output $\bot$. Otherwise, output $(\dbadvone{1,i}{}(ivk || z))$.
            \item $\ora^i_2(ivk, y_0, y_1, z)$: If $(ivk, y_0, y_1, z) \not\in \helplistone{2,i}{}$, output $\bot$. Otherwise, output $\dbadvone{2,i}{}(z)$.
       
        \end{itemize}
            \end{itemize}
        
        Let $\advintstatenosub', S$ be the output.
        \item Output $\advintstatenosub', S$.
    \end{enumerate}
\end{mdframed}

\paragraph{$\plastadvnoind$-Adversaries}
For $\hybindex \in [\numrounds+1]$, we define the adversaries $\plastadv{1}{\hybindex}$,$\plastadv{2}{\hybindex}$. 
\begin{mdframed}
 {\bf \underline{$\plastadv{1}{\hybindex}^{(\ora_0, \ora_1, \ora_2, \ora_{\oss.\sign},   \ora_3, \ora_4)}(\newleak{},\leakmes{})$}}

    \begin{enumerate}
 \item Set $\helplistone{sign}$ to be the empty set.
 \item $\qhelpstr{} \samp  \estadv{1}{\hybindex}^{(\ora_0, \ora_1, \ora_2, \ora_{\oss.\sign},   \ora_3, \ora_4)}(\newleak{},\leakmes{})$.
        \item For $j \in [\lambda^8\cdot \hybindex^2 \cdot (\numfourqueries)^4 \cdot (n(\lambda))^2 \cdot(t(\lambda)\cdot {\numrounds}\cdot \origwinpoly(\lambda))^2\cdot (\lambda\cdot {\numrounds}\cdot \origwinpoly(\lambda))^2 ]$:
        \begin{enumerate}[label=\arabic*.]
            \item Sample a random index $ind$ between $1$ and the number of oracle queries by $\estadv{2}{\ell}$ to $\ora_{\oss.\sign}$.
            \item Simulate $\estadv{2}{\hybindex}^{\ora_{\oss.\sign},   \ora_3, \ora_4}(\qhelpstr{})$ until right before its $ind$-th oracle query to $\ora_{\oss.\sign}$ is executed and measure the query input register instead of executing the query. Let $vk, x$ be the measurement outcome.  Check if $\ora_{\oss.\sign}(vk, x) \neq 0$ and if $x \not\in \mathsf{Span}(\{v: (vk, v) \in \helplistone{sign}\})$ and if so add $(vk, x)$ to $\helplistone{sign}$.
        \end{enumerate}
        \item Output $(\qhelpstr{}, \helplistone{sign})$.
    \end{enumerate}
\end{mdframed}

\begin{mdframed}
 {\bf \underline{$\plastadv{2}{\hybindex}^{\ora_3,\ora_4}(\plastadvout{})$}}

    \begin{enumerate}
    \item Parse $(\qhelpstr{}, \helplistone{sign}) = \plastadvout{}.$
    \item Define the oracle $\ora_{sign}'$: On input $vk, z$, output $1$ if $z \not\in\mathsf{Span}(\{v: (vk, v) \in \helplistone{sign}\}) \setminus \{0\}$. Otherwise, output $0$.
\item  Simulate $\estadv{2}{\hybindex}^{\ora_{sign}', \ora_3, \ora_4}(\qhelpstr{})$ to obtain $\advintstatenosub, \advetworeplynoind$.
\item Output $\advintstatenosub, \advetworeplynoind$.
    \end{enumerate}
\end{mdframed}

\paragraph{$\lastadvnoind$-Adversaries}
For $\hybindex \in [\numrounds+1]$, we define the adversaries $\lastadv{1}{\hybindex}$,$\lastadv{2}{\hybindex}$.
\begin{mdframed}
 {\bf \underline{$\lastadv{1}{\hybindex}^{(\ora_0, \ora_1, \ora_2, \ora_{\oss.\sign},   \ora_3, \ora_4)}(\newleak{},\leakmes{})$}}

    \begin{enumerate}
 \item Set $\helplistone{3}$ to be the empty list.
 \item Set $\dbadvone{3}$ to be empty database.
 \item $\plastadvout{} \samp  \plastadv{1}{\hybindex}^{(\ora_0, \ora_1, \ora_2, \ora_{\oss.\sign},   \ora_3, \ora_4)}(\newleak{},\leakmes{})$.
              \item For $i \in [512\cdot\lambda\cdot (t^2(\lambda)\cdot \ell)\cdot\lambda^4\cdot (\numrounds\cdot \origwinpoly(\lambda)\cdot\numqueries)^2]$:
        \begin{enumerate}[label=\arabic*.]
            \item Sample a random index $ind$ between $1$ and the number of oracle queries by $\plastadv{2}{\ell}$ to $\ora_{3}$.
            
            \item Simulate $\plastadv{2}{\hybindex}^{\ora_3, \ora_4}(\plastadvout{}{})$ until right before its $ind$-th oracle query to $\ora_{3}$ is executed and measure the query input register instead of executing the query. Let $ivk,y_0,y_1,m$ be the measurement outcome. If $\ora_3(ivk,y_0,y_1,m) \neq \bot$, add $(ivk,y_0,y_1,m)$ to $\helplistone{3}$ and set $\dbadvone{3}(m) = \ora_3(ivk,y_0,y_1,m)$.
        \end{enumerate}
        \item Output $(\plastadvout{}{}, (\helplistone{3},\dbadvone{3}))$.
    \end{enumerate}
\end{mdframed}

\begin{mdframed}
 {\bf \underline{$\lastadv{2}{\hybindex}^{\ora_4}(\lastadvout{})$}}

    \begin{enumerate}
    \item Parse $(\plastadvout{}{}, (\helplistone{3}, \dbadvone{3})) = \lastadvout{}.$
    \item Define the oracle $\ora_3'$: On input $ivk,y_0,y_1,m$, output $\dbadvone{3}(m)$ if $ivk,y_0,y_1,m \in \helplistone{3}$, otherwise output $\bot$. 
\item  Simulate $\plastadv{2}{\hybindex}^{\ora_3', \ora_4}(\plastadvout{})$ to obtain $\advintstatenosub, \advetworeplynoind$.
\item Output $\advintstatenosub, \advetworeplynoind$.
    \end{enumerate}
\end{mdframed}

\subsection{Defining the Hybrids}\label{sec:hybdef}
In this section, we define the hybrids for our proof, each of which is constructed by modifying the previous one. 

    \paragraph{\underline{$\hyb_{0}$:}} The original game $\lrsiggame{\adve}(1^\lambda)$ (\cref{def:qfireorasec}). For ease of reference, we write it out in full, with the notation introduced above.

\begin{mdframed}
 {\bf \underline{$\hyb_{0}$}}
 
    \begin{enumerate}
    \item The challenger samples $sn^*, \regis{flame} \samp \qkeyfiresign.\setup(1^\lambda)$.
    \item Parse $\fireora= (\ora_0^*, \ora_1^*, \ora_2^*, \ora^*_{\oss.\sign}, \ora^*_3, \ora^*_4) = sn^*$.
    \item Set $\advintstate{1}{0} = \regis{flame}$ and $\advetworeply{0} = \bot$.
    \item Set $\advintstate{2}{0} = \bot$.
    \item For $i \in [\numrounds]$:
    \begin{enumerate}[label=\arabic*.]
        \item $\roundadve{1}{i}^{\fireora}$ receives $\advintstate{1}{i-1}$ and $\advetworeply{i-1}$, and it outputs an internal state register $\advintstate{1}{i}$ and a leakage message $\leakmes{i}$.
        \item $\roundadve{2}{i}^{\fireora}$ receives $\advintstate{2}{i-1}$ and $\leakmes{i}$, and it outputs an internal state register $\advintstate{2}{i}$ and a message $\advetworeply{i}$.
    \end{enumerate}
    \item The challenger samples a random message $m^* \samp \messpa$.
    \item The adversary $\adve_{2,\finaladvsuffix}^{\fireora}$ receives $\advintstate{2}{\numrounds}, m^*$, and it outputs $sig^*$.
    \item Output $\ora^*_4(m^*, sig^*)$.
    \end{enumerate}

\end{mdframed}

\paragraph{\underline{$\hyb_{1}$:}} In this hybrid, we simply convert the challenge phase of the experiment into an additional round by using a special choice of adversaries (that is $\adve_{1,\hybindex+1},\adve_{2,\hybindex+1}$) for this new round.
\begin{mdframed}
 {\bf \underline{$\hyb_{1}$}}
 
    \begin{enumerate}
\item The challenger samples $sn^*, \regis{flame} \samp \qkeyfiresign.\setup(1^\lambda)$.
\item Parse $\fireora= (\ora_0^*, \ora_1^*, \ora_2^*, \ora^*_{\oss.\sign}, \ora^*_3, \ora^*_4) = sn^*$.
    \item Set $\advintstate{1}{0} = \regis{flame}$ and $\advetworeply{0} = \bot$.
    \item Set $\advintstate{2}{0} = \bot$.
    \item For $\textcolor{red}{i \in [\numrounds+1]}$:
    \begin{enumerate}[label=\arabic*.]
        \item $\roundadve{1}{i}^{\fireora}$ receives $\advintstate{1}{i-1}$ and $\advetworeply{i-1}$, and it outputs an internal state register $\advintstate{1}{i}$ and a leakage message $\leakmes{i}$.
        \item $\roundadve{2}{i}^{\fireora}$ receives $\advintstate{2}{i-1}$ and $\leakmes{i}$, and it outputs an internal state register $\advintstate{2}{i}$ and a message $\advetworeply{i}$.
    \end{enumerate}
    \item \textcolor{red}{Parse $(m^*, sig^*) = S^{\numrounds+1}$}.
    \item Output $\ora^*_4(m^*, sig^*)$.
    \end{enumerate}

\end{mdframed}
\paragraph{}

For the subhybrids $\hyb_{2,\hybindex,c}$ below, each index $\hybindex$ corresponds to a round of the leakage protocol between $\adve_1$ and $\adve_2$.

\paragraph{\underline{$\hyb_{2,\hybindex, 0}$ for $\hybindex \in [\numrounds + 2]$:}} In this hybrid, the first $\hybindex - 1$ rounds of the leakage protocol of $\adve_1, \adve_2$ are essentially collapsed into (i.e. simulated by) a single message between the two \emph{new} adversaries $\newadv{1}{\hybindex-1}, \newadv{2}{\hybindex-2}$  (where the new receiver adversary $\newadv{2}{\hybindex-2}$ does not make any oracle calls), and then rounds $\hybindex$ through $\numrounds+1$ are executed as before. The very first hybrid $\hyb_{2,1,0}$ will be equivalent to $\hyb_1$. We also label some parts of the experiment for ease of reference later on.
\begin{mdframed}
 {\bf \underline{$\hyb_{2,\hybindex, 0}$}}
 
    \begin{enumerate}
    \item The challenger samples $sn^*, \regis{flame} \samp \qkeyfiresign.\setup(1^\lambda)$.
    \item Parse $\fireora= (\ora_0^*, \ora_1^*, \ora_2^*, \ora^*_{\oss.\sign}, \ora^*_3, \ora^*_4) = sn^*$.
    \item[]\noindent\textbf{\textcolor{blue}{Round-collapsed Simulation for First $\hybindex-1$ Rounds:}}
    \item Run $\newadv{1}{\hybindex-1}^{\fireora}(\regis{flame})$ to get $\advintstate{1}{\hybindex-1}$ and $(\newleak{\hybindex-1}, S^{\hybindex-1})$.
    \item Run $\newadv{2}{\hybindex-1}(\newleak{\hybindex-1})$ to get $\advintstate{2}{\hybindex-1}$.
    \item[]\noindent\textbf{\textcolor{blue}{Original Protocol Continuation:}}
    \item For $i \in \{\hybindex,\dots, \numrounds+1\}$:
    \begin{enumerate}[label=\arabic*.]
        \item $\roundadve{1}{i}^{\fireora}$ receives $\advintstate{1}{i-1}$ and $\advetworeply{i-1}$, and it outputs an internal state register $\advintstate{1}{i}$ and a leakage message $\leakmes{i}$.
        \item $\roundadve{2}{i}^{\fireora}$ receives $\advintstate{2}{i-1}$ and $\leakmes{i}$, and it outputs an internal state register $\advintstate{2}{i}$ and a message $\advetworeply{i}$.
    \end{enumerate}
    \item {Parse $(m^*, sig^*) = S^{\numrounds+1}$}.
    \item Output $\ora^*_4(m^*, sig^*)$.
    \end{enumerate}

\end{mdframed}

\paragraph{\underline{$\hyb_{2,\hybindex, 1}$ for $\hybindex \in [\numrounds + 1]$:}} In this hybrid, we remove the oracle calls to $\ora_0^*, \ora_1^*,\ora_2^*$ from the receiver adversary during round $\hybindex$, by instead simulating it with a new leakage and a new receiver adversary $\estadv{2}{\hybindex}$.
\begin{mdframed}
 {\bf \underline{$\hyb_{2,\hybindex, 1}$}}
 
    \begin{enumerate}
    \item The challenger samples $sn^*, \regis{flame} \samp \qkeyfiresign.\setup(1^\lambda)$.
\item Parse $\fireora= (\ora_0^*, \ora_1^*, \ora_2^*, \ora^*_{\oss.\sign}, \ora^*_3, \ora^*_4) = sn^*$.
    \item[]\noindent\textbf{\textcolor{blue}{Round-collapsed Simulation for First $\hybindex-1$ Rounds:}}
    \item Run $\newadv{1}{\hybindex-1}^{\fireora}(\regis{flame})$ to get $\advintstate{1}{\hybindex-1}$ and $(\newleak{\hybindex-1}, S^{\hybindex-1})$.
    \item[]\noindent\textbf{Round $\hybindex$:}
    \item \textcolor{red}{ $\roundadve{1}{\hybindex}^{\fireora}$ receives $\advintstate{1}{\hybindex-1}$ and $\advetworeply{\hybindex-1}$, and it outputs an internal state register $\advintstate{1}{\hybindex}$ and a leakage message $\leakmes{\hybindex}$}
    \item \textcolor{red}{$\estadv{1}{\hybindex}^{\fireora}$ receives $\newleak{\hybindex-1},\leakmes{\hybindex}$ and produces a message $\qhelpstr{\hybindex}$}.
    \item \textcolor{red}{Run $\estadv{2}{\hybindex}^{(\ora_{\oss.\sign}^*, \ora_3^*, \ora_4^*)}(\qhelpstr{\hybindex})$ to obtain $(\advintstate{2}{\hybindex}, \advetworeply{\hybindex})$}
    \item[]\noindent\textbf{\textcolor{blue}{Original Protocol Continuation:}}
    \item For $i \in \{\textcolor{red}{\hybindex+1},\dots, \numrounds+1\}$:
    \begin{enumerate}[label=\arabic*.]
        \item $\roundadve{1}{i}^{\fireora}$ receives $\advintstate{1}{i-1}$ and $\advetworeply{i-1}$, and it outputs a leakage message $\leakmes{i}$ and an internal state register $\advintstate{1}{i}$.
        \item $\roundadve{2}{i}^{\fireora}$ receives $\advintstate{2}{i-1}$ and $\leakmes{i}$, and it outputs a message $\advetworeply{i}$ and an internal state register $\advintstate{2}{i}$.
    \end{enumerate}
    \item {Parse $(m^*, sig^*) = S^{\numrounds+1}$}.
    \item Output $\ora_4^*(m^*, sig^*)$.
    \end{enumerate}
\end{mdframed}

\paragraph{\underline{$\hyb_{2,\hybindex, 2}$ for $\hybindex \in [\numrounds+1]$:}} We further remove the calls to  $\ora_{\oss.\sign}^*$  made by the receiver adversary during this round, instead simulating it with a new leakage and new receiver adversary $\plastadv{2}{\hybindex}$.
\begin{mdframed}
 {\bf \underline{$\hyb_{2,\hybindex, 2}$}}
 
    \begin{enumerate}
    \item The challenger samples $sn^*, \regis{flame} \samp \qkeyfiresign.\setup(1^\lambda)$.
    \item Parse $\fireora= (\ora_0^*, \ora_1^*, \ora_2^*, \ora^*_{\oss.\sign}, \ora^*_3, \ora^*_4) = sn^*$.
    \item[]\noindent\textcolor{blue}{\textbf{Round-collapsed Simulation for First $\hybindex-1$ Rounds:}}
    \item Run $\newadv{1}{\hybindex-1}^{\fireora}(\regis{flame})$ to get $\advintstate{1}{\hybindex-1}$ and $(\newleak{\hybindex-1}, S^{\hybindex-1})$.
    \item[]\noindent\textbf{Round $\hybindex$:}
    \item { $\roundadve{1}{\hybindex}^{\fireora}$ receives $\advintstate{1}{\hybindex-1}$ and $\advetworeply{\hybindex-1}$, and it outputs an internal state register $\advintstate{1}{\hybindex}$ and a leakage message $\leakmes{\hybindex}$}
   
    \item \textcolor{red}{$\plastadv{1}{\hybindex}^{\fireora}$ receives $\newleak{\hybindex-1}, \leakmes{\hybindex}$ and produces a message $\plastadvout{\hybindex}$}.
    \item \textcolor{red}{Run $\plastadv{2}{\hybindex}^{\ora^*_3, \ora_4^*}(\plastadvout{\hybindex})$ to obtain $(\advintstate{2}{\hybindex}, \advetworeply{\hybindex})$}
    \item[]\noindent\textcolor{blue}{\textbf{Original Protocol Continuation:}}
    \item For $i \in \{\hybindex+1,\dots, \numrounds+1\}$:
    \begin{enumerate}[label=\arabic*.]
        \item $\roundadve{1}{i}^{\fireora}$ receives $\advintstate{1}{i-1}$ and $\advetworeply{i-1}$, and it outputs a leakage message $\leakmes{i}$ and an internal state register $\advintstate{1}{i}$.
        \item $\roundadve{2}{i}^{\fireora}$ receives $\advintstate{2}{i-1}$ and $\leakmes{i}$, and it outputs a message $\advetworeply{i}$ and an internal state register $\advintstate{2}{i}$.
    \end{enumerate}
    \item {Parse $(m^*, sig^*) = S^{\numrounds+1}$}.
    \item Output $\ora^*_4(m^*, sig^*)$.
    \end{enumerate}
\end{mdframed}

\paragraph{\underline{$\hyb_{2,\hybindex, 3}$ for $\hybindex \in [\numrounds+1]$:}} We further remove the calls to $\ora^*_3$ made by the receiver adversary during this round, instead simulating it with a new leakage and new receiver adversary $\lastadv{2}{\hybindex}$.
\begin{mdframed}
 {\bf \underline{$\hyb_{2,\hybindex, 3}$}}
 
    \begin{enumerate}
    \item The challenger samples $sn^*, \regis{flame} \samp \qkeyfiresign.\setup(1^\lambda)$.
\item Parse $\fireora= (\ora_0^*, \ora_1^*, \ora_2^*, \ora^*_{\oss.\sign}, \ora^*_3, \ora^*_4) = sn^*$.
    \item[]\noindent\textcolor{blue}{\textbf{Round-collapsed Simulation for First $\hybindex-1$ Rounds:}}
    \item Run $\newadv{1}{\hybindex-1}^{\fireora}(\regis{flame})$ to get $\advintstate{1}{\hybindex-1}$ and $(\newleak{\hybindex-1}, S^{\hybindex-1})$.
    \item[]\noindent\textbf{Round $\hybindex$:}
    \item { $\roundadve{1}{\hybindex}^{\fireora}$ receives $\advintstate{1}{\hybindex-1}$ and $\advetworeply{\hybindex-1}$, and it outputs an internal state register $\advintstate{1}{\hybindex}$ and a leakage message $\leakmes{\hybindex}$}
   
    \item \textcolor{red}{$\lastadv{1}{\hybindex}^{\fireora}$ receives $\newleak{\hybindex-1}, \leakmes{\hybindex}$ and produces a message $\lastadvout{\hybindex}$}.
    \item \textcolor{red}{Run $\lastadv{2}{\hybindex}^{\ora_4^*}(\lastadvout{\hybindex})$ to obtain $(\advintstate{2}{\hybindex}, \advetworeply{\hybindex})$}
    \item[]\noindent\textcolor{blue}{\textbf{Original Protocol Continuation:}}
    \item For $i \in \{\hybindex+1,\dots, \numrounds+1\}$:
    \begin{enumerate}[label=\arabic*.]
        \item $\roundadve{1}{i}^{\fireora}$ receives $\advintstate{1}{i-1}$ and $\advetworeply{i-1}$, and it outputs a leakage message $\leakmes{i}$ and an internal state register $\advintstate{1}{i}$.
        \item $\roundadve{2}{i}^{\fireora}$ receives $\advintstate{2}{i-1}$ and $\leakmes{i}$, and it outputs a message $\advetworeply{i}$ and an internal state register $\advintstate{2}{i}$.
    \end{enumerate}
    \item {Parse $(m^*, sig^*) = S^{\numrounds+1}$}.
    \item Output $\ora^*_4(m^*, sig^*)$.
    \end{enumerate}
\end{mdframed}

\paragraph{\underline{$\hyb_{2,\hybindex, 4}$ for $\hybindex \in [\numrounds+1]$:}} We further remove the calls to $\ora_4^*$ made by the receiver adversary during this round, and instead we simulate it with a new leakage.
\begin{mdframed}
 {\bf \underline{$\hyb_{2,\hybindex, 4}$}}
 
    \begin{enumerate}
    \item The challenger samples $sn^*, \regis{flame} \samp \qkeyfiresign.\setup(1^\lambda)$.
\item Parse $\fireora= (\ora_0^*, \ora_1^*, \ora_2^*, \ora^*_{\oss.\sign}, \ora^*_3, \ora^*_4) = sn^*$.
    \item[]\noindent\textcolor{blue}{\textbf{Round-collapsed Simulation for First $\hybindex-1$ Rounds:}}
    \item Run $\newadv{1}{\hybindex-1}^{\fireora}(\regis{flame})$ to get $\advintstate{1}{\hybindex-1}$ and $(\newleak{\hybindex-1}, S^{\hybindex-1})$.
    \item[]\noindent\textbf{Round $\hybindex$:}
    \item { $\roundadve{1}{\hybindex}^{\fireora}$ receives $\advintstate{1}{\hybindex-1}$ and $\advetworeply{\hybindex-1}$, and it outputs an internal state register $\advintstate{1}{\hybindex}$ and a leakage message $\leakmes{\hybindex}$}
    \item {$\lastadv{1}{\hybindex}^{\fireora}$ receives $\newleak{\hybindex-1}, \leakmes{\hybindex}$ and produces a message $\lastadvout{\hybindex}$}.
                \item \textcolor{red}{Set $\helplistnew{4}{\hybindex}$ to be the empty set.}
        \item \textcolor{red}{For $j \in [512^2\cdot(t(\lambda))^6\cdot \lambda^8 \cdot (u(\lambda))^4 \cdot (q^*(\lambda))^4 \cdot (\ell)^2]$:}
        \begin{enumerate}[label=\arabic*.]
            \item \textcolor{red}{Sample a random index $ind$ between $1$ and the number of oracle queries made by $\lastadv{2}{\hybindex}$.}
            \item \textcolor{red}{Simulate $\lastadv{2}{\hybindex}^{\ora^*_4}(\lastadvout{\hybindex})$ until right before its $ind$-th oracle query is executed and measure the query input register instead of executing the query. Let $m, y$ be the measurement outcome.}
            \item \textcolor{red}{Check if $\ora^*_4(m,y)$ is $1$. If so, add $(m,y)$ to $\helplistnew{4}{\hybindex}$.}
        \end{enumerate}
        \item \textcolor{red}{{Simulate $\lastadv{2}{\hybindex}(\lastadvout{\hybindex})$ with the following oracle $\ora'$: On input $m,y$, output $1$ if $(m,y)\in \helplistnosub$, otherwise output $0$.  Let $(\advintstate{2}{\hybindex}, \advetworeply{\hybindex})$ be the output.}}
    \item[]\noindent\textcolor{blue}{\textbf{Original Protocol Continuation:}}
    \item For $i \in \{\hybindex+1,\dots, \numrounds+1\}$:
    \begin{enumerate}[label=\arabic*.]
        \item $\roundadve{1}{i}^{\fireora}$ receives $\advintstate{1}{i-1}$ and $\advetworeply{i-1}$, and it outputs a leakage message $\leakmes{i}$ and an internal state register $\advintstate{1}{i}$.
        \item $\roundadve{2}{i}^{\fireora}$ receives $\advintstate{2}{i-1}$ and $\leakmes{i}$, and it outputs a message $\advetworeply{i}$ and an internal state register $\advintstate{2}{i}$.
    \end{enumerate}
    \item {Parse $(m^*, sig^*) = S^{\numrounds+1}$}.
    \item Output $\ora^*_4(m^*, sig^*)$.
    \end{enumerate}
\end{mdframed}
\paragraph{\underline{$\hyb_{3}$:}} We define this hybrid to be the same as $\hyb_{2,\numrounds+2, 0}$.
\begin{mdframed}
 {\bf \underline{$\hyb_{3}$}}
 
    \begin{enumerate}
    \item The challenger samples $sn^*, \regis{flame} \samp \qkeyfiresign.\setup(1^\lambda)$.
    \item Parse $\fireora= (\ora_0^*, \ora_1^*, \ora_2^*, \ora^*_{\oss.\sign}, \ora^*_3, \ora^*_4) = sn^*$.
    \item Run $\newadv{1}{\numrounds+1}^{\fireora}(\regis{flame})$ to get $\advintstate{1}{\hybindex-1}$ and $(\newleak{\numrounds+1}, S^{\numrounds+1})$.
    \item Run $\newadv{2}{\numrounds+1}(\newleak{\numrounds+1})$ to get $\advintstate{2}{\numrounds+1}$.
   
    \item {Parse $(m^*, sig^*) = S^{\numrounds+1}$}.
    \item Output $\ora^*_4(m^*, sig^*)$.
    \end{enumerate}

\end{mdframed}

\subsection{Hybrid Indistinguishability (Proof of \cref{thm:hybridindist})}\label{sec:hybrid-indist}
In this section, we prove the indistinguishability of the first and the last hybrid. That is, we prove \cref{thm:hybridindist}.

First we give the hybrid indistinguishability lemmata.
\begin{lemma}
    $\hyb_0 \equiv \hyb_1$.
\end{lemma}
\begin{proof}
    This is straightforward by the definition of $\adve_{1,\numrounds+1},\adve_{2,\numrounds+1}$. 
\end{proof}

\begin{lemma}
    $\hyb_1 \equiv \hyb_{2,1,0}$.
\end{lemma}
\begin{proof}
    This is straightforward by the definition of $\newadv{1}{0}, \newadv{2}{0}$.
\end{proof}

\begin{lemma}\label{lem:bighybrid}
   For all $\hybindex \in [\numrounds+1]$, $\statdist{\hyb_{2,\hybindex,0}}{\hyb_{2,\hybindex,1}} \leq \frac{1}{\lambda\cdot \numrounds\cdot \origwinpoly(\lambda)}$.
\end{lemma}
We prove this lemma in \cref{sec:bighybrid}.

\begin{lemma}\label{lem:twotwonew}
   For all $\hybindex \in [\numrounds+1]$, $\statdist{\hyb_{2,\hybindex,1}}{\hyb_{2,\hybindex,2}} \leq \frac{1}{\lambda\cdot \numrounds\cdot \origwinpoly(\lambda)}$.
\end{lemma}
We prove this lemma in \cref{sec:twotwonew}.

\begin{lemma}\label{lem:hyb2l2to2l3}
   For all $\hybindex \in [\numrounds+1]$, $\statdist{\hyb_{2,\hybindex,2}}{\hyb_{2,\hybindex,3}} \leq \frac{1}{\lambda\cdot \numrounds\cdot \origwinpoly(\lambda)}$.
\end{lemma}
We prove this lemma in \cref{sec:hyb2l2to2l3}.

\begin{lemma}\label{lem:twoltwotwolthree}
   For all $\hybindex \in [\numrounds+1]$, $\statdist{\hyb_{2,\hybindex,3}}{\hyb_{2,\hybindex,4}} \leq \frac{1}{\lambda\cdot \numrounds\cdot \origwinpoly(\lambda)}$.
\end{lemma}
We prove this lemma in \cref{sec:hybtwo23}.

\begin{lemma}
   For all $\hybindex \in [\numrounds+1]$, ${\hyb_{2,\hybindex,4}}\equiv{\hyb_{2,\hybindex+1,0}}$.
\end{lemma}
\begin{proof}
    Observe that, for the internal state, $\newadv{2}{\hybindex+1}$ produces the exact same mixed state as the one produced by $\lastadv{2}{\hybindex}$ in ${\hyb_{2,\hybindex,4}}$ since it is conditioning on the same output message.
\end{proof}

\begin{lemma}
    $\hyb_{2,\numrounds+2,0} \equiv \hyb_3$.
\end{lemma}
\begin{proof}
    By definition.
\end{proof}

Finally, combining these lemmas gives us
\begin{equation*}
    \statdist{\hyb_0}{\hyb_{3}} \leq \frac{1}{4\cdot \origwinpoly(\lambda)}
\end{equation*}
which proves \cref{thm:hybridindist}.
\subsubsection{Proof of \cref{lem:twotwonew}}\label{sec:twotwonew}
In this section, we prove \cref{lem:twotwonew}. That is, we prove that for all $\hybindex \in [\numrounds+1]$, $\statdist{\hyb_{2,\hybindex,1}}{\hyb_{2,\hybindex,2}} \leq \frac{1}{\lambda\cdot \numrounds\cdot \origwinpoly(\lambda)}$. Throughout the rest of the proof, fix any $\hybindex \in [\numrounds+1]$. 

Note that the difference between the hybrids $\hyb_{2,\hybindex,1},\hyb_{2,\hybindex,2}$ is that in ${\hyb_{2,\hybindex,1}}$ we run $\estadv{2}{\ell}$ with the actual oracle $\ora_{\oss.\sign}^*$ whereas in $\hyb_{2,\hybindex,2}$, we use $\plastadv{2}{\ell}$ which simulates $\estadv{2}{\ell}$ with an estimated (by $\plastadv{\ell}{1}$) oracle. We will use the incompressibility property of  $\oss$ (\cref{def:ossincomprnew}) to show that this only makes a difference of $\frac{1}{\lambda\cdot \numrounds\cdot \origwinpoly(\lambda)}$ statistical difference. In this proof whenever we refer to incompressibility we will be referring to $\ora_{\oss.\sign}$ incompressibility of $\oss$. We make two notes about this hybrid. First, note that this hybrid is only needed for the LOCC leakage-resilience security: it can be removed in case of non-interactive security. Second, this hybrid indistinguishability can be proven directly using strong $\ora_{\oss.\sign}$ incompressibility of any $\oss$ (the proof would be essentially the same as \cref{sec:bighyblem3}), but here we use regular incompressibility and the properties of our actual $\oss$ scheme, rather than strong incompressibility.

Let $\gamma_{vk, z}^{\fireora, \qhelpstr{\hybindex}}$ be the total query weight of $\estadv{2}{\hybindex}(\qhelpstr{\hybindex})$ to $\ora^*_{\oss.\sign}$ in  ${\hyb_{2,\hybindex,1}}$ on the input $vk, z$. 
We also define the following.
\begin{align*}
    &\alpha_{vk,z}^{\fireora, \qhelpstr{\hybindex}} = \begin{cases}
        &\gamma_{vk, z}^{\fireora, \qhelpstr{\hybindex}},\text{ if } \ora^*_{\oss.\sign}(vk, z) \neq 0 \\
        &0, \text{ if } \ora^*_{\oss.\sign}(vk,z) = 0
    \end{cases}
\end{align*}

We also define $W^{\fireora, \qhelpstr{\hybindex},  \helplistone{sign}}$. Here,  let $\helplistone{sign}$ denote the values sampled during the execution of $\plastadv{1}{\ell}$, and let $L_{sign}$ denote its projection onto its elements' first components (i.e. $L_{sign} = \{vk: (vk, z) \in  \helplistone{sign}\}$). Similarly, for any $vk$, let $S(vk)$ denote $\mathsf{Span}(\{v: (vk, v) \in \helplistone{sign}\}) \setminus \{0\}$.
\begin{align*}
    &W^{\fireora, \qhelpstr{\hybindex},  \helplistone{sign}} := \sum_{\substack{vk, z: \\ z \not\in S(vk) }} \alpha^{\fireora, \qhelpstr{\hybindex}}_{vk,z}
\end{align*}

We now claim the following.

\begin{claim}\label{claim:hybnew}
For all sufficiently large $\lambda > 0$,
\begin{align*}
&\E_{\fireora, \qhelpstr{\hybindex},  \helplistone{sign}}[W^{\fireora, \qhelpstr{\hybindex},  \helplistone{sign}}]  \leq\frac{1}{128\cdot \numfourqueries\cdot(\lambda\cdot {\numrounds}\cdot \origwinpoly(\lambda))^2}.
 \end{align*}
\end{claim}
\begin{proof}
We will prove this using the incompressibility property of $\oss$: Consider the following adversaries $\adhocadv_1$ and $\adhocadv_2$ for it.  $\adhocadv_1$ simulates the hybrid $\hyb_{2,\hybindex,1}$ until $\estadv{1}{\hybindex}$ produces its output $\qhelpstr{\hybindex}$, and then outputs $\qhelpstr{\hybindex}$. Then, $\adhocadv_2$ receives this value and runs  $\estadv{2}{\hybindex}$ with it. When considering $\adhocadv_1, \adhocadv_2$ for incompressibility of $\oss$, we assume that $\adhocadv_1, \adhocadv_2$ are only oracle-adversaries for $\oss$, and the other oracles are just simulated, e.g. they sample $H_0$ in advance and both use this sample to simulate the oracle outputs of $\ora_0^*$. Observe that both $\adhocadv_1$ and $\adhocadv_2$ can perform their simulations with their respective oracles (e.g. $\adhocadv_2$ only queries $\ora_{\sign}^*, \ora_{\ver}^*$ and does not query $\ora_{\oss.\genqkey}^*$).

Let $\mathfrak{L}$ be the incompressibility list for the signing oracle as in \cref{def:ossincomprnew}, and we know $|\mathfrak{L}| = \promisedlistsize{OSS}{\hybindex} = 32768\cdot \lambda^5\cdot (\hybindex \cdot \numfourqueries)^2 \cdot(t(\lambda)\cdot {\numrounds}\cdot \origwinpoly(\lambda))^2$ since $\adhocadv_1$ makes at most $\promisedlistsize{OSS}{\hybindex}/2$ queries to $\ora_{\oss.\genqkey}^*$. The list depends on the oracle itself and the \emph{leakage} $\qhelpstr{\hybindex}$, but we will not write this dependence explicitly for conciseness. 

\begin{align*}
   &W_1^{\fireora, \qhelpstr{\hybindex},  \helplistone{sign}, \mathfrak{L} } := \sum_{\substack{vk, z: \\ z \not\in S(vk) \\ vk \not\in \mathfrak{L} }} \alpha^{\fireora, \qhelpstr{\hybindex}}_{vk,z}\\
    &W_2^{\fireora, \qhelpstr{\hybindex},  \helplistone{sign}, \mathfrak{L} } := \sum_{\substack{vk, z: \\ z \not\in S(vk) \\ vk \in \mathfrak{L} }} \alpha^{\fireora, \qhelpstr{\hybindex}}_{vk,z}
    \end{align*}

    Now we claim  \begin{equation*}
        \E_{\fireora, \qhelpstr{\hybindex},  \helplistone{sign},\mathfrak{L}}[W_1^{\fireora, \qhelpstr{\hybindex},  \helplistone{sign}, \mathfrak{L} }] \leq \negl(\lambda).
    \end{equation*}

To see this, consider the value 
\begin{equation*}
   A = \sum_{\substack{vk, z: \\ vk \not\in \mathfrak{L} }} \alpha^{\fireora, \qhelpstr{\hybindex}}_{vk,z}
\end{equation*}
Observe that with probability $1$, this is greater than or equal to $W_1^{\fireora, \qhelpstr{\hybindex},  \helplistone{sign}, \mathfrak{L} }$. Then, observe that $\E[A] \leq \negl(\lambda)$ by the incompressibility property of $\oss$.

    Now we claim that for all sufficiently large $\lambda > 0$ \begin{equation*}
  \E_{\fireora, \qhelpstr{\hybindex},  \helplistone{sign},\mathfrak{L}}[W_2^{\fireora, \qhelpstr{\hybindex},  \helplistone{sign}, \mathfrak{L} }]\leq\frac{1}{256\cdot\numfourqueries \cdot(\lambda\cdot {\numrounds}\cdot \origwinpoly(\lambda))^2}.
    \end{equation*}

    To prove this, we first define for $i \in [\promisedlistsize{OSS}{\hybindex}]$,
\begin{align*}
    Q_i^{\fireora, \qhelpstr{\hybindex},  \helplistone{sign}, \mathfrak{L} } := \sum_{\substack{vk, z: \\ z \not\in S(vk) \\ vk = \mathsf{Item}_i(\mathfrak{L}) }} \alpha^{\fireora, \qhelpstr{\hybindex}}_{vk,z}
\end{align*}
We claim $\E[Q_i] \leq \frac{1}{256\cdot \promisedlistsize{OSS}{\hybindex} \cdot\numfourqueries \cdot(\lambda\cdot {\numrounds}\cdot \origwinpoly(\lambda))^2}$ for all $i \in  [\promisedlistsize{OSS}{\hybindex}]$. Observe that with probability $1$ we have $W = W_1 + \sum_{i\in[\promisedlistsize{OSS}{\hybindex} ]}Q_i$, thus, proving this claim will complete our proof.

Now fix some $i^* \in [\promisedlistsize{OSS}{\hybindex}]$. Fix any value $\fireora, \qhelpstr{\hybindex}$. We will prove the above for any such fixed values. Note that fixing these also fixes $\mathfrak{L}$ and thus $\mathsf{Item}_{i^*}(\mathfrak{L})$ - denote this as $vk^*$. There are  $\numestbyind{OSS}{\ell} = \lambda^8\cdot \hybindex^2 \cdot (\numfourqueries)^4 \cdot (n(\lambda))^2 \cdot(t(\lambda)\cdot {\numrounds}\cdot \origwinpoly(\lambda))^2\cdot (\lambda\cdot {\numrounds}\cdot \origwinpoly(\lambda))^2 $  iterations during the run of $\plastadv{\ell}{1}$, and we will consider these iterations in $n(\lambda) + 1$ \emph{groups}: we have $\frac{\numestbyind{OSS}{\ell}}{n(\lambda)+1}$ iterations in each group. Now let $\mathsf{GOOD}$ denote the event that $Q_{i^*} \leq\frac{1}{256\cdot\lambda\cdot \promisedlistsize{OSS}{\hybindex} \cdot\numfourqueries \cdot(\lambda\cdot {\numrounds}\cdot \origwinpoly(\lambda))^2}$. We will prove that $\Pr[\mathsf{GOOD}] \geq 1 - 2\exp{-\lambda}$. For each $j \in [n+1]$, define the event $E_j$ to be the event that during the $j$-th group, a new element is added to $\helplistone{sign}$, and define $E'_j$ to be the event that $\sum_{\substack{vk, z: \\ z \not\in S_j(vk^*) }} \alpha^{\fireora, \qhelpstr{\hybindex}}_{vk^*,z} > \frac{1}{256\cdot\lambda\cdot \promisedlistsize{OSS}{\hybindex} \cdot\numfourqueries \cdot(\lambda\cdot {\numrounds}\cdot \origwinpoly(\lambda))^2}$ where $S_j$ denotes $\mathsf{Span}(\{v: (vk, v) \in \helplistone{sign}\}) \setminus \{0\}$ for the value of $\helplistone{sign}$ at the beginning of the $j$-th group. It is easy to see that $\left(\vee_{j = 1}^{n+1} \overline{E'_j}\right) \implies \mathsf{GOOD}$. Similarly, it is easy to see that $E_1 \wedge \dots \wedge E_n \implies \overline{E'_{n+1}}$, since the subspace $\mathsf{Colspan}(A_{vk^*})^\perp$ is of dimension at most $n$, so, once $n$ linearly independent elements are added, there cannot be any vectors outside $\mathsf{Colspan}(A_{vk^*})^\perp$. Note that combining these two, we also get $E_1 \wedge \dots \wedge E_n \implies \mathsf{GOOD}$. Finally, we claim that $\Pr[E_j | E'_j \wedge E] \geq 1-\exp{-\lambda\cdot 2n}$ for all $j \in [n]$ and any event $E$ such that $\Pr[E] > 0$. For now, we will assume this and complete the proof, and later we will prove it. First, observe that $\Pr[\bigwedge_{j=1}^{n} E_j | \bigwedge_{j=1}^{n} E_j'] \geq (1 - \exp{-\lambda\cdot 2n})^n \geq  1 - \exp{-\lambda}$ by above. This also means $\Pr[\mathsf{GOOD} | \bigwedge_{j=1}^{n} E_j'] \geq 1 - \exp{-\lambda}$. Also recall that $\overline{\bigwedge_{j=1}^{n} E_j'} \implies \mathsf{GOOD}$. Combining these two gives us $\Pr[\mathsf{GOOD}] \geq 1 - 2\exp{-\lambda}$. 

Finally, we prove $\Pr[E_j | E'_j \wedge E] \geq 1-\exp{-\lambda\cdot 2n}$ for all $j \in [n]$ and any event $E$ such that $\Pr[E] > 0$. To see this, observe that during the $j$-th group, we measure the query of the adversary for $\frac{\numestbyind{OSS}{\ell}}{n(\lambda)+1}$ times. Since the weight on \emph{good elements} (i.e. elements that would be added to our list and thus cause $E_j$ to occur) is at least $\frac{1}{256\cdot\lambda\cdot \promisedlistsize{OSS}{\hybindex} \cdot\numfourqueries \cdot(\lambda\cdot {\numrounds}\cdot \origwinpoly(\lambda))^2}$ when conditioned on $E'_j$, we get the desired result since $\numestbyind{OSS}{\ell} \geq 1024 \cdot (n^2(\lambda)\cdot t(\lambda)) \cdot\lambda^2\cdot \promisedlistsize{OSS}{\hybindex} \cdot\numfourqueries \cdot(\lambda\cdot {\numrounds}\cdot \origwinpoly(\lambda))^2$.
\end{proof}

Now we claim that the lemma follows from \cref{claim:hybnew}. To see this, observe that the two hybrids only differ on the oracles they use for $\estadv{\ell}{2}$: $\hyb_{2,\hybindex,1}$ uses $\ora_{\oss.\sign}^*$ whereas $\hyb_{2,\hybindex,2}$ uses the estimated oracle created by $\plastadv{\ell}{1}$. Observe that these oracles can differ only on inputs $vk, z$ such that $\ora_{\oss.\sign}^*(vk,z) \neq 0$ (i.e. $z \in \mathsf{Colspan}(A_{vk})^\perp \setminus \{0\}$) but $z \not\in S(vk)$. Now observe that $W^{\fireora, \qhelpstr{\hybindex},  \helplistone{sign}}$ is an upper bound on the total weight on the inputs on which the oracles might differ. Thus, the lemma follows by the Jensen's inequality, the fact that $\estadv{2}{\ell}$ makes at most $\numfourqueries$ queries to $\ora_{\oss.\sign}^*$ and \cref{prethm:bbbv97}.
\subsubsection{Proof of \cref{lem:hyb2l2to2l3}}\label{sec:hyb2l2to2l3}
 In this section, we prove \cref{lem:hyb2l2to2l3}. That is, we prove that for all $\hybindex \in [\numrounds+1]$, $\statdist{\hyb_{2,\hybindex,2}}{\hyb_{2,\hybindex,3}} \leq \frac{1}{\lambda\cdot \numrounds\cdot \origwinpoly(\lambda)}$. Throughout the rest of the proof, fix any $\hybindex \in [\numrounds+1]$.

Observe that the difference between the two hybrids is that in $\hyb_{2,\hybindex,2}$ we run $\plastadv{2}{\hybindex}$ directly, whereas in $\hyb_{2,\hybindex,3}$ we instead run $\lastadv{2}{\hybindex}$  which simulates $\plastadv{2}{\hybindex}$ while using an estimated oracle for its queries to $\ora^*_3$. We will use the incompressibility property of $H_0, H_1$ to show that this only makes a difference of $\frac{1}{\lambda\cdot \numrounds\cdot \origwinpoly(\lambda)}$ in statistical distance. 

Let $\gamma_{ivk,y_0,y_1,z}^{\fireora, \plastadvout{\hybindex}}$ denote the total query weight of $\plastadv{2}{\hybindex}(\plastadvout{\hybindex})$ to the oracle $\ora_2^*$ on input $ivk,y_0,y_1,z$. We also define the following.

\begin{align*}
    &\alpha_{ivk,z}^{\fireora, \plastadvout{\hybindex}} = \sum_{y_0,y_1:  \ora^*_2(ivk, y_0, y_1, z) \neq \bot} \gamma_{ivk,y_0,y_1,z}^{\fireora, \plastadvout{\hybindex}}
\end{align*}

Now let $(\helplist{0}{\hybindex'}{j'})_{\hybindex' \in [\hybindex], j' \in [\numqueries]}$, $(\helplist{1}{\hybindex'}{j'})_{\hybindex' \in [\hybindex], j' \in [\numqueries]}$, $(\dbadv{0}{\hybindex'}{j'})_{\hybindex' \in [\hybindex^*], j' \in [i^*]}$, $(\dbadv{1}{\hybindex'}{j'})_{\hybindex' \in [\hybindex^*], j' \in [i^*]}$ be the estimated lists and databases in $\lastadvout{\hybindex}$ (or equivalently, $\plastadvout{\hybindex}$) when it is fully unwrapped and parsed. Let $\helplistone{3}, \dbadvone{3}$ be the estimated list and database inside $\lastadvout{\ell}$ that is sampled by $\lastadv{1}{\hybindex}$. With abuse of notation, sometimes we will refer to the projection of the sets $\helplist{\cdot}{\hybindex'}{j'}$ onto their elements' (which are tuples) subcomponents also as  $\helplist{\cdot}{\hybindex'}{j'}$ (e.g. we will write $\helplist{2}{\hybindex'}{j'}$ to mean the set of pairs $(ivk, z)$ obtained by projecting each tuple in the actual set $\helplist{2}{\hybindex'}{j'}$ to its first and fourth elements) - however, the meaning will clear from context. Now we claim the following.

\begin{claim}\label{claim:lastnewtrivnew}
\begin{equation*}
    \Pr\left[\left(\bigcup_{\hybindex' \in [\hybindex], j' \in [\numqueries]} \helplist{0}{\hybindex'}{j'} \right) \bigcap \left(\bigcup_{\hybindex' \in [\hybindex], j' \in [\numqueries]} \helplist{1}{\hybindex'}{j'} \right)  = \emptyset \right] \geq 1 - \negl(\lambda).
\end{equation*}
\end{claim}
\begin{proof}
    Observe that if this intersection is not empty, then there is $ivk$ such that $ivk \in  \helplist{0}{\hybindex'}{j'}$ and $ivk \in  \helplist{1}{\hybindex''}{j''}$ for some $\hybindex', \hybindex'', j', j''$. However, by construction we know that this implies that we can obtain $isig_0, isig_1$ from $\lastadvout{\hybindex}$ such that $\ora_{\oss.\ver^*}(ivk, 0, isig_0)=\ora_{\oss.\ver^*}(ivk, 1, isig_1) = 1$. Since $\lastadvout{\hybindex}$ is obtained by making polynomially many queries to the oracles of $\oss$, the claim follows by the one-shot security of $\oss$.
\end{proof}

Now we define the following (some dependencies are implicit to keep the notation concise).
\begin{align*}
    &W_0^{\fireora, \plastadvout{\hybindex}, \helplistone{3}} = \sum_{\substack{ivk,z:\\ ivk \not\in\left(\bigcup_{\hybindex' \in [\hybindex], j' \in [\numqueries]} \helplist{0}{\hybindex'}{j'} \right) \\
    (ivk,z) \not\in \helplistone{3}}} \alpha_{ivk,z}\\
    &W_1^{\fireora, \plastadvout{\hybindex}, \helplistone{3}} = \sum_{\substack{ivk,z:\\ ivk \not\in\left(\bigcup_{\hybindex' \in [\hybindex], j' \in [\numqueries]} \helplist{1}{\hybindex'}{j'} \right) \\
    (ivk,z) \not\in \helplistone{3}}} \alpha_{ivk,z}\\
    &W_2^{\fireora, \plastadvout{\hybindex}, \helplistone{3}} = \sum_{\substack{ivk,z:\\ ivk \not\in\left(\bigcup_{\hybindex' \in [\hybindex], j' \in [\numqueries]} \helplist{0}{\hybindex'}{j'} \right) \bigcap \left(\bigcup_{\hybindex' \in [\hybindex], j' \in [\numqueries]} \helplist{1}{\hybindex'}{j'} \right) }} \alpha_{ivk,z}\\
\end{align*}

Now we make the following claims.
\begin{claim}
    \begin{equation}
        \E[W_2^{\fireora, \plastadvout{\hybindex}, \helplistone{3}}]\leq \negl(\lambda).
    \end{equation}
\end{claim}
\begin{proof}
    This simply follows from the definition of $W_2$ and \cref{claim:lastnewtrivnew}.
\end{proof}
\begin{claim}
For all sufficiently large $\lambda > 0$.
\begin{align*}
    &\E[W_0^{\fireora, \plastadvout{\hybindex}, \helplistone{3}}] \leq \frac{1}{128\cdot\lambda^4\cdot (\numrounds\cdot \origwinpoly(\lambda)\cdot\numqueries)^2} \\
    &\E[W_1^{\fireora, \plastadvout{\hybindex}, \helplistone{3}}] \leq \frac{1}{128\cdot\lambda^4\cdot (\numrounds\cdot \origwinpoly(\lambda)\cdot\numqueries)^2}
\end{align*}
\end{claim}
\begin{proof}
We will only prove the first one (using incompressibility of the random oracle $H_0$), and the second bound follows by the same argument (but this time relying on the incompressibility of $H_1$).

Now we will use the incompressibility property (\cref{thm:randoraincomp}) of the random oracle $H_0$ with the predicate $P$ defined as $P(ivk || z) = 1$ if $ivk \in S$, where $S = \left(\bigcup_{\hybindex' \in [\hybindex], j' \in [\numqueries]} \helplist{0}{\hybindex'}{j'} \right)$. Note that this set $S$ is exactly the set of prefixes on which our adversary has \emph{section queries} (otherwise it only makes normal queries and verification queries). In more detail, let $\adhocadv_{1}$ be the adversary that simulates $\hyb_{2, \ell, 2}$ until the point where $\plastadv{1}{\ell}$ outputs its message, and it also outputs $S$ as its predicate. Then $\mathcal{P}_2$ simulates $\plastadv{2}{\hybindex}$. We invoke \cref{thm:randoraincomp} for this pair of adversaries, and let $\mathfrak{L}$ denote the incompressibility list that is promised to exist, and we know $|\mathfrak{L}| = \ell\cdot t(\lambda)$ since $\adhocadv_1$ makes at most $\ell\cdot t(\lambda)$ normal queries. We then split $W_0$ into two summands $U_0, U_1$:

\begin{align*}
    &U_0^{\fireora, \plastadvout{\hybindex}, \helplistone{3}} = \sum_{\substack{ivk,z:\\ ivk \not\in\left(\bigcup_{\hybindex' \in [\hybindex], j' \in [\numqueries]} \helplist{0}{\hybindex'}{j'} \right) \\
    (ivk,z) \not\in \helplistone{3} \\ (ivk,z)\not\in \mathfrak{L}  }}\alpha_{ivk,isig}\\
   &U_1^{\fireora, \plastadvout{\hybindex}, \helplistone{3}} = \sum_{\substack{ivk,z:\\ ivk \not\in\left(\bigcup_{\hybindex' \in [\hybindex], j' \in [\numqueries]} \helplist{0}{\hybindex'}{j'} \right) \\
    (ivk,z) \not\in \helplistone{3} \\ (ivk,z)\in \mathfrak{L}  }}\alpha_{ivk,isig}
\end{align*}

By the incompressibility lemma, we get $\E[U_0] \leq \negl(\lambda)$. Now, we will focus on bounding $U_1$.

    First we claim that, for all $\fireora,\plastadvout{\hybindex}$, with overwhelming probability we have that any $ivk,z$ that satisfies 
    \begin{equation*}
        \alpha_{ivk,z} \geq \frac{1}{512\cdot (t(\lambda)\cdot \ell)\cdot\lambda^4\cdot (\numrounds\cdot \origwinpoly(\lambda)\cdot\numqueries)^2}
    \end{equation*}
    is contained in  $\helplistone{3}$. To see this, observe that  $\helplistone{3}$  is created by measuring a random query of $\plastadv{\hybindex}{2}$ to $\ora_3$  for $512\cdot\lambda\cdot (t^2(\lambda)\cdot \ell)\cdot\lambda^4\cdot (\numrounds\cdot \origwinpoly(\lambda)\cdot\numqueries)^2$ times. Thus, for any $ivk, z$ such that $\alpha_{ivk,z}$ satisfies the above condition, the probability that $(ivk,z)\not\in\helplistone{3}$ is at most $\exp{-\lambda}$.  Now define $\mathsf{GOOD}$ to be the event that all $(ivk,z)\in\mathfrak{L}$ that satisfies \begin{equation*}
      \alpha_{ivk,z} \geq \frac{1}{512\cdot (t(\lambda)\cdot \ell)\cdot\lambda^4\cdot (\numrounds\cdot \origwinpoly(\lambda)\cdot\numqueries)^2}
    \end{equation*}
    are included in $\helplistone{3}$. Since this is true with probability $1 - \exp\{-\lambda\}$ for each $ivk,z \in \mathfrak{L}$ individually, and since $|\mathfrak{L}|$ is polynomial, we have that $\Pr[\mathsf{GOOD}] \geq 1 - \negl(\lambda)$.

Now we claim \begin{equation*}
     \E[U_1 \big| \mathsf{GOOD}] \leq \frac{1}{512\cdot\lambda^4\cdot (\numrounds\cdot \origwinpoly(\lambda)\cdot\numqueries)^2}
\end{equation*}
To see this, first by linearity of expectation we have
\begin{align*}
   &\E[U_1 \big| \mathsf{GOOD}]\\ &=\sum_{\substack{ivk,z:\\ ivk \not\in\left(\bigcup_{\hybindex' \in [\hybindex], j' \in [\numqueries]} \helplist{0}{\hybindex'}{j'} \right) \\
    (ivk,z) \not\in \helplistone{3} \\ (ivk,z)\in \mathfrak{L}  }}\E[\alpha_{ivk,z} | \mathsf{GOOD}]
\end{align*}
However, by definition of $\mathsf{GOOD}$, when $\mathsf{GOOD}$ occurs, we have
\begin{equation*}
    ivk,z \in \mathfrak{L} \setminus \helplistone{3} \implies  \alpha_{ivk,z}\leq  \frac{1}{512\cdot (t(\lambda)\cdot \ell)\cdot\lambda^4\cdot (\numrounds\cdot \origwinpoly(\lambda)\cdot\numqueries)^2}
\end{equation*}
Since $|\mathfrak{L}| =  \ell\cdot t(\lambda)$, we obtain the desired upper bound  on expectation conditioned on $\mathsf{GOOD}$. Since $\Pr[\mathsf{GOOD}] \geq 1 - \negl(\lambda)$, we also get \begin{equation*}
    \E[U_1] \leq \frac{1}{256\cdot\lambda^4\cdot (\numrounds\cdot \origwinpoly(\lambda)\cdot\numqueries)^2}
\end{equation*}
as desired.
\end{proof}

Observe that \cref{lem:bighyblem2} follows from the above claims. To see this, observe that $W_0+W_1+W_2$ is an upper bound on the total weight on which the two different oracles in the hybrids might differ. Thus the result follows by the linearity of expectation, \cref{prethm:bbbv97} and Jensen's inequality.
\subsubsection{Proof of \cref{lem:twoltwotwolthree}}\label{sec:hybtwo23}
In this section, we prove \cref{lem:twoltwotwolthree}. That is, we prove that for all $\hybindex \in [\numrounds+1]$, $\statdist{\hyb_{2,\hybindex,3}}{\hyb_{2,\hybindex,4}} \leq \frac{1}{\lambda\cdot \numrounds\cdot \origwinpoly(\lambda)}$. Throughout the rest of the proof, fix any $\hybindex \in [\numrounds+1]$.

First, note that the difference between the hybrids $\hyb_{2,\hybindex,3},\hyb_{2,\hybindex,4}$ is that in ${\hyb_{2,\hybindex,3}}$ we run $\lastadv{2}{\ell}$ with the actual oracle $\ora_4^*$ whereas in ${\hyb_{2,\hybindex,4}}$ we use the simulated oracle $\ora'$. We will use the incompressibility property of the random oracle $\hsig$ (\cref{thm:randoraincomp}) to show that this only makes a difference of $\frac{1}{\lambda\cdot \numrounds\cdot \origwinpoly(\lambda)}$ statistical difference. In this proof whenever we refer to incompressibility we will be referring to that of $\hsig$.

For the incompressibility property, consider the following adversaries $\adhocadv_1$ and $\adhocadv_2$.  $\adhocadv_1$ simulates the hybrid $\hyb_{2,\hybindex,3}$ until $\lastadv{1}{\hybindex}$ produces its output $\lastadvout{\hybindex}$, and then outputs $\lastadvout{\hybindex}$. Then, $\adhocadv_2$ receives this value and runs $\lastadv{2}{\hybindex}$ with it. When considering $\adhocadv_1, \adhocadv_2$ for incompressibility of $\hsig$, we assume that $\adhocadv_1, \adhocadv_2$ are only oracle-adversaries for $\hsig$ and the verifier (as in \cref{thm:randoraincomp}) of $\hsig$ respectively, and the other oracles are just simulated, e.g. they sample $H_0$ in advance and both use this sample to simulate the oracle calls to $\ora_0^*$. Observe that both $\adhocadv_1$ and $\adhocadv_2$ can perform their simulations with their respective oracles, e.g., $\ora_4^*$ is simply the verification oracle of $\hsig$ and $\adhocadv_2$ does not query $\hsig$.

Let $\mathfrak{L}$ be the incompressibility list as in \cref{thm:randoraincomp}, and we know $|\mathfrak{L}| = \promisedlistsize{4}{\hybindex} = 512\cdot\lambda\cdot (t^2(\lambda)\cdot \ell^2(\lambda))\cdot\lambda^4\cdot (\numrounds\cdot \origwinpoly(\lambda)\cdot\numqueries)^2$ since that many queries to $\hsig$ are made by $\adhocadv_1$. The list depends on the oracle itself and the \emph{leakage} $\lastadvout{\hybindex}$, but we will not write this dependence explicitly for conciseness. 

Let $\gamma_{m,y}^{\fireora, \lastadvout{\hybindex}}$ be the total query weight of $\lastadv{2}{\hybindex}(\lastadvout{\hybindex})$ to $\ora_4^*$ in  ${\hyb_{2,\hybindex,2}}$ on the input $m,y$. Define $\alpha_m$ as 
\begin{equation*}
    \alpha^{\fireora, \lastadvout{\hybindex}}_m =  \sum_{\substack{y: \ora_4^*(m,y)=1}} \gamma^{\fireora, \lastadvout{\hybindex}}_{m,y}.
\end{equation*}

Now define $W^{\fireora, \lastadvout{\hybindex},  \helplistnew{4}{\hybindex}}, W^{\fireora, \lastadvout{\hybindex},  \helplistnew{4}{\hybindex}, \mathfrak{L}}_1, W^{\fireora, \lastadvout{\hybindex},  \helplistnew{4}{\hybindex}, \mathfrak{L}}_2$ as 
\begin{align*}
    &W^{\fireora, \lastadvout{\hybindex},  \helplistnew{4}{\hybindex}} := \sum_{\substack{m: \\ m \not\in \helplistnew{4}{\hybindex} }} \alpha^{\fireora, \lastadvout{\hybindex}}_{m}\\
    &W_1^{\fireora, \lastadvout{\hybindex},  \helplistnew{4}{\hybindex}, \mathfrak{L}}  := \sum_{\substack{m: \\ m \not\in \helplistnew{4}{\hybindex} \\ m \not\in \mathfrak{L}}} \alpha^{\fireora, \lastadvout{\hybindex}}_{m} \\
    &W_2^{\fireora, \lastadvout{\hybindex},  \helplistnew{4}{\hybindex}, \mathfrak{L}}  := \sum_{\substack{m: \\ m \not\in \helplistnew{4}{\hybindex} \\ m \in \mathfrak{L}}} \alpha^{\fireora, \lastadvout{\hybindex}}_{m}
\end{align*}
In the definition above, and in the rest of the proof, by abuse of notation, we write $m \in \helplistnew{4}{\hybindex}$ to denote that there exists a pair $(m,y) \in \helplistnew{4}{\hybindex}$ for some $y$.

We claim the following.
\begin{claim}\label{claim:twoltwotwolthreeone}
For all sufficiently large $\lambda > 0$,
\begin{equation*}
 \E_{\fireora, \lastadvout{\hybindex},  \helplistnew{4}{\hybindex},\mathfrak{L}}[W^{\fireora, \lastadvout{\hybindex},  \helplistnew{4}{\hybindex}}]  \leq\frac{1}{128\cdot \numfourqueries\cdot(\lambda\cdot {\numrounds}\cdot \origwinpoly(\lambda))^2}.
 \end{equation*}
\end{claim}

Observe that once we prove this, the lemma follows by \cref{prethm:bbbv97} and by the Jensen's inequality since the oracles $\ora_4^*$ and $\ora'$ in these hybrids can possibly differ only on inputs $(m,y)$ such that $(m,y) \not\in \helplistnew{4}{\hybindex}$ and $\ora_4^*(m,y)=1$, and $W^{\fireora, \lastadvout{\hybindex},  \helplistnew{4}{\hybindex}}$ is an upper bound on the total weight on such inputs and  $\lastadv{2}{\ell}$ makes at most $\numfourqueries$ queries to the oracle $\ora_4^*$.

To prove \cref{claim:twoltwotwolthreeone}, we make the following two other claims.
\begin{claim}\label{claim:twoltwotwolthreetwo}
\begin{equation*}
\E_{\fireora, \lastadvout{\hybindex},  \helplistnew{4}{\hybindex},\mathfrak{L}}[W_1^{\fireora, \lastadvout{\hybindex},  \helplistnew{4}{\hybindex}, \mathfrak{L}}]  \leq \negl(\lambda).
 \end{equation*}
\end{claim}
\begin{proof}
Consider the value.
\begin{equation*}
    A = \sum_{\substack{m: \\ m \not\in \mathfrak{L}}} \alpha^{\fireora, \lastadvout{\hybindex}}_{m}
\end{equation*}
Observe that this value is always greater than or equal to $W_1^{\fireora, \lastadvout{\hybindex},  \helplistnew{4}{\hybindex}, \mathfrak{L}}$.
Also observe that $\E[A] \leq\negl(\lambda)$ by the incompressibility of $\hsig$.
\end{proof}

\begin{claim}\label{claim:twoltwotwolthreethree}
  For all sufficiently large $\lambda > 0$:
\begin{equation*}
\E_{\fireora, \lastadvout{\hybindex},  \helplistnew{4}{\hybindex},\mathfrak{L}}[W_2^{\fireora, \lastadvout{\hybindex},  \helplistnew{4}{\hybindex}, \mathfrak{L}}]  \leq \frac{1}{256\cdot\numfourqueries \cdot(\lambda\cdot {\numrounds}\cdot \origwinpoly(\lambda))^2}.
 \end{equation*}
\end{claim}
\begin{proof}
    First we claim that, for all ${\fireora, \lastadvout{\hybindex}}$, with overwhelming probability we have that any $m \in \messpa$ that satisfies
\begin{equation*}
    \alpha_m^{\fireora, \lastadvout{\hybindex}} \geq \frac{1}{512\cdot \numfourqueries \cdot(\lambda\cdot {\numrounds}\cdot \origwinpoly(\lambda))^2\cdot \promisedlistsize{4}{\ell}}
\end{equation*}
is contained in $\helplistnew{4}{\hybindex}$. To see this, observe that $\helplistnew{4}{\hybindex}$ is obtained by measuring a random query of $\lastadv{2}{\hybindex}(\lastadvout{\hybindex})$ to $\ora_4^*$ for $\numestbyind{4}{\ell} = 512^2\cdot(t(\lambda))^6\cdot \lambda^8 \cdot (u(\lambda))^4 \cdot (q^*(\lambda))^4 \cdot (\ell)^2$ times and adding to the set if the pair $m,y$ obtained from the measurement is \emph{valid} (i.e. $\ora^*_4(m,y)=1$). Thus, for any $m \in \messpa$ that satisfies the above condition, the probability that $m \not\in \helplistnew{4}{\hybindex}$ is at most $\exp{-\numestbyind{4}{\ell}\cdot\frac{\alpha_{m}^{\fireora, \lastadvout{\hybindex}}}{\numfourqueries}}$. Since we have 
\begin{equation*}
1 - \exp{\numestbyind{4}{\ell}\cdot\frac{1}{ \numfourqueries \cdot 512\cdot \numfourqueries \cdot(\lambda\cdot {\numrounds}\cdot \origwinpoly(\lambda))^2\cdot \promisedlistsize{4}{\ell}} }= 1 - \exp{-\lambda}
\end{equation*} 
 this claim follows.

Now define $\mathsf{GOOD}$ to be the event that all $m \in \mathfrak{L}$ that satisfies  \begin{equation*}
    \alpha_m^{\fireora, \lastadvout{\hybindex}} \geq \frac{1}{512\cdot \numfourqueries \cdot(\lambda\cdot {\numrounds}\cdot \origwinpoly(\lambda))^2\cdot \promisedlistsize{4}{\ell}}
\end{equation*}
are included in $\helplistnew{4}{\hybindex}$. Since this is true with probability $1 - \exp{-\lambda}$ for each $m \in \mathfrak{L}$ and since $\mathfrak{L}$ is of polynomial size, we have that $\Pr[\mathsf{GOOD}] \geq 1 - \negl(\lambda)$.

Now we claim \begin{equation*}
    \E_{\fireora, \lastadvout{\hybindex},  \helplistnew{4}{\hybindex},\mathfrak{L}}[W_2^{\fireora, \lastadvout{\hybindex},  \helplistnew{4}{\hybindex}, \mathfrak{L}} \big| \mathsf{GOOD}]  \leq \frac{1}{512\cdot\numfourqueries \cdot(\lambda\cdot {\numrounds}\cdot \origwinpoly(\lambda))^2}.
\end{equation*}
To see this, first by linearity of expectation we have
\begin{equation*}
   \E_{\fireora, \lastadvout{\hybindex},  \helplistnew{4}{\hybindex},\mathfrak{L}}[W_2^{\fireora, \lastadvout{\hybindex},  \helplistnew{4}{\hybindex}, \mathfrak{L}} \big| \mathsf{GOOD}] =\sum_{\substack{m: \\ m \not\in \helplistnew{4}{\hybindex}  \\ m \in \mathfrak{L}}} \E[\alpha^{\fireora, \lastadvout{\hybindex}}_{m} \big| \mathsf{GOOD}].
\end{equation*}
However, by definition of $\mathsf{GOOD}$, when $\mathsf{GOOD}$ occurs, we have
\begin{equation*}
    m \in \mathfrak{L} \setminus \helplistnew{4}{\hybindex} \implies \alpha^{\fireora, \lastadvout{\hybindex}}_{m} \leq \frac{1}{512\cdot \numfourqueries \cdot(\lambda\cdot {\numrounds}\cdot \origwinpoly(\lambda))^2\cdot \promisedlistsize{4}{\ell}}
\end{equation*}
Since $|\mathfrak{L}| =  \promisedlistsize{4}{\ell}$, we obtain the desired upper bound on $ \E_{\fireora, \lastadvout{\hybindex},  \helplistnew{4}{\hybindex},\mathfrak{L}}[W_2^{\fireora, \lastadvout{\hybindex},  \helplistnew{4}{\hybindex}, \mathfrak{L}} \big| \mathsf{GOOD}]$. Since $\Pr[\mathsf{GOOD}] \geq 1 - \negl(\lambda)$, we get \begin{equation*}
    \E_{\fireora, \lastadvout{\hybindex},  \helplistnew{4}{\hybindex},\mathfrak{L}}[W_2^{\fireora, \lastadvout{\hybindex},  \helplistnew{4}{\hybindex}, \mathfrak{L}}]  \leq \frac{1}{256\cdot\numfourqueries \cdot(\lambda\cdot {\numrounds}\cdot \origwinpoly(\lambda))^2}.
\end{equation*}
\end{proof}
Now observe that \cref{claim:twoltwotwolthreeone} follows immediately from \cref{claim:twoltwotwolthreetwo} and \cref{claim:twoltwotwolthreethree} since $W^{\fireora, \lastadvout{\hybindex},  \helplistnew{4}{\hybindex}} = W_1^{\fireora, \lastadvout{\hybindex},  \helplistnew{4}{\hybindex}, \mathfrak{L}}  + W_2^{\fireora, \lastadvout{\hybindex},  \helplistnew{4}{\hybindex}, \mathfrak{L}}$ for all ${\fireora, \lastadvout{\hybindex},  \helplistnew{4}{\hybindex}, \mathfrak{L}}$ (i.e. with probability $1$).  This completes the proof of \cref{claim:twoltwotwolthreeone}, and then the lemma follows as discussed above.

\subsection{Proof of \cref{thm:finalhybridsecure}}\label{sec:finalhybridsecure}
In this section, we prove security in the last hybrid. That is, we prove \cref{thm:finalhybridsecure}, which says $\Pr[\hyb_{3}=1]\leq\negl(\lambda)$.

First we prove the following lemma.
\begin{theorem}\label{lem:qromunlearn}
Let $H : \{0,1\}^{v(\lambda)} \rightarrow \{0,1\}^{\lambda}$ be modeled as a quantum random oracle,
and let $v(\lambda)$ be a superlogarithmic function.

For any adversaries $(\mathcal{A}_1,\mathcal{A}_2)$,
consider the following experiment.

\begin{mdframed}
\textbf{$\mathsf{Exp}_{\mathcal{A}_1,\mathcal{A}_2}(\lambda)$}
\begin{enumerate}
\item Sample a random function $H : \{0,1\}^{v(\lambda)} \rightarrow \{0,1\}^{\lambda}$.

\item $\mathcal{A}_1$ is given quantum oracle access to $H$ and outputs a message register $\regi$.

\item Sample $m^* \leftarrow \{0,1\}^{v(\lambda)}$.

\item $\mathcal{A}_2$ does not have access to $H$, and it receives $(\regi,m^*)$ and outputs $y^*$.

\item Output $1$ if $y^* = H(m^*)$, and $0$ otherwise.
\end{enumerate}
\end{mdframed}

\noindent
Then, for any query-bounded $\adve_1$,
\[
\Pr\!\left[
\mathsf{Exp}_{\mathcal{A}_1,\mathcal{A}_2}(\lambda)=1
\right]
\le \mathsf{negl}(\lambda).
\]
\end{theorem}
\begin{proof}
Since $m^*$ is unpredictable to $\adve_1$, the result follows by \cref{prethm:compor}.
\end{proof}

Observe that in $\hyb_{3}$, to win the game, the adversary needs to output a value $sig^*$ such that $\hsig(m^*) = sig^*$. Also observe that in this hybrid, the receiver adversary  $\newadv{2}{\numrounds+1}$ does not have access to any oracles, and the sender adversary $\newadv{1}{\numrounds+1}^{\fireora}$ makes polynomially many queries to the random oracle $\hsig$ and to $\ora_4^*$ (which can be simulated by querying $\hsig$). Thus the result follows by \cref{lem:qromunlearn}.

\newpage\section{Proof of \cref{lem:bighybrid}}\label{sec:bighybrid}
In this section, we prove \cref{lem:bighybrid}. Throughout this section, fix any $\hybindex^* \in [\numrounds+1]$. Therefore, we will prove $\statdist{\hyb_{2,\hybindex^*,0}}{\hyb_{2,\hybindex^*,1}} \leq \frac{1}{\lambda\cdot \numrounds\cdot \origwinpoly(\lambda)}$.   The proof will be done through a sequence of hybrids, each of which is constructed by modifying the previous one. We will denote our hybrids as $\hyb{\textcolor{blue}{'}}_i$ (\emph{hybrid prime}), and their indices will be independent of the indices of the hybrids $\hyb_j$ from \cref{sec:proofsec}. 

We first define our hybrids.

\paragraph{\underline{$\hyb'_{j, 0}$ for $j \in [\numqueries]$:}} 
\begin{mdframed}
 {\bf \underline{$\hyb'_{j, 0}$}}
 
    \begin{enumerate}
    \item The challenger samples $sn^*, \regis{flame} \samp \qkeyfiresign.\spark(1^\lambda)$.
    \item Parse $\fireora= (\ora_0^*, \ora_1^*, \ora_2^*, \ora^*_{\oss.\sign}, \ora^*_3, \ora^*_4) = sn^*$.
    \item[]\noindent\textbf{Round-collapsed Simulation for First $\hybindex^*-1$ Rounds:}
    \item Run $\newadv{1}{\hybindex^*-1}^{\fireora}(\regis{flame})$ to get $\advintstate{1}{\hybindex^*-1}$ and $(\newleak{\hybindex^*-1}, S^{\hybindex^*-1})$.
    \item[]\noindent\textbf{Round $\hybindex^*$:}
    \item { $\roundadve{1}{\hybindex^*}^{\fireora}$ receives $\advintstate{1}{\hybindex^*-1}$ and $\advetworeply{\hybindex^*-1}$, and it outputs an internal state register $\advintstate{1}{\hybindex^*}$ and a leakage message $\leakmes{\hybindex^*}$}
    \item \textcolor{red}{$\estadvsub{1}{\hybindex^*}{j}{0}^{\fireora}$ receives $\newleak{\hybindex^*-1},\leakmes{\hybindex^*}$ and produces a message $\qhelpstr{\hybindex^*,j,0}$}.
    \item \textcolor{red}{Run $\estadvsub{2}{\hybindex^*}{j}{0}^{(\ora_{\oss.\sign}^*,\ora_3^*, \ora_4^*)}(\qhelpstr{\hybindex^*,j,0})$ to obtain $(\advintstate{2}{\hybindex^*}, \advetworeply{\hybindex^*})$}
    \item[]\noindent\textbf{Original Protocol Continuation:}
    \item For $i \in \{{\hybindex^*+1},\dots, \numrounds+1\}$:
    \begin{enumerate}[label=\arabic*.]
        \item $\roundadve{1}{i}^{\fireora}$ receives $\advintstate{1}{i-1}$ and $\advetworeply{i-1}$, and it outputs a leakage message $\leakmes{i}$ and an internal state register $\advintstate{1}{i}$.
        \item $\roundadve{2}{i}^{\fireora}$ receives $\advintstate{2}{i-1}$ and $\leakmes{i}$, and it outputs a message $\advetworeply{i}$ and an internal state register $\advintstate{2}{i}$.
    \end{enumerate}
    \item {Parse $(m^*, sig^*) = S^{\numrounds+1}$}.
    \item Output $\ora_4^*(m^*, sig^*)$.
    \end{enumerate}
\end{mdframed}

\paragraph{\underline{$\hyb'_{j, 1}$ for $j \in [\numqueries]$:}} 
\begin{mdframed}
 {\bf \underline{$\hyb'_{j, 1}$}}
 
    \begin{enumerate}
    \item The challenger samples $sn^*, \regis{flame} \samp \qkeyfiresign.\spark(1^\lambda)$.
\item Parse $\fireora= (\ora_0^*, \ora_1^*, \ora_2^*, \ora^*_{\oss.\sign}, \ora^*_3, \ora^*_4) = sn^*$.
    \item[]\noindent\textbf{Round-collapsed Simulation for First $\hybindex^*-1$ Rounds:}
    \item Run $\newadv{1}{\hybindex^*-1}^{\fireora}(\regis{flame})$ to get $\advintstate{1}{\hybindex^*-1}$ and $(\newleak{\hybindex^*-1}, S^{\hybindex^*-1})$.
    \item[]\noindent\textbf{Round $\hybindex^*$:}
    \item { $\roundadve{1}{\hybindex^*}^{\fireora}$ receives $\advintstate{1}{\hybindex^*-1}$ and $\advetworeply{\hybindex^*-1}$, and it outputs an internal state register $\advintstate{1}{\hybindex^*}$ and a leakage message $\leakmes{\hybindex^*}$}
    \item \textcolor{red}{$\estadvsub{1}{\hybindex^*}{j}{1}^{\fireora}$ receives $\newleak{\hybindex^*-1},\leakmes{\hybindex^*}$ and produces a message $\qhelpstr{\hybindex^*,j,1}$}.
    \item \textcolor{red}{Run $\estadvsub{2}{\hybindex^*}{j}{1}^{(\ora_{\oss.\sign}^*, \ora_3^*, \ora_4^*)}(\qhelpstr{\hybindex^*,j,1})$ to obtain $(\advintstate{2}{\hybindex^*}, \advetworeply{\hybindex^*})$}
    \item[]\noindent\textbf{Original Protocol Continuation:}
    \item For $i \in \{{\hybindex^*+1},\dots, \numrounds+1\}$:
    \begin{enumerate}[label=\arabic*.]
        \item $\roundadve{1}{i}^{\fireora}$ receives $\advintstate{1}{i-1}$ and $\advetworeply{i-1}$, and it outputs a leakage message $\leakmes{i}$ and an internal state register $\advintstate{1}{i}$.
        \item $\roundadve{2}{i}^{\fireora}$ receives $\advintstate{2}{i-1}$ and $\leakmes{i}$, and it outputs a message $\advetworeply{i}$ and an internal state register $\advintstate{2}{i}$.
    \end{enumerate}
    \item {Parse $(m^*, sig^*) = S^{\numrounds+1}$}.
    \item Output $\ora_4^*(m^*, sig^*)$.
    \end{enumerate}
\end{mdframed}

\paragraph{\underline{$\hyb'_{j, 2}$ for $i \in \{0,\dots,\numqueries\}$:}} 
\begin{mdframed}
 {\bf \underline{$\hyb'_{j, 2}$}}
 
    \begin{enumerate}
    \item The challenger samples $sn^*, \regis{flame} \samp \qkeyfiresign.\spark(1^\lambda)$.
\item Parse $\fireora= (\ora_0^*, \ora_1^*, \ora_2^*, \ora^*_{\oss.\sign}, \ora^*_3, \ora^*_4) = sn^*$.
    \item[]\noindent\textbf{Round-collapsed Simulation for First $\hybindex^*-1$ Rounds:}
    \item Run $\newadv{1}{\hybindex^*-1}^{\fireora}(\regis{flame})$ to get $\advintstate{1}{\hybindex^*-1}$ and $(\newleak{\hybindex^*-1}, S^{\hybindex^*-1})$.
    \item[]\noindent\textbf{Round $\hybindex^*$:}
    \item { $\roundadve{1}{\hybindex^*}^{\fireora}$ receives $\advintstate{1}{\hybindex^*-1}$ and $\advetworeply{\hybindex^*-1}$, and it outputs an internal state register $\advintstate{1}{\hybindex^*}$ and a leakage message $\leakmes{\hybindex^*}$}
    \item \textcolor{red}{$\estadvsub{1}{\hybindex^*}{j}{2}^{\fireora}$ receives $\newleak{\hybindex^*-1},\leakmes{\hybindex^*}$ and produces a message $\qhelpstr{\hybindex^*,j,2}$}.
    \item \textcolor{red}{Run $\estadvsub{2}{\hybindex^*}{j}{2}^{(\ora_{\oss.\sign}^*,\ora_3^*, \ora_4^*)}(\qhelpstr{\hybindex^*,j,2})$ to obtain $(\advintstate{2}{\hybindex^*}, \advetworeply{\hybindex^*})$}
    \item[]\noindent\textbf{Original Protocol Continuation:}
    \item For $i \in \{{\hybindex^*+1},\dots, \numrounds+1\}$:
    \begin{enumerate}[label=\arabic*.]
        \item $\roundadve{1}{i}^{\fireora}$ receives $\advintstate{1}{i-1}$ and $\advetworeply{i-1}$, and it outputs a leakage message $\leakmes{i}$ and an internal state register $\advintstate{1}{i}$.
        \item $\roundadve{2}{i}^{\fireora}$ receives $\advintstate{2}{i-1}$ and $\leakmes{i}$, and it outputs a message $\advetworeply{i}$ and an internal state register $\advintstate{2}{i}$.
    \end{enumerate}
    \item {Parse $(m^*, sig^*) = S^{\numrounds+1}$}.
    \item Output $\ora_4^*(m^*, sig^*)$.
    \end{enumerate}
\end{mdframed}

\begin{lemma}
    $\hyb'_{0,2} \equiv \hyb_{2,\hybindex^*, 0}$
\end{lemma}
\begin{proof}
This follows by the definition of the adversaries $\estadvsub{1}{\ell}{0}{2}$,$\estadvsub{2}{\ell}{0}{2}$.
\end{proof}

\begin{lemma}
    $\hyb'_{\numqueries,2} \equiv \hyb_{2,\hybindex^*, 1}$.
\end{lemma}
\begin{proof}
    Follows by the definition of the adversary $\estadv{1}{\ell}, \estadv{2}{\ell}$ which are defined to be $\estadvsub{1}{\ell}{\numqueries}{2}$,$\estadvsub{2}{\ell}{\numqueries}{2}$.
\end{proof}

\begin{lemma}\label{lem:bighyblem3}
    For $i \in [\numqueries]$, $\statdist{\hyb'_{i-1, 2}}{\hyb'_{i, 0}} \leq \frac{1}{\lambda^2\cdot \numrounds\cdot \origwinpoly(\lambda)\cdot\numqueries}$ and $\statdist{\hyb'_{i, 0}}{\hyb'_{i, 1}} \leq \frac{1}{\lambda^2\cdot \numrounds\cdot \origwinpoly(\lambda)\cdot\numqueries}$.
\end{lemma}
We prove this lemma in \cref{sec:bighyblem3}.

\begin{lemma}\label{lem:bighyblem2}
    For $i \in [\numqueries]$, $\statdist{\hyb'_{i, 1}}{\hyb'_{i, 2}} \leq \frac{1}{\lambda^2\cdot \numrounds\cdot \origwinpoly(\lambda)\cdot\numqueries}$
\end{lemma}
We prove this lemma in \cref{sec:bighyblem2}.

\paragraph{}
Finally, \cref{lem:bighybrid} follows by simply combining these lemmata.

\subsection{Proof of \cref{lem:bighyblem3}}\label{sec:bighyblem3}\label{sec:bighyblem1}

In this section, we prove \cref{lem:bighyblem3}. We will only explicitly prove the first part, and the second part follows similarly.

Fix any $i^* \in \{1,\dots,\numqueries\}$. Thus, we will prove $\statdist{\hyb'_{i^*-1, 2}}{\hyb'_{i^*, 0}} \leq \frac{1}{\lambda^2\cdot \numrounds\cdot \origwinpoly(\lambda)\cdot\numqueries}$. Observe that the difference between the two hybrids is that in ${\hyb'_{i^*-1, 2}}$ we run $\estadvsub{2}{\hybindex^*}{i^*-1}{2}$ directly, whereas in ${\hyb'_{i^*, 0}}$ we instead run $\estadvsub{2}{\hybindex^*}{i^*}{0}$  which simulates $\estadvsub{2}{\hybindex^*}{i^*-1}{2}$ while using an estimated oracle for its first query to $\ora^*_0$. We will use the strong incompressibility property of $\oss$ to show that this only makes a difference of $\frac{1}{\lambda^2\cdot \numrounds\cdot \origwinpoly(\lambda)\cdot\numqueries}$ in statistical distance. In this proof, whenever we refer to incompressibility, we will be referring to the strong incompressibility of $\oss$.
For the strong incompressibility property of $\oss$, consider the following adversaries $\adhocadv_1, \adhocadv_2$. $\adhocadv_1$ simulates $\hyb'_{i^*-1, 2}$ until $\qhelpstr{\hybindex^*,i^*,1}$ is produced, and outputs $\qhelpstr{\hybindex^*,i^*,1}$. Then, $\adhocadv_2$ receives this value and runs $\estadvsub{\hybindex^*}{2}{i^*-1}{2}$ with it until right after first query to $\ora_0^*$. When considering $\adhocadv_1, \adhocadv_2$ for incompressibility of $\oss$, we assume that $\adhocadv_1, \adhocadv_2$ are only oracle-adversaries for $\oss$, and the other oracles are just simulated, e.g. they sample $H_0, H_1, \hsig$ in advance and both use this sample to simulate the oracle calls to relevant oracles. 

Now let $\mathfrak{L}$ denote the strong incompressibility list for these adversaries, and we know $|\mathfrak{L}|$ = $\promisedlistsize{OSS}{\hybindex^*} = 2048\cdot\lambda\cdot (t^2(\lambda)\cdot \ell^*)\cdot\lambda^4\cdot (\numrounds\cdot \origwinpoly(\lambda)\cdot\numqueries)^2\cdot (\hybindex^* - 1) + 2\cdot i^*$ since $\mathcal{P}_1$ makes at most $\frac{\promisedlistsize{OSS}{\hybindex^*}}{2}$ calls to $\ora^*_{\oss.\genqkey}$. The list depends on the oracle itself and the \emph{leakage} $\qhelpstr{\hybindex^*,i^*,1}$, but we will
not write this dependence explicitly for conciseness.

Let $\gamma_{ivk,isig,z}^{\fireora, \qhelpstr{\hybindex^*,i^*-1,2}}$ denote the query amplitude of $\estadvsub{\hybindex^*}{2}{i^*-1}{2}(\qhelpstr{\hybindex^*,i^*-1,2})$ on input $ivk,isig,z$ during its first query to $\ora^*_0$. We also define the following. Here, $\helplist{0}{\hybindex^*}{i^*}$ is the list sampled during execution of $\estadvsub{1}{\hybindex^*}{i^*}{0}$.

\begin{align*}
    &\alpha_{ivk,isig}^{\fireora, \qhelpstr{\hybindex^*,i^*-1,2}} = \begin{cases} &|\gamma_{ivk,isig}^{\fireora, \qhelpstr{\hybindex^*,i^*-1,2}}|^2, \text{ if } \ora^*_{\oss.\ver}(ivk, 0, isig) =  1 \\
    &0, \text{ if } \ora^*_{\oss.\ver}(ivk, 0, isig) = 0
    \end{cases}\\
    &W^{\fireora, \qhelpstr{\hybindex^*,i^*-1,2},\helplist{0}{\hybindex^*}{i^*}} = \sum_{\substack{ivk,isig:\\ (ivk,isig)\not\in\helplist{0}{\hybindex^*}{i^*}}} \alpha_{ivk,isig}^{\fireora, \qhelpstr{\hybindex^*,i^*-1,2}}\\
    &W_1^{\fireora, \qhelpstr{\hybindex^*,i^*-1,2},\helplist{0}{\hybindex^*}{i^*},\mathfrak{L}} = \sum_{\substack{ivk,isig:\\ (ivk,isig)\not\in\helplist{0}{\hybindex^*}{i^*}\\ (ivk,isig)\not\in\mathfrak{L}}} \alpha_{ivk,isig}^{\fireora, \qhelpstr{\hybindex^*,i^*-1,2}}\\
    &W_2^{\fireora, \qhelpstr{\hybindex^*,i^*-1,2},\helplist{0}{\hybindex^*}{i^*},\mathfrak{L}} = \sum_{\substack{ivk,isig:\\ (ivk,isig)\not\in\helplist{0}{\hybindex^*}{i^*}\\ (ivk,isig)\in\mathfrak{L}}} \alpha_{ivk,isig}^{\fireora, \qhelpstr{\hybindex^*,i^*-1,2}}
\end{align*}

Now we make the following claims.
\begin{claim}
    \begin{equation*}
        \E_{\fireora, \qhelpstr{\hybindex^*,i^*-1,2},\helplist{0}{\hybindex^*}{i^*},\mathfrak{L}}[W_1^{\fireora, \qhelpstr{\hybindex^*,i^*-1,2},\helplist{0}{\hybindex^*}{i^*},\mathfrak{L}}] \leq \negl(\lambda).
    \end{equation*}
\end{claim}
\begin{proof}
Observe that with probability $1$, we have
\begin{align*}
&W_1^{\fireora, \qhelpstr{\hybindex^*,i^*-1,2},\helplist{0}{\hybindex^*}{i^*},\mathfrak{L}} \leq \sum_{\substack{ivk,isig:\\ (ivk,isig)\not\in\mathfrak{L}}} \alpha_{ivk,isig}^{\fireora, \qhelpstr{\hybindex^*,i^*-1,2}}
\end{align*}
However, expectation of the value on the right is negligible by incompressibility of $\oss$ (with the adversaries $\adhocadv_1, \adhocadv_2$).
\end{proof}

\begin{claim}
For all sufficiently large $\lambda > 0$.
\begin{equation*}
    \E_{\fireora, \qhelpstr{\hybindex^*,i^*-1,2},\helplist{0}{\hybindex^*}{i^*},\mathfrak{L}}[W_2^{\fireora, \qhelpstr{\hybindex^*,i^*-1,2},\helplist{0}{\hybindex^*}{i^*},\mathfrak{L}}] \leq \frac{1}{256\cdot\lambda^4\cdot (\numrounds\cdot \origwinpoly(\lambda)\cdot\numqueries)^2}
\end{equation*}
\end{claim}
\begin{proof}
    First we claim that, for all $\fireora, \qhelpstr{\hybindex^*,i^*-1,2}$,  with overwhelming probability all $ivk,isig$ that satisfies 
    \begin{equation*}
        \alpha_{ivk,isig}^{\fireora, \qhelpstr{\hybindex^*,i^*-1,2}}  \geq \frac{1}{512\cdot \promisedlistsize{OSS}{\hybindex^*}\cdot\lambda^4\cdot (\numrounds\cdot \origwinpoly(\lambda)\cdot\numqueries)^2}
    \end{equation*}
    is contained in  $\helplist{0}{\hybindex^*}{i^*}$. To see this, observe that $\helplist{0}{\hybindex^*}{i^*}$ is created by measuring the first query of $\estadvsub{\hybindex^*}{2}{i^*-1}{2}$ to $\ora_0$  for $\numestbyind{OSS}{\hybindex^*} = 4194304\cdot \lambda^{10} \cdot t^2(\lambda)\cdot (\hybindex^*)^5\cdot (\numrounds\cdot \origwinpoly(\lambda)\cdot\numqueries)^4$ times. Thus, for any $ivk, isig$, the probability that $(ivk,isig)\not\in\helplist{0}{\hybindex^*}{i^*}$ is at most $\exp\{-\numestbyind{OSS}{\hybindex^*}\cdot \alpha_{ivk,isig}\}$. Since we have
    \begin{equation*}
        1 - \exp\{-\numestbyind{OSS}{\hybindex^*}\cdot \frac{1}{512\cdot \promisedlistsize{OSS}{\hybindex^*}\cdot\lambda^4\cdot (\numrounds\cdot \origwinpoly(\lambda)\cdot\numqueries)^2}\} = 1 - \exp{-\lambda}
    \end{equation*}
     the claim follows. Now define $\mathsf{GOOD}$ to be the event that all $(ivk,isig)\in\mathfrak{L}$ that satisfies \begin{equation*}
        \alpha_{ivk,isig}^{\fireora, \qhelpstr{\hybindex^*,i^*-1,2}}  \geq \frac{1}{512\cdot \promisedlistsize{OSS}{\hybindex^*}\cdot\lambda^4\cdot (\numrounds\cdot \origwinpoly(\lambda)\cdot\numqueries)^2}
    \end{equation*}
    are included in $\helplist{0}{\hybindex^*}{i^*}$. Since this is true with probability $1 - \exp\{-\lambda\}$ for each $ivk,isig \in \mathfrak{L}$ individually, and since $|\mathfrak{L}|$ is polynomial, we have that $\Pr[\mathsf{GOOD}] \geq 1 - \negl(\lambda)$.

Now we claim \begin{equation*}
     \E_{\fireora, \qhelpstr{\hybindex^*,i^*-1,2},\helplist{0}{\hybindex^*}{i^*},\mathfrak{L}}[W_2^{\fireora, \qhelpstr{\hybindex^*,i^*-1,2},\helplist{0}{\hybindex^*}{i^*},\mathfrak{L}} \big| \mathsf{GOOD}] \leq \frac{1}{512\cdot\lambda^4\cdot (\numrounds\cdot \origwinpoly(\lambda)\cdot\numqueries)^2}
\end{equation*}
To see this, first by linearity of expectation we have
\begin{align*}
   &\E_{\fireora, \qhelpstr{\hybindex^*,i^*-1,2},\helplist{0}{\hybindex^*}{i^*},\mathfrak{L}}[W_2^{\fireora, \qhelpstr{\hybindex^*,i^*-1,2},\helplist{0}{\hybindex^*}{i^*},\mathfrak{L}} \big| \mathsf{GOOD}]\\ &=\sum_{\substack{ivk,isig:\\ (ivk,isig)\not\in\helplist{0}{\hybindex^*}{i^*}\\ (ivk,isig)\in\mathfrak{L}}} \E_{\fireora, \qhelpstr{\hybindex^*,i^*-1,2}}\left[\alpha_{ivk,isig}^{\fireora, \qhelpstr{\hybindex^*,i^*-1,2}}\big|\mathsf{GOOD}\right].
\end{align*}
However, by definition of $\mathsf{GOOD}$, when $\mathsf{GOOD}$ occurs, we have
\begin{equation*}
    ivk,isig \in \mathfrak{L} \setminus \helplist{0}{\hybindex^*}{i^*} \implies  \alpha_{ivk,isig}^{\fireora, \qhelpstr{\hybindex^*,i^*-1,2}}  \leq \frac{1}{512\cdot \promisedlistsize{OSS}{\hybindex^*}\cdot\lambda^4\cdot (\numrounds\cdot \origwinpoly(\lambda)\cdot\numqueries)^2}
\end{equation*}
Since $|\mathfrak{L}| =  \promisedlistsize{OSS}{\hybindex^*}$, we obtain the desired upper bound on $$\E_{\fireora, \qhelpstr{\hybindex^*,i^*-1,2},\helplist{0}{\hybindex^*}{i^*},\mathfrak{L}}[W_2^{\fireora, \qhelpstr{\hybindex^*,i^*-1,2},\helplist{0}{\hybindex^*}{i^*},\mathfrak{L}} \big| \mathsf{GOOD}].$$ Since $\Pr[\mathsf{GOOD}] \geq 1 - \negl(\lambda)$, we get \begin{equation*}
    \E_{\fireora, \qhelpstr{\hybindex^*,i^*-1,2},\helplist{0}{\hybindex^*}{i^*},\mathfrak{L}}[W_2^{\fireora, \qhelpstr{\hybindex^*,i^*-1,2},\helplist{0}{\hybindex^*}{i^*},\mathfrak{L}}] \leq \frac{1}{256\cdot\lambda^4\cdot (\numrounds\cdot \origwinpoly(\lambda)\cdot\numqueries)^2}
\end{equation*}
\end{proof}

Now, observe that \cref{lem:bighyblem3} follows from these two lemmata. To see this, observe that $W^{\fireora, \qhelpstr{\hybindex^*,i^*-1,2},\helplist{0}{\hybindex^*}{i^*}}$ is an upper bound on the total weight on inputs at which the two oracles might differ. Since we have $W^{\fireora, \qhelpstr{\hybindex^*,i^*-1,2},\helplist{0}{\hybindex^*}{i^*}} = W_1^{\fireora, \qhelpstr{\hybindex^*,i^*-1,2},\helplist{0}{\hybindex^*}{i^*},\mathfrak{L}}  + W_2^{\fireora, \qhelpstr{\hybindex^*,i^*-1,2},\helplist{0}{\hybindex^*}{i^*}\mathfrak{L}}$ with probability $1$, the result follows by the linearity of expectation, \cref{prethm:bbbv97} and Jensen's inequality.
\subsection{Proof of \cref{lem:bighyblem2}}\label{sec:bighyblem2}
Fix any $i^* \in \{1,\dots,\numqueries\}$. Thus, we will prove $\statdist{\hyb'_{i^*,1}}{\hyb'_{i^*, 2}} \leq \frac{1}{\lambda^2\cdot \numrounds\cdot \origwinpoly(\lambda)\cdot\numqueries}$. The proof will follow similarly to \cref{lem:hyb2l2to2l3}.

Observe that the difference between the two hybrids is that in ${\hyb'_{i^*, 1}}$ we run $\estadvsub{2}{\hybindex^*}{i^*}{1}$ directly, whereas in ${\hyb'_{i^*, 2}}$ we instead run $\estadvsub{2}{\hybindex^*}{i^*}{2}$  which simulates $\estadvsub{2}{\hybindex^*}{i^*}{1}$ while using an estimated oracle for its first query to $\ora^*_2$. We will use the incompressibility property of $H_0, H_1$ to show that this only makes a difference of $\frac{1}{\lambda^2\cdot \numrounds\cdot \origwinpoly(\lambda)\cdot\numqueries}$ in statistical distance. 

Let $\gamma_{ivk,y_0,y_1,z}^{\fireora, \qhelpstr{\hybindex^*,i^*,1}}$ denote the query amplitude of $\estadvsub{\hybindex^*}{2}{i^*}{1}(\qhelpstr{\hybindex^*,i^*,1})$ on input $ivk,y_0,y_1,z$ during its first query to $\ora^*_2$. We also define the following.

\begin{align*}
    &\alpha_{ivk,z}^{\fireora, \qhelpstr{\hybindex^*,i^*,1}} = \sum_{y_0,y_1:  \ora^*_2(ivk, y_0, y_1, z) \neq \bot} |\gamma_{ivk,y_0,y_1,z}^{\fireora, \qhelpstr{\hybindex^*,i^*,1}}|^2
\end{align*}

Now let $(\helplist{0}{\hybindex'}{j'})_{\hybindex' \in [\hybindex^*], j' \in [i^*]}$, $(\helplist{1}{\hybindex'}{j'})_{\hybindex' \in [\hybindex^*], j' \in [i^*]}$, $(\helplist{2}{\hybindex'}{j'})_{\hybindex' \in [\hybindex^*], j' \in [i^*]-1}, (\dbadv{0}{\hybindex'}{j'})_{\hybindex' \in [\hybindex^*], j' \in [i^*]}$, $(\dbadv{1}{\hybindex'}{j'})_{\hybindex' \in [\hybindex^*], j' \in [i^*]}$, $(\dbadv{2}{\hybindex'}{j'})_{\hybindex' \in [\hybindex^*], j' \in [i^*]-1}$ be the estimated lists and databases in $\qhelpstr{\hybindex^*,i^*,1}$ when it is parsed and fully unwrapped. With abuse of notation, sometimes we will refer to the projection of the sets $\helplist{\cdot}{\hybindex'}{j'}$ onto their elements' (which are tuples) subcomponents also as  $\helplist{\cdot}{\hybindex'}{j'}$ (e.g. we will write $\helplist{2}{\hybindex'}{j'}$ to mean the set of pairs $(ivk, z)$ obtained by projecting each tuple in the actual set $\helplist{2}{\hybindex'}{j'}$ to its first and fourth elements) - however, the meaning will clear from context. Now we claim the following.

\begin{claim}\label{claim:lastnewtriv}
\begin{equation*}
    \Pr\left[\left(\bigcup_{\hybindex' \in [\hybindex^*], j' \in [i^*]} \helplist{0}{\hybindex'}{j'} \right) \bigcap \left(\bigcup_{\hybindex' \in [\hybindex^*], j' \in [i^*]} \helplist{1}{\hybindex'}{j'} \right)  = \emptyset \right] \geq 1 - \negl(\lambda).
\end{equation*}
\end{claim}
\begin{proof}
    Observe that if this intersection is not empty, then there is $ivk$ such that $ivk \in  \helplist{0}{\hybindex'}{j'}$ and $ivk \in  \helplist{1}{\hybindex''}{j''}$ for some $\hybindex', \hybindex'', j', j''$. However, by construction we know that this implies that we can obtain $isig_0, isig_1$ from $\qhelpstr{\hybindex^*,i^*,1}$ such that $\ora_{\oss.\ver^*}(ivk, 0, isig_0)=\ora_{\oss.\ver^*}(ivk, 1, isig_1) = 1$. Since $\qhelpstr{\hybindex^*,i^*,1}$ is obtained by making polynomially many queries to the oracles of $\oss$, the claim follows by the one-shot security of $\oss$.
\end{proof}

Now we define the following (some dependencies are implicit to keep the notation concise).
\begin{align*}
    &W_0^{\fireora, \qhelpstr{\hybindex^*,i^*,1},\helplist{2}{\hybindex^*}{i^*}} = \sum_{\substack{ivk,z:\\ ivk \not\in\left(\bigcup_{\hybindex' \in [\hybindex^*], j' \in [i^*]} \helplist{0}{\hybindex'}{j'} \right) \\
    (ivk,z) \not\in \helplist{2}{\hybindex^*}{i^*}}} \alpha_{ivk,z}\\
    &W_1^{\fireora, \qhelpstr{\hybindex^*,i^*,1},\helplist{2}{\hybindex^*}{i^*}} = \sum_{\substack{ivk,z:\\ ivk \not\in\left(\bigcup_{\hybindex' \in [\hybindex^*], j' \in [i^*]} \helplist{1}{\hybindex'}{j'} \right) \\
    (ivk,z) \not\in \helplist{2}{\hybindex^*}{i^*}}} \alpha_{ivk,z}\\
    &W_2^{\fireora, \qhelpstr{\hybindex^*,i^*,1},\helplist{2}{\hybindex^*}{i^*}} = \sum_{\substack{ivk,z:\\ ivk \in\left(\bigcup_{\hybindex' \in [\hybindex^*], j' \in [i^*]} \helplist{0}{\hybindex'}{j'} \right) \bigcap \left(\bigcup_{\hybindex' \in [\hybindex^*], j' \in [i^*]} \helplist{1}{\hybindex'}{j'} \right) }} \alpha_{ivk,z}\\
\end{align*}

Now we make the following claims.
\begin{claim}
    \begin{equation}
        \E[W_2^{\fireora, \qhelpstr{\hybindex^*,i^*,1},\helplist{2}{\hybindex^*}{i^*}}]\leq \negl(\lambda).
    \end{equation}
\end{claim}
\begin{proof}
    This simply follows from the definition of $W_2$ and \cref{claim:lastnewtriv}.
\end{proof}
\begin{claim}
For all sufficiently large $\lambda > 0$.
\begin{align*}
    &\E[W_0^{\fireora, \qhelpstr{\hybindex^*,i^*,1},\helplist{2}{\hybindex^*}{i^*}}] \leq \frac{1}{128\cdot\lambda^4\cdot (\numrounds\cdot \origwinpoly(\lambda)\cdot\numqueries)^2} \\
    &\E[W_1^{\fireora, \qhelpstr{\hybindex^*,i^*,1},\helplist{2}{\hybindex^*}{i^*}}] \leq \frac{1}{128\cdot\lambda^4\cdot (\numrounds\cdot \origwinpoly(\lambda)\cdot\numqueries)^2}
\end{align*}
\end{claim}
\begin{proof}
We will only prove the first one (using incompressibility of the random oracle $H_0$), and the second bound follows by the same argument (but this time relying on the incompressibility of $H_1$).

Now we will use the incompressibility property (\cref{thm:randoraincomp}) of the random oracle $H_0$ with the predicate $P$ defined as $P(ivk || z) = 1$ if $ivk \in S$, where $S = \left(\bigcup_{\hybindex' \in [\hybindex^*], j' \in [i^*]} \helplist{0}{\hybindex'}{j'} \right)$. Note that this set $S$ is exactly the set of prefixes on which our adversary has \emph{section queries} (otherwise it only makes normal queries and verification queries). In more detail, let $\adhocadv_{1}$ be the adversary that simulates $\hyb'_{i^*, 1}$ until the point where $\estadvsub{1}{\hybindex^*}{i^*}{2}$ outputs its message, and it also outputs $S$ as its predicate. Then $\mathcal{P}_2$ simulates $\estadvsub{2}{\hybindex^*}{i^*}{1}$  until right after its first query to $\ora_2^*$. We invoke \cref{thm:randoraincomp} for this pair of adversaries, and let $\mathfrak{L}$ denote the incompressibility list that is promised to exist, and we know $|\mathfrak{L}| = \ell^*\cdot t(\lambda)$ since $\adhocadv_1$ makes at most $\ell^*\cdot t(\lambda)$ normal queries. We then split $W_0$ into two summands $U_0, U_1$:
\begin{align*}
    &U_0^{\fireora, \qhelpstr{\hybindex^*,i^*,1},\helplist{2}{\hybindex^*}{i^*}} = \sum_{\substack{ivk,z:\\ ivk \not\in\left(\bigcup_{\hybindex' \in [\hybindex^*], j' \in [i^*]} \helplist{0}{\hybindex'}{j'} \right) \\
    (ivk,z) \not\in \helplist{2}{\hybindex^*}{i^*} \\ (ivk,z)\not\in \mathfrak{L} }} \alpha_{ivk,z}\\
    &U_1^{\fireora, \qhelpstr{\hybindex^*,i^*,1},\helplist{2}{\hybindex^*}{i^*}} = \sum_{\substack{ivk,z:\\ ivk \not\in\left(\bigcup_{\hybindex' \in [\hybindex^*], j' \in [i^*]} \helplist{0}{\hybindex'}{j'} \right) \\
    (ivk,z) \not\in \helplist{2}{\hybindex^*}{i^*} \\ (ivk,z)\in \mathfrak{L} }} \alpha_{ivk,z}
\end{align*}

It is easy to see that $\E[U_0] \leq \negl(\lambda)$ by the incompressibility of $H_0$, thus, we will focus on bounding $U_1$.

    First we claim that, for all $\fireora, \qhelpstr{\hybindex^*,i^*,1}$, with overwhelming probability we have that any $ivk,z$ that satisfies 
    \begin{equation*}
        \alpha_{ivk,z} \geq \frac{1}{512\cdot (t(\lambda)\cdot \ell^*)\cdot\lambda^4\cdot (\numrounds\cdot \origwinpoly(\lambda)\cdot\numqueries)^2}
    \end{equation*}
    is contained in  $\helplist{2}{\hybindex^*}{i^*}$. To see this, observe that $\helplist{2}{\hybindex^*}{i^*}$ is created by measuring the first query of $\estadvsub{\hybindex^*}{2}{i^*}{1}$ to $\ora_2$  for $512\cdot\lambda\cdot (t(\lambda)\cdot \ell^*)\cdot\lambda^4\cdot (\numrounds\cdot \origwinpoly(\lambda)\cdot\numqueries)^2$ times. Thus, for any $ivk, z$ such that $\alpha_{ivk,z}$ satisfies the above condition, the probability that $(ivk,z)\not\in\helplist{2}{\hybindex^*}{i^*}$ is at most $\exp{-\lambda}$.  Now define $\mathsf{GOOD}$ to be the event that all $(ivk,z)\in\mathfrak{L}$ that satisfies \begin{equation*}
        \alpha_{ivk,z}  \geq  \frac{1}{512\cdot (t(\lambda)\cdot \ell^*)\cdot\lambda^4\cdot (\numrounds\cdot \origwinpoly(\lambda)\cdot\numqueries)^2}
    \end{equation*}
    are included in $\helplist{2}{\hybindex^*}{i^*}$. Since this is true with probability $1 - \exp\{-\lambda\}$ for each $ivk,z \in \mathfrak{L}$ individually, and since $|\mathfrak{L}|$ is polynomial, we have that $\Pr[\mathsf{GOOD}] \geq 1 - \negl(\lambda)$.

Now we claim \begin{equation*}
     \E[U_1 \big| \mathsf{GOOD}] \leq \frac{1}{512\cdot\lambda^4\cdot (\numrounds\cdot \origwinpoly(\lambda)\cdot\numqueries)^2}
\end{equation*}
To see this, first by linearity of expectation we have
\begin{align*}
   &\E[U_1 \big| \mathsf{GOOD}]\\ &=\sum_{\substack{ivk,z:\\ ivk \not\in\left(\bigcup_{\hybindex' \in [\hybindex^*], j' \in [i^*]} \helplist{0}{\hybindex'}{j'} \right) \\
    (ivk,z) \not\in \helplist{2}{\hybindex^*}{i^*} \\ (ivk,z)\in \mathfrak{L} }} \E[\alpha_{ivk,z} | \mathsf{GOOD}]
\end{align*}
However, by definition of $\mathsf{GOOD}$, when $\mathsf{GOOD}$ occurs, we have
\begin{equation*}
    ivk,z \in \mathfrak{L} \setminus \helplist{2}{\hybindex^*}{i^*} \implies  \alpha_{ivk,z}\leq  \frac{1}{512\cdot (t(\lambda)\cdot \ell^*)\cdot\lambda^4\cdot (\numrounds\cdot \origwinpoly(\lambda)\cdot\numqueries)^2}
\end{equation*}
Since $|\mathfrak{L}| =   \ell^*\cdot t(\lambda)$, we obtain the desired upper bound on expectation conditioned on $\mathsf{GOOD}$. Since $\Pr[\mathsf{GOOD}] \geq 1 - \negl(\lambda)$, we also get \begin{equation*}
    \E[U_1] \leq \frac{1}{256\cdot\lambda^4\cdot (\numrounds\cdot \origwinpoly(\lambda)\cdot\numqueries)^2}
\end{equation*}
which completes the proof.
\end{proof}

Observe that \cref{lem:bighyblem2} follows from the above claims. To see this, observe that $W_0+W_1+W_2$ is an upper bound on the total weight on inputs at which the two different oracles in the hybrids might differ. Thus the result follows by the linearity of expectation, \cref{prethm:bbbv97} and Jensen's inequality.

\newpage\addrefs

\newpage\appendix
\part*{Appendix}
\addtocontents{toc}{\protect\setcounter{tocdepth}{-10}}
\section{Alternative Equivalent Definitions for Quantum Protection Notions}\label{appn:alternativedefnslr}
We recall the definition of LOCC leakage-resilience for the case of signing and decryption. Note that this is a special case of the general definition (\cref{predef:ublrfunc}), and we write the specialized version explicitly for ease of exposition.

\begin{definition}[Signature Scheme with LOCC Leakage-Resilience \cite{TCC:CGLR24,cryptoeprint:2024/1876}]\label{predef:ublrsig}
Consider the following game between the challenger and an adversary $\adve$.

\paragraph{\underline{$\lrsiggame{\adve}(1^\lambda)$}}
\begin{enumerate}
    \item The challenger samples $vk, \regis{key} \samp \digsig.\setup(1^\lambda)$ and submits $vk$ to $\adve$.
    \item The adversary obtains leakage on $\regis{key}$ adaptively over any number of rounds as in \cref{predef:ublrfunc}.
    \item The challenger samples $m^* \samp \messpa$ and submits $m^*$ to $\adve$.
    \item $\adve$ outputs a forged signature $sig$.
    \item The challenger outputs $1$ if and only if $\digsig.\ver(vk, m^*, sig)$.
\end{enumerate}
A digital signature scheme $\digsig$ is said to satisfy \emph{LOCC leakage-resilience} if for any QPT adversary $\adve$,
\begin{equation*}
    \Pr[\lrsiggame{\adve}(1^\lambda) = 1] \leq \frac{1}{|\messpa|} + \negl(\lambda).
\end{equation*}
\end{definition}

\begin{definition}[Public-Key Encryption with LOCC Leakage-Resilience \cite{TCC:CGLR24,cryptoeprint:2024/1876}]\label{predef:ublrpke}
Consider the following game between the challenger and an adversary $\adve$.

\paragraph{\underline{$\lrpkegame{\adve}(1^\lambda)$}}
\begin{enumerate}
    \item The challenger samples $pk, \regis{key} \samp \pke.\mathsf{Setup}(1^\lambda)$ and submits $pk$ to $\adve$.
    \item The adversary obtains leakage on $\regis{key}$ adaptively over any number of rounds as in \cref{predef:ublrfunc}.
    \item The adversary $\adve$ outputs two messages $m_0, m_1$.
    \item The challenger samples $b \samp \zo$ and $ct \samp \pke.\mathsf{Enc}(pk, m_b)$ and submits $ct$ to $\adve$.
    \item $\adve$ outputs a guess $b' \in \zo$.
    \item The challenger outputs $1$ if and only if $b' = b$.
\end{enumerate}
A public-key encryption scheme $\pke$ is said to satisfy \emph{LOCC leakage-resilience} if for any QPT adversary $\adve$,
\begin{equation*}
    \Pr[\lrpkegame{\adve}(1^\lambda) = 1] \leq \frac{1}{2} + \negl(\lambda).
\end{equation*}
\end{definition}

\addtocontents{toc}{\protect\setcounter{tocdepth}{-10}}
\section{On \cite{vinodhuang}}\label{appn:vh}
The initial version (v1) of our paper on arXiV contained a small gap in  last hybrid of our quantum key-fire proof, which was communicated to us by the authors of \cite{vinodhuang} and at the same time they put their paper online. Within a day of receiving their communication, we fixed the small gap in our proof and let them know. Current version of our paper does not have this gap (and it has additional results, such as \emph{LOCC leakage-resilience security} for key-fire and unbounded-message-length incompressibility for OSS) however, we include a discussion here for historical reference. 

In \cite{vinodhuang}, the authors claimed the following (about the first version of our paper).
\begin{enumerate}
    \item They claimed that the oracles $\mathcal{D}, \mathcal{D}_0$ that we set to be the signing oracle ($\ora_{\oss.\sign}$) in our incompressible one-shot signature (OSS) scheme construction are not sufficient for signing, and they claimed that thus our claim about existence of $1$-bit incompressible OSS was incorrect.
    
    \item They claimed that incompressible OSS for multi-bit  messages does not follow directly from incompressible OSS for $1$-bit messages.
\end{enumerate}

In the following sections we address both of these points in detail. However, a quick summary is given below:
\begin{itemize}
    \item
    The first point mentioned was a misunderstanding by the authors of \cite{vinodhuang} regarding the prior work of \cite{C:ShmZha25} and our construction. \cite{vinodhuang} claimed that the oracles $\mathcal{D}, \mathcal{D}_0$ are not sufficient for our signing algorithm, then proceeded to change our signing algorithm (they did not ask us for a clarification) by giving additional access to $\mathcal{P}, \mathcal{P}^{-1}$, and finally, claimed that our argument is flawed since incompressibility cannot hold given access to oracles $\mathcal{P}, \mathcal{P}^{-1}$. However, this is incorrect: It is indeed possible to sign using only the oracles $\mathcal{D}, \mathcal{D}_0$ as shown by various works (\cite{STOC:AGKZ20,C:Shmueli22,C:ShmZha25}).
    
    \item
    The second point was correct and indeed pointed to a small gap in the first version of our key-fire proof: We had implicitly assumed incompressible OSS for multi-bit messages follows in a straightforward manner from incompressible OSS for 1-bit messages, thus we had only given a 1-bit incompressible OSS scheme explicitly.
    
    However, first, our quantum fire security proof (as opposed to key-fire, which is a slightly stronger primitive) had already relied on only incompressible OSS for 1-bit messages, and thus this gap did not effect our original quantum fire security proof.
    Second, for proving key-fire, we had relied on incompressible OSS for multi-bit messages only in the last hybrid of the proof. Thus, this was a localized gap and can be fixed it with a small change to make it rely on only 1-bit incompressible OSS. Overall, it did not affect the main ideas and techniques of our paper. We communicated this small fix to them within a day of receiving their email. 
\end{itemize}

More details now follow.

\subsection{Signing Oracle}
In our incompressible OSS scheme, we have four oracles $\mathcal{P}, \mathcal{P}^{-1}, \mathcal{D}, \mathcal{D}_0$. We set $\mathcal{D}$ and $\mathcal{D}_0$ as the signing and signature verification oracles. We proved that in the presence of the signing and verification oracle (that is $\mathcal{D}, \mathcal{D}_0$), our scheme satisfies an incompressibility property.

The authors of \cite{vinodhuang} claimed that $\mathcal{D}, \mathcal{D}_0$ are not sufficient for signing, and they claimed that we must include $\mathcal{P}, \mathcal{P}^{-1}$ to the signing oracle $\ora_{\sign}$. Then they point out that given $\mathcal{P}$, $\mathcal{P}^{-1}$, the incompressibility property is broken. However, indeed the oracles $\mathcal{D}, \mathcal{D}_0$ are sufficient for signing and we do not need to include  $\mathcal{P}, \mathcal{P}^{-1}$. In this version, we added a part that recalls the details of the signing algorithm (which is entirely from the previous work) to prevent any future confusion.

To explain in more detail, in the one-shot signature literature, there are two main/well-known signing algorithms, the so-called Grover-style signing algorithm (from \cite{STOC:AGKZ20}), and the generic signing algorithm from \cite{dalls23}. The confusion from \cite{vinodhuang} might be due to the fact that we said that our signing algorithm is as in \cite{C:ShmZha25}, however \cite{C:ShmZha25} does not explicitly write out its signing algorithm but rather it is implicit that the signing algorithm of \cite{C:ShmZha25} is the Grover-style signing algorithm. In the one-shot signature literature, it is well-known that the Grover style signing algorithm is used when the quantum signature token takes the specific form of a coset state (i.e., a uniform superposition over vectors from some coset), which is the case for \cite{C:ShmZha25}. This algorithm (with small modifications) has been used in many OSS works such as \cite{STOC:AGKZ20} and \cite{C:Shmueli22}, and it is well understood that this is the signing algorithm for coset based one-shot signature schemes like \cite{C:ShmZha25}. Indeed, another follow-up work to \cite{C:ShmZha25} also mentions this Grover-style signing algorithm as the signing algorithm of the \cite{C:ShmZha25} construction: See Section 2.3, Page 11 in \cite{bartusek25} where it is also clear that the only oracle being used for signing is $\mathcal{D}$. Note that \cite{bartusek25} was put on arXiv on October 10th, which is prior to the publication of \cite{vinodhuang}.

Perhaps another point of the confusion from \cite{vinodhuang} is because \cite{C:ShmZha25} essentially constructs both a one-shot signature scheme and a stronger primitive called a non-collapsing collision-resistant hash (non-collapsing CRH). Specifically, \cite{C:ShmZha25} cites a known result by \cite{dalls23}, which says that a non-collapsing CRH generically implies an OSS. However, the context is to claim a stronger primitive (non-collapsing CRH over OSS), and they do not not mean that the specific construction at hand needs to use this generic transformation to get an OSS. Indeed, in the specific case of the construction from \cite{C:ShmZha25}, their construction is both an OSS directly and a non-collapsing CRH, and the well-known Grover-style signing is to be used for signing when the goal is to obtain OSS directly.

\subsection{Multi-bit Incompressible OSS}
We showed the existence of incompressible OSS for 1-bit messages. 1-bit OSS is sufficient for most of our proof, but in the first version of our paper, in the proof of our last hybrid in key-fire security, we had invoked incompressible OSS for multi-bit messages. We had implicitly assumed incompressible OSS for multi-bit messages follow in a straightforward manner from incompressible OSS for 1-bit messages, thus we had only given a 1-bit incompressible OSS scheme explicitly. However in a nice observation, \cite{vinodhuang} point out that parallel repetition of a 1-bit OSS scheme to obtain multi-bit OSS scheme, does not preserve incompressibility.

This was a localized issue, and did not affect the rest of our paper. Moreover, this affected only key-fire: Without any changes, our quantum fire security (as opposed to key-fire, which is a slightly stronger primitive) in the first version of our paper still holds using only 1-bit OSS. For key-fire security, by making a small change to our construction and to the last hybrid,  we made the proof work using only 1-bit incompressible OSS. When the gap was first pointed out to us by the authors of \cite{vinodhuang}, we communicated this fix to them within a day of receiving their email.
\newpage
\fi

\bibliographystyle{alpha}
\end{document}